\documentclass{aa}  

\usepackage{graphicx}
\usepackage{txfonts}
\usepackage{color}

\usepackage{amstext}

\begin{document} 

\titlerunning{Cold molecular gas and PAH emission in Seyfert galaxies}

\title{Cold molecular gas and PAH emission in the nuclear and circumnuclear
  regions of Seyfert galaxies}


\author{A. Alonso-Herrero\inst{\ref{inst1}}
         \and
          M. Pereira-Santaella\inst{\ref{inst2}}
          \and
          D. Rigopoulou\inst{\ref{inst3}}
          \and
          I. Garc\'{\i}a-Bernete\inst{\ref{inst3}}
          \and
          S. Garc\'{\i}a-Burillo\inst{\ref{inst4}}
          \and
          A. J. Dom\'{\i}nguez-Fern\'andez\inst{\ref{inst4}}
          \and
          F. Combes\inst{\ref{inst5}}
          \and
          R. I. Davies\inst{\ref{inst6}}
          \and
          T. D\'{\i}az-Santos\inst{\ref{inst7},\ref{inst8}}
          \and
          D. Esparza-Arredondo\inst{\ref{inst9}}
          \and
          O. Gonz\'alez-Mart\'{\i}n\inst{\ref{inst9}}
          \and
          A. Hern\'an-Caballero\inst{\ref{inst10}}
          \and
          E. K. S. Hicks\inst{\ref{inst11}}
          \and
          S. F. H\"onig\inst{\ref{inst12}}
          \and
          N. A. Levenson\inst{\ref{inst13}}
          \and
          C. Ramos Almeida\inst{\ref{inst14},\ref{inst15}}
          \and
          P. F. Roche\inst{\ref{inst3}}
          \and
          D. Rosario\inst{\ref{inst16}}
          }
   \institute{Centro de Astrobiolog\'{\i}a (CAB, CSIC-INTA), ESAC
     Campus, E-28692 Villanueva de la Ca\~nada, Madrid, Spain\\
              \email{aalonso@cab.inta-csic.es}\label{inst1}
      \and
      Centro de Astrobiolog\'{\i}a (CAB, CSIC-INTA), Carretera de
      Torrej\'on a Ajalvir, 
      E-28880 Torrej\'on de Ardoz, Madrid, Spain\label{inst2}
      \and
      Department of Physics, University of Oxford, Keble Road, Oxford OX1 3RH, UK\label{inst3}   
      \and
      Observatorio de Madrid, OAN-IGN, Alfonso XII, 3, E-28014 Madrid, Spain\label{inst4}
   \and
   LERMA, Obs. de Paris, PSL Research Univ., Coll\`ege de France, CNRS, Sorbonne Univ., UPMC, Paris, France\label{inst5}
   \and
   Max Planck Institut fuer extraterrestrische Physik Postfach 1312, D-85741 Garching bei M\"unchen, Germany\label{inst6}
   \and
    N\'ucleo de Astronom\'{\i}a, Facultad de Ingenier\'{\i}a y Ciencias, Universidad Diego Portales, Ej\'ercito Libertador 441, Santiago, 8320000, Chile\label{inst7}
    \and
    Chinese Academy of Sciences South America Center for Astronomy, National Astronomical Observatories, CAS, Beijing 100101, China\label{inst8}
   \and
   Instituto de Radioastronom\'{\i}a y Astrof\'{\i}sica (IRyA-UNAM), 3-72 (Xangari), 8701, Morelia, Mexico\label{inst9}
   \and
   Centro de Estudios de Física del Cosmos de Arag\'on, Unidad Asociada al CSIC, Plaza San Juan 1, E-44001 Teruel, Spain\label{inst10}
   \and
   Department of Physics \& Astronomy, University of Alaska Anchorage, 99508-4664, USA\label{inst11}
   \and
   Department of Physics \& Astronomy, University of Southampton, Hampshire SO17 1BJ, Southampton, UK\label{inst12}
   \and
   Space Telescope Science Institute, 3700 San Martin Drive, Baltimore, MD 21218, USA\label{inst13}
   \and
   Instituto de Astrof\'{\i}sica de Canarias, Calle vía Láctea, s/n, E-38205 La Laguna, Tenerife, Spain\label{inst14}
   \and
   Departamento de Astrof\'{\i}sica, Universidad de La Laguna, E-38205 La Laguna, Tenerife, Spain\label{inst15}
   \and
   Centre for Extragalactic Astronomy, Durham University, South Road, Durham DH1 3LE, UK\label{inst16}}
   \date{Received:  --; accepted ---}

 
  \abstract
{We investigate the relation between the detection of the $11.3\,\mu$m polycyclic aromatic hydrocarbon (PAH)
  feature in the nuclear ($\sim 24-230\,$pc)
        regions of 22 nearby Seyfert galaxies and the properties of the cold
        molecular gas. For the former we use ground-based
        ($0.3-0.6\arcsec$ resolution) mid-infrared (mid-IR) spectroscopy. The cold molecular gas is
        traced by ALMA and NOEMA high ($0.2-1.1\arcsec$) angular resolution observations
        of  the CO(2-1) transition. Galaxies with a nuclear detection of the $11.3\,\mu$m
        PAH feature contain more cold molecular gas (median $1.6\times 10^7\,M_\odot$) and have higher
        column densities ($N({\rm H}_2) = 2 \times 10^{23}\,{\rm
          cm}^{-2}$) over the regions sampled by the mid-IR slits than those without a detection.
This suggests that molecular gas plays a role in shielding the
PAH molecules in the harsh environments of Seyfert nuclei.
Choosing the PAH molecule
naphthalene as an illustration, we compute its half-life in the nuclear regions of
our sample when exposed to
2.5\,keV hard X-ray photons. 
We estimate shorter  half-lives for naphthalene in nuclei without a $11.3\,\mu$m
 PAH detection than in those with a detection. The {\it Spitzer}/IRS PAH
 ratios on circumnuclear scales ($\sim 4\arcsec \sim 0.25-1.3\,$kpc)
 are in between model predictions for neutral and partly ionized
 PAHs. However, Seyfert galaxies in our sample with the highest nuclear
H$_2$  column densities are not generally closer to the neutral PAH
tracks. This is because in the majority of our sample galaxies,
the CO(2-1) emission in the inner $\sim 4\arcsec$ is  not centrally
peaked and in some galaxies traces circumnuclear sites of strong
star formation activity. Spatially resolved observations with the MIRI medium-resolution spectrograph (MRS) on the James Webb Space Telescope will 
be able to distinguish the effects of an active galactic nucleus (AGN)
and star formation on the
PAH emission in nearby AGN.}  


   \maketitle

%


\begin{table*}
\caption{Properties of the sample.}             
\label{tab:Sample}      
\centering                          
\begin{tabular}{c c c c c c c}        
\hline\hline                 
Galaxy      & Type & Seyfert & Dist  & $1\arcsec$  &log $L$(2-10keV) & Ref\\
            &  & class & (Mpc)& (pc) & (erg s$^{-1}$) &\\
\hline
IC~4518W    &  Sc pec         &2 & 67.9      &319  &42.58    &1\\
Mrk~1066    & (R)SB0\^\,+(s) &2 & 47.2      &224  &42.92    &2\\
NGC~1068    & (R)SA(rs)b     &2 &14.0      &67   &42.80    &1\\
NGC~1320    & Sa? edge-on    &2 &34.5      &164  &42.7-42.3    &3\\
NGC~1365    & SB(s)b         &1.8&18.3      &102  &42.09    &1\\
NGC~1386    & SB0\^\,+(s)    &1¡&18.0      &102  &40.9-41.7   &3\\
NGC~1808    & (R)SAB(s)a     &2&14.0      &67   &40.40    &3\\
NGC~2110    & SAB0\^\,-      &1¡&33.1      &158  &42.68    &1\\
NGC~2273    & SB(r)a?        &2&25.8      &124  &42.70    &2\\
NGC~2992    & Sa pec  &1¡&36.6      &174  &42.09    &1\\
NGC~3081    &(R)SAB0/a(r) &2 &37.7      &179  &42.81    &1\\
NGC~3227    &SAB(s)a pec &1.5&20.4      &98   &42.27    &1\\
NGC~4253    & (R')SB(s)a?  &1.5&57.6      &272  &42.74    &1\\
NGC~4388    & SA(s)b? edge-on &1.9&18.1      &102  &42.45    &1\\
NGC~5135    & SB(s)ab &2&60.9      &287  &43.00    &4\\
NGC~5643    & SAB(rs)c  &2&19.2      &92   &42.53    &1\\
NGC~7130    & Sa pec &2 & 63.6      &299  &42.05    &1\\
NGC~7172    & Sa pec edge-on &2&32.1      &153  &42.62    &1\\
NGC~7213    & SA(s)a? &1&21.2      &102  &41.81    &1\\
NGC~7465    & (R')SB0\^\,0?(s) &2&21.9      &105  &41.74    &1\\
NGC~7469    & (R')SAB(rs)a &1.5 &62.6      &295  &43.09    &1\\
NGC~7582    & (R')SB(s)ab   &1¡ &18.3      &88   &43.31    &1\\
\hline
\end{tabular}
\tablefoot{Morphological types are from the Revised Catalog 3 (RC3) Catalogue \citep{deVaucouleurs1991}.
  The Seyfert class is taken from \cite{VeronCetty2006}
  except for
   NGC~1808, NGC~3081, and
  NGC~7213, which are taken from \cite{VeronCetty1985}, \cite{Phillips1983},
  and 
  \cite{Phillips1979}, respectively. The ``Sy1¡'' classification means
  that broad hydrogen recombination lines have been detected in the
  near-IR. Luminosity distances and angular
    scales are taken from the NASA/IPAC Extragalactic Database (NED) for
    $H_0=73\,{\rm km\,s}^{-1}\,{\rm Mpc}^{-1}$,
    $\Omega_M=0.27$ and $\Omega_V=0.30$.
NGC~1386 and NGC~1365 are in the Fornax Cluster.
NGC~4388 is in the Virgo Cluster. The hard X-ray luminosities are intrinsic
(i.e., corrected for absorption), and the last column indicates the reference.}\\
\tablebib{1. The BAT AGN Spectroscopic Survey (BASS) http://www.bass-survey.com/, \cite{Koss2017},
  \cite{Ricci2017}, 2. \cite{Marinucci2012}
3. \cite{Brightman2011}, 4. \cite{PereiraSantaella2011}.}

\end{table*}

   \section{Introduction}

The balance between  gas inflows and outflows in the nuclear regions
of active galaxies is fundamental for understanding both
the growth of supermassive black holes \citep[see the review by][]{Alexander2012}
and the maintenance of the
obscuring molecular dusty torus \citep{Elitzur2006, RamosAlmeida2017,
  Hoenig2019}.
Moreover, theoretical models and numerical simulations  predict that
the accumulation of molecular gas due to the inflow process will also
trigger star formation (SF) 
activity in the nuclear or circumnuclear regions of active galaxies
\citep[see, e.g.,][]{Kawakatu2008, Vollmer2008, Hopkins2010}.
Understanding the SF properties, including the timescales for
gas consumption and SF-driven
feedback, has important implications for the final amount of gas
available for fuelling the active galactic nucleus (AGN) and perhaps for contributing to momentum loss in the nuclear regions \citep[see, e.g.,][and references therein]{Izumi2016}.
However, measuring the SF
properties in
the nuclear regions of active galaxies is particularly challenging because
many of the traditional  SF indicators are contaminated by emission from
the AGN.

The emission from polycyclic aromatic hydrocarbon (PAH) features has been
proposed
as a good indicator of the star formation rate (SFR) over timescales of
up to a few hundred million years \citep{Peeters2004, 
  Shipley2016, Xie2019}.  The majority of
nearby Seyfert galaxies show PAH emission especially on kiloparsec scales,
generally with smaller equivalent
widths (EW) than star-forming galaxies 
\citep[see, e.g.,][]{Roche1991, Clavel2000, Wu2009,  Gallimore2010,
  DiamondStanic2012}.
On nuclear scales (tens to hundreds of parsecs), the $11.3\,\mu$m
PAH feature is detected in a significant number of Seyfert galaxies,
but with even smaller EW \citep{Hoenig2010, GonzalezMartin2013, AlonsoHerrero2014, AlonsoHerrero2016, RamosAlmeida2014, Jensen2017,
  EsparzaArredondo2018}. It is not clear whether the
  reduced EW are due  to destruction of
  PAH molecules, increased continuum contribution from the AGN, and/or
  changes in  PAH excitation conditions.
In local Seyfert galaxies, the $11.3\,\mu$m PAH
emission is well correlated  
with the [Ne\,{\sc ii}]$12.8\,\mu$m emission,  indicating
that it may potentially be a good SFR indicator for AGN
\citep{DiamondStanic2010,
  DiamondStanic2012, Esquej2014}. On the other hand,
\cite{Jensen2017} proposed that PAH molecules might also be excited
by AGN photons, making it necessary to evaluate this component before
using the PAH emission as an SFR indicator.

Theoretical works argued that the PAH molecules may be destroyed in
the immediate vicinity of the AGN due to dust evaporation by
X-ray irradiation \citep{Voit1991, Siebenmorgen2004}. However, 
\cite{Voit1992} postulated that when PAH emission is detected in the
nuclear or circumnuclear regions of AGN, the PAH molecules must be shielded
by high column densities of X-ray absorbing gas. Recently, \cite{Monfredini2019}
presented  experimental work to evaluate the photoionization and
photodissociation properties of some simple (number of carbon atoms between
10 and 16) PAH molecules when exposed
to X-ray photons. They demonstrated that this process results in dissociation of the
molecules and  enrichment of multiply
charged ions caused by X-ray photoselection.
 They also
 estimated relatively short half-lives for these PAH molecules in a few AGN
 when compared to the PAH injection times
in the interstellar medium (ISM) by evolved stars.

\begin{table*}
\caption{Ground-based mid-IR spectroscopic observations.}             
\label{tab:midIR}      
\centering                          
\begin{tabular}{c c c c c c c }        
\hline\hline                 
Galaxy   &Instrument & Slit width &PA$_{\rm slit}$ &  PAH det  & $r_{\rm AGN}$& Ref \\  
         &            & ($\arcsec$)     & (deg) & & (pc)\\
\hline
IC~4518W   &T-ReCS     &0.72       &5        &no       &115   &1\\  
Mrk~1066   &CanariCam  &0.52       &315      &yes      &58    &2\\   
NGC~1068  &Michelle   &0.36       &0        &yes      &12    &3\\  
NGC~1320  &CanariCam  &0.52       &315      &no       &43    &2\\   
NGC~1365  &T-ReCS     &0.35       &15       &no       &18    &4, 5\\  
NGC~1386  &T-ReCS     &0.31       &0        &no       &16    &4, 6\\  
NGC~1808  &T-ReCS     &0.35       &45       &yes      &12    &4, 6, 7\\  
NGC~2110  &VISIR      &0.75       &55       &no       &59    &8\\   
NGC~2273  &CanariCam  &0.52       &290      &yes      &32    &2\\   
NGC~2992  &CanariCam  &0.52       &30       &no       &45    &2, 9\\    
NGC~3081  &T-ReCS     &0.65       &0        &no       &58    &4, 6\\   
NGC~3227  &CanariCam  &0.52       &0        &yes      &25    &2\\   
NGC~4253  &CanariCam  &0.52       &285      &yes      &71    &2\\   
NGC~4388  &CanariCam  &0.52       &90       &no       &27    &2\\    
NGC~5135  &T-ReCS     &0.72       &30       &yes       &103   &1, 4\\  
NGC~5643  &T-ReCS     &0.35       &80       &yes      &16    &6, 10\\  
NGC~7130  &T-ReCS     &0.72       &348      &yes      &108   &1\\    
NGC~7172  &T-ReCS     &0.35       &90       &no       &27    &4, 6\\  
NGC~7213  &VISIR      &0.75       &300      &no       &38    &8\\  
NGC~7465  &CanariCam  &0.52       &330      &yes      &27    &2\\   
NGC~7469  &VISIR      &0.75       &80       &yes      &111   &8\\  
NGC~7582  &T-ReCS     &0.70       &0        &yes      &31    &4, 6\\

\hline
\end{tabular}
\tablefoot{PA$_{\rm slit}$ indicates the position angle of the slit measured
  from the north counterclock rotation.
  NGC~7172 was observed on several nights, but the majority of
  the observations were taken with a slit PA=90deg. ``PAH det''
  indicates whether
  the $11.3\,\mu$m PAH feature was detected in the nuclear spectrum.
  The distance from
  the AGN  $r_{\rm AGN}$ for the PAH detection (or nondetection)
  is computed as half the slit width. The last column provides
the reference of the published mid-IR spectrum for each galaxy.}\\
\tablebib{ 1. \cite{DiazSantos2010}, 2. \cite{AlonsoHerrero2016},
  3. \cite{Mason2006}, 
  4. \cite{Esquej2014}, 5. \cite{AlonsoHerrero2012}, 6. \cite{GonzalezMartin2013},
  7. \cite{Sales2013}, 8. \cite{Hoenig2010}, 9. \cite{GarciaBernete2015}, 
  10. \cite{EsparzaArredondo2018}.}

\end{table*}


The main goal of this paper is to investigate the role of cold molecular
gas in shielding PAH molecules from the AGN radiation field. To do so,
we study a sample of nearby Seyfert galaxies. 
  The sample (see Table~\ref{tab:Sample})
is based on AGN with previously published ground-based
  mid-infrared (mid-IR) spectra
  that would cover the $11.3\,\mu$m PAH feature.
We focus on observations obtained with 8-10\,m class telescopes, so that the
slit widths ($0.3-0.7$\arcsec, see Table~\ref{tab:midIR})
cover the nuclear regions  on physical scales between 22 and
230\,pc. We also use  archival {\it Spitzer} observations to study
the PAH emission from the circumnuclear regions.
To trace the cold molecular gas, we use high ($\simeq 0.2-1.1\arcsec$)
angular resolution observations of
the CO(2-1) transition taken with both the NOrthern
Extended Millimeter Array (NOEMA) in the northern hemisphere
and the Atacama Large Millimeter Array (ALMA) in the southern hemisphere. The high angular
resolutions provided by ALMA and NOEMA allow us to probe the molecular
gas properties on nuclear and circumnuclear scales in Seyfert galaxies.

The paper is organized as follows. Section~\ref{sec:observations} presents
the observations and data reduction. Section~\ref{sec:COmeasurements} describes
the analysis of the  CO(2-1) observations. In
Section~\ref{sec:PAHdet} we investigate the dependence of
the nuclear $11.3\,\mu$m PAH detection with 
the molecular gas and AGN. In Section~\ref{sec:circumnuclear} we
analyze the circumnuclear PAH emission in our sample of Seyfert galaxies.
Section~\ref{sec:discussion} summarizes the main results and briefly discusses future prospects with the James Webb Space Telescope ({\it JWST}).

\section{Observations and data reduction}\label{sec:observations}
\subsection{CO(2-1) and continuum 1.3mm observations}

The NOEMA observations and data reduction for the galaxies
in our sample (see Table~\ref{tab:CO21obs}) are described in detail in  
\cite{DominguezFernandez2020}.
The majority of the ALMA observations used in this work
are either from our own programs or are taken from the ALMA archive.
Of the remaining galaxies in our sample, NGC~1068 and NGC~2110 have
published ALMA CO(2-1) observations
\citep[see][respectively]{GarciaBurillo2019, Rosario2019}. 
For NGC~1320 and NGC~2992, which belong to the $12\,\mu$m ALMA
program (PI: M. Malkan), the nuclear CO(2-1) fluxes were provided to us
prior to publication (see Section~\ref{sec:COmeasurements}).

We downloaded the  band 6 CO(2-1) and adjacent continuum
raw data from the ALMA archive.
For galaxies observed by multiple programs, we selected the observations
with the highest angular resolution. The raw data were then calibrated
using the standard pipeline calibration within the Common Astronomy
Software Applications (CASA) v5.1
\citep[][]{McMullin2007}. This calibration includes
the flagging of the data, system
temperature calibration, bandpass calibration, and phase and amplitude
calibrations. Then, we subtracted the continuum using the line-free
channels in the uv visibility data and used the CASA {\sc clean} task to
generate the CO(2-1) data cubes. We used a Briggs weighting
\citep{Briggs1995}, with a robustness parameter of $b = 0.5$.
For the continuum images, we combined the line-free channels in all the observed
spectral windows of each observation. We applied
  a primary beam correction, although the effect on the regions
covered by the nuclear slits is negligible. For simplicity, we refer
to the wavelength of the band 6 continuum images as 1.3\,mm.
We list the resulting synthesized beam sizes and position angles
(PA$_{\rm beam}$) for the
CO(2-1) and 1.3\,mm continuum maps in Tables~\ref{tab:CO21obs} and
\ref{tab:1p3mm}, respectively.

From the fully reduced ALMA datacubes, we produced maps of
the integrated
CO(2-1) intensity as the zeroth-order moment using
pixels at all frequencies with detections  $>4\sigma$. For galaxies with
fainter emission (NGC~1365, NGC~3081, and NGC~7213),
we also generated CO(2-1) maps with a lower detection threshold ($2\sigma$), 
  which we used for the aperture photometry (see Section~\ref{sec:COmeasurements}).
We used the NOEMA CO(2-1) maps generated for detections $> 3\sigma$ from
\cite{DominguezFernandez2020}.
Figure~\ref{fig:CO21maps} shows the maps of the
integrated CO(2-1) line emission for
the galaxies analyzed in this work
(see Section~\ref{sec:COmeasurements}). We chose a field of view (FoV)
of $4''\times 4''$ to
show the approximate region covered by the extracted
the InfraRed Spectrograph \citep[IRS, ][]{Houck2004}
 short-low (SL)
spectra  where the brightest mid-IR PAH features are
(see Section~\ref{subsec:IRSobs}).

\begin{figure*}[!ht]
  \centering
  \includegraphics[width=6cm]{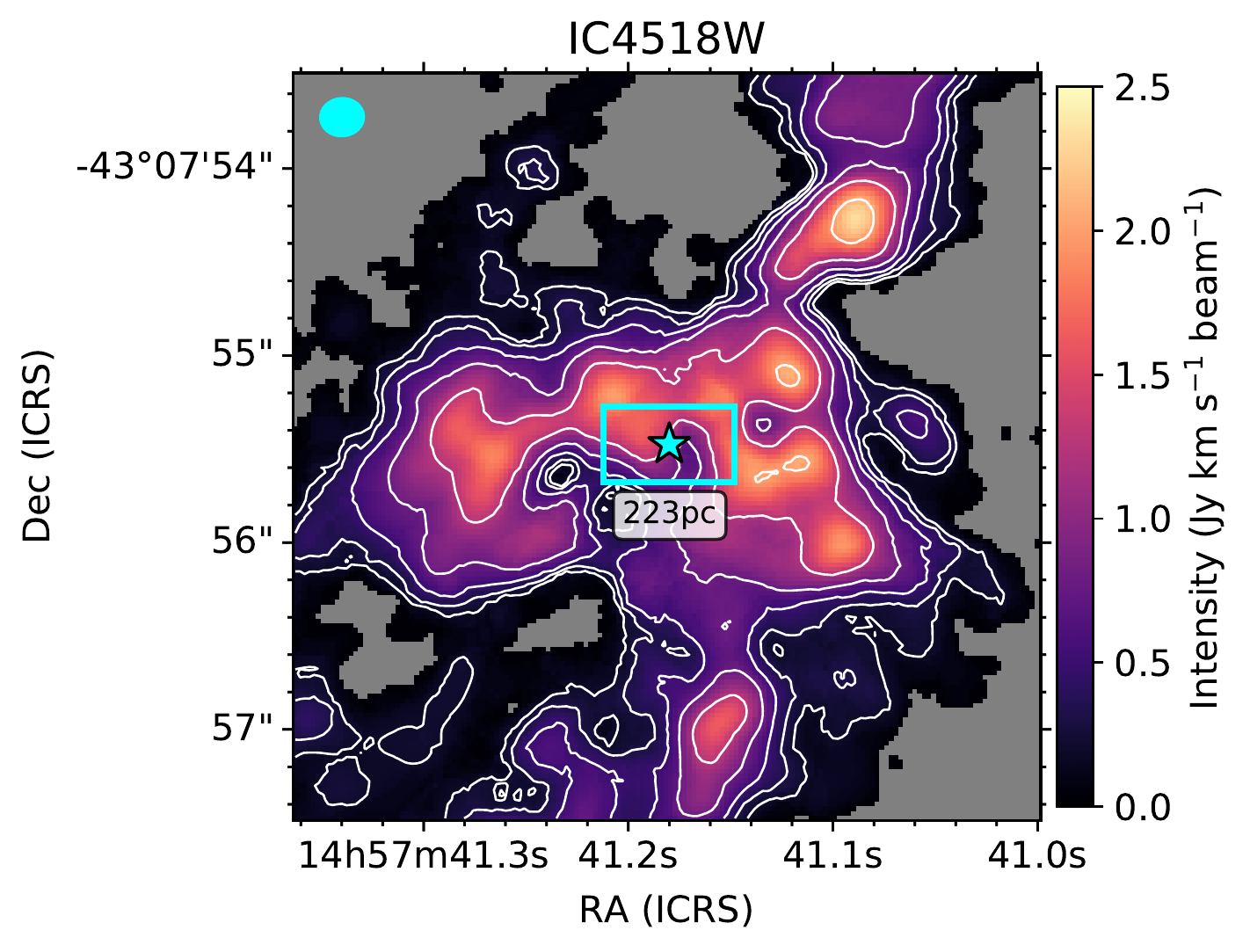}
  \includegraphics[width=6cm]{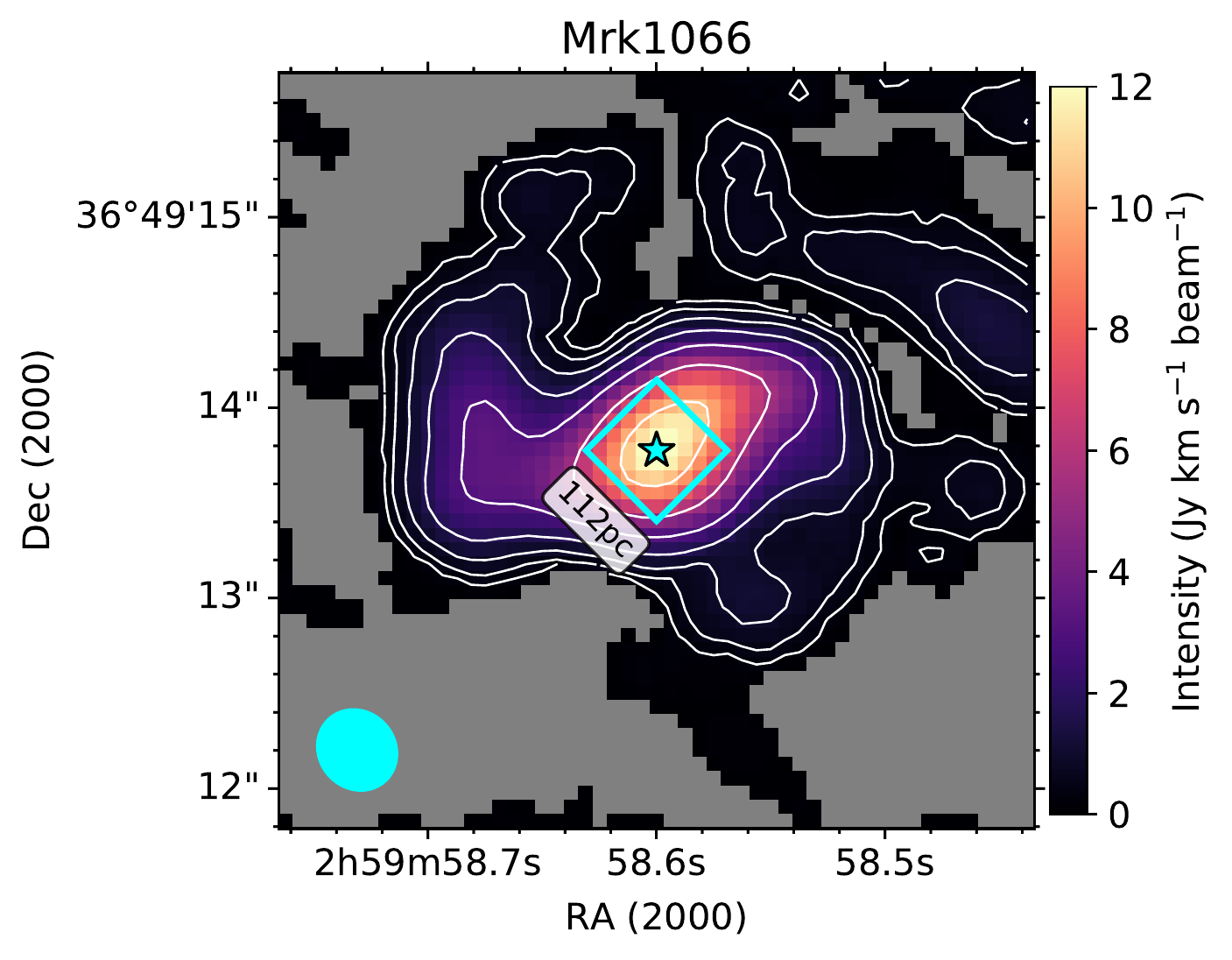}
  \includegraphics[width=6cm]{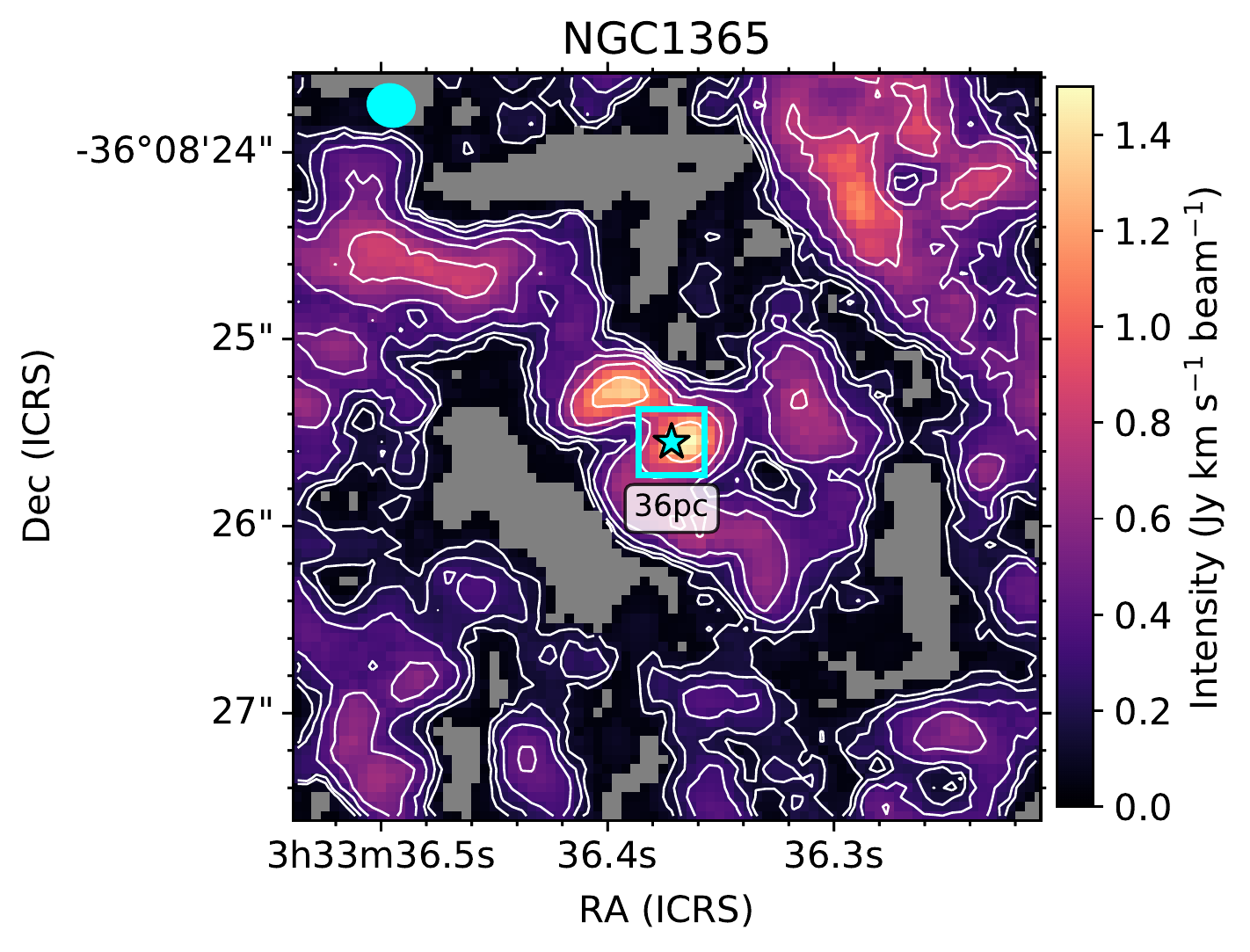}

  \includegraphics[width=6cm]{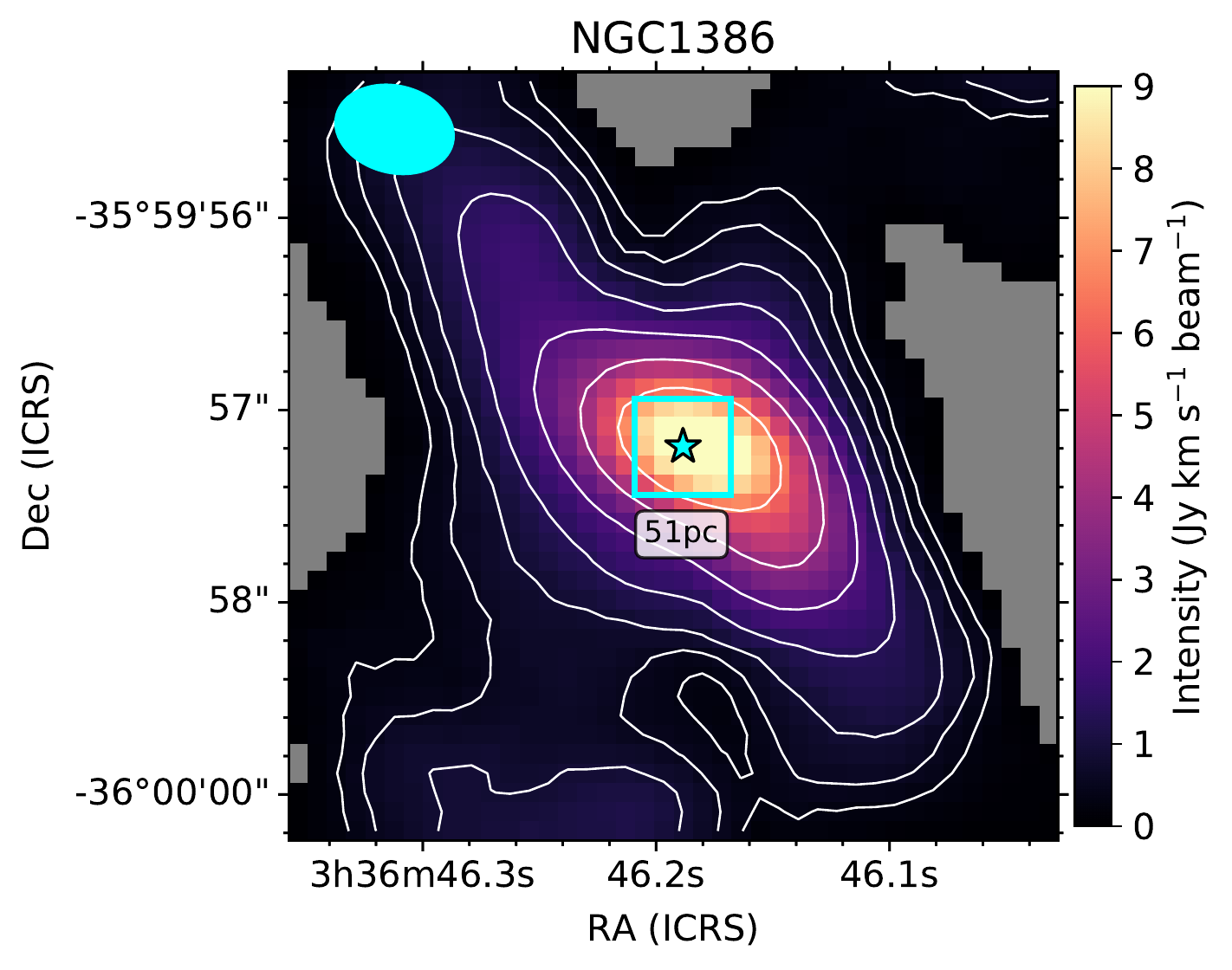}
  \includegraphics[width=6cm]{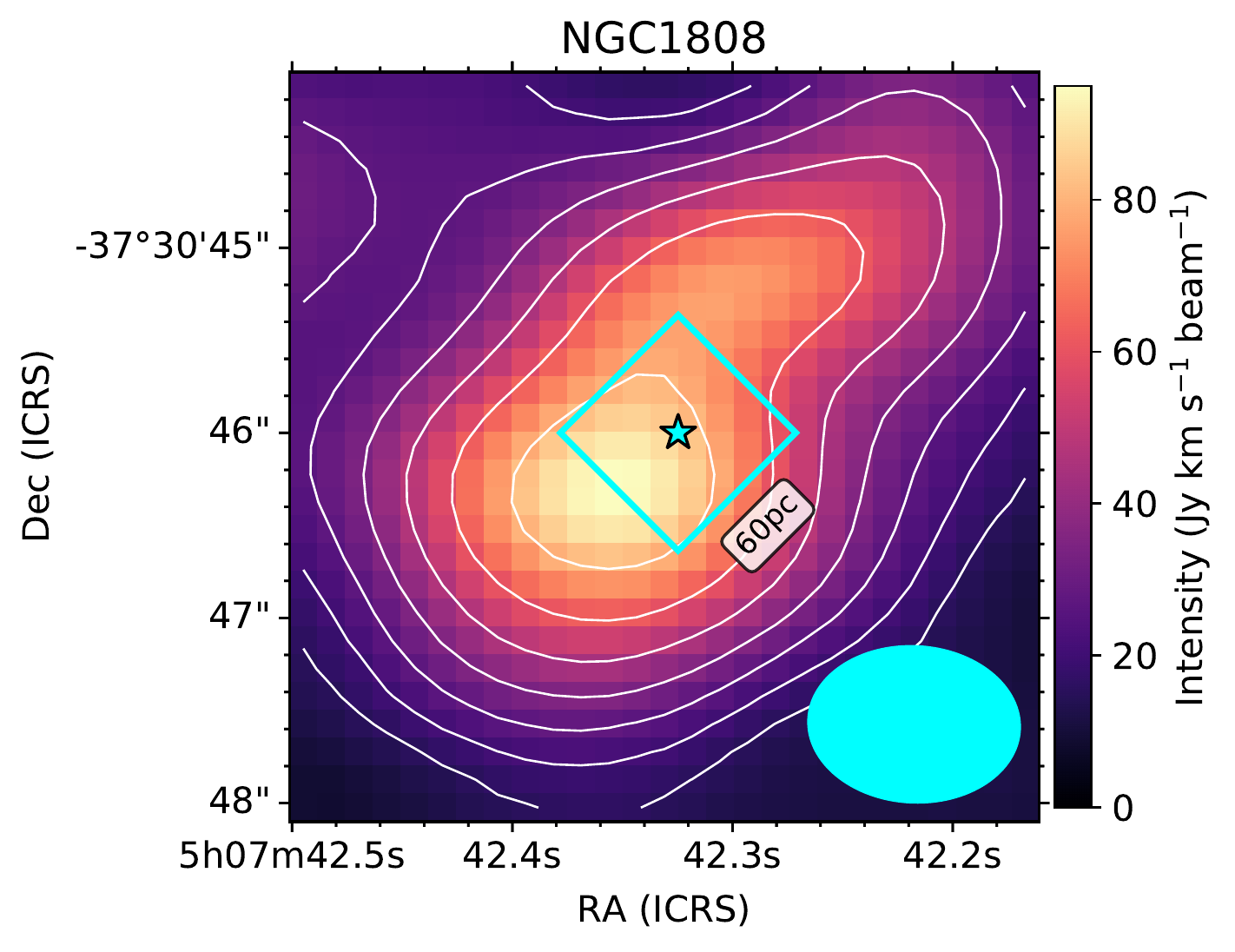}
  \includegraphics[width=6cm]{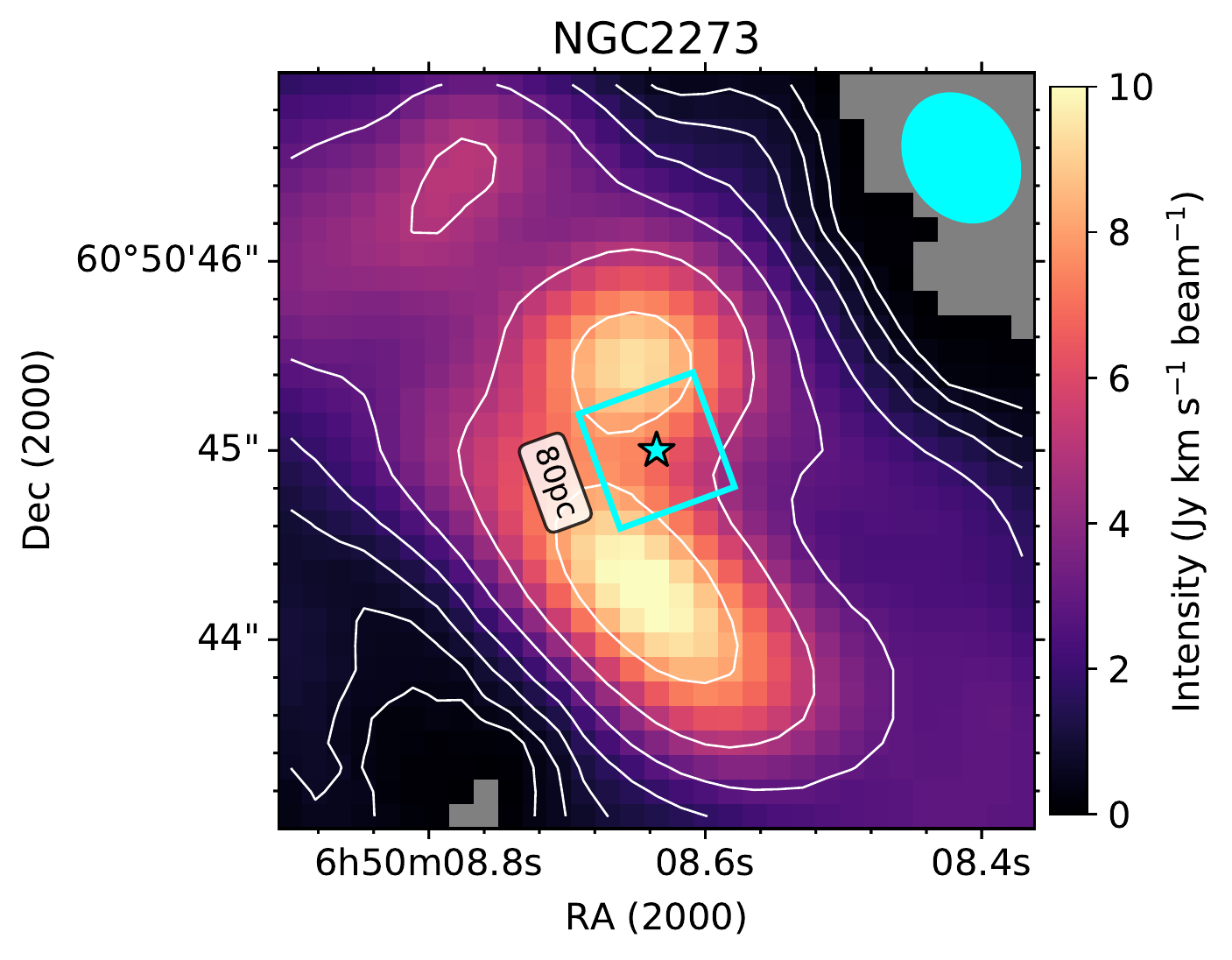}

  \includegraphics[width=6cm]{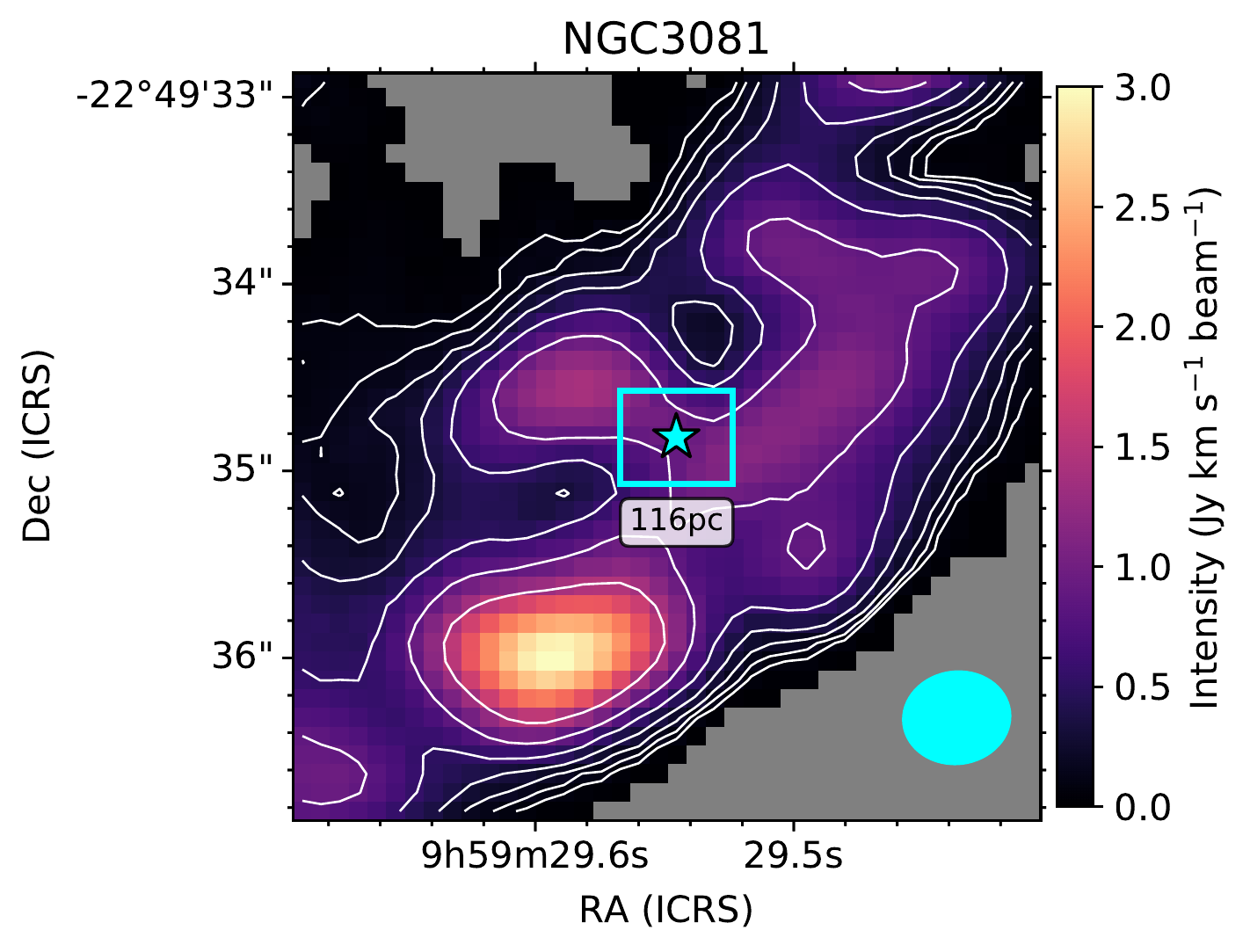}
  \includegraphics[width=6cm]{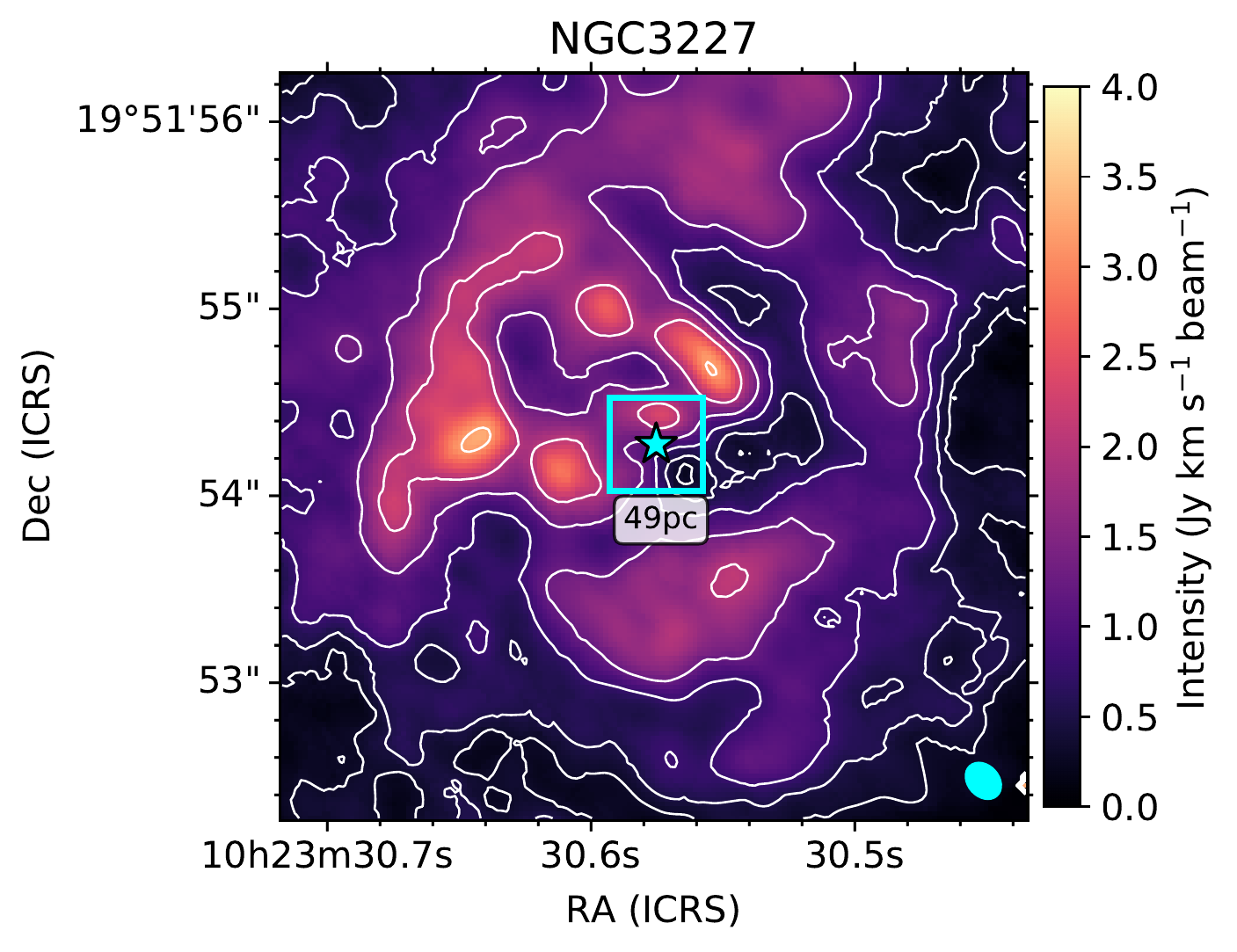}
  \includegraphics[width=6cm]{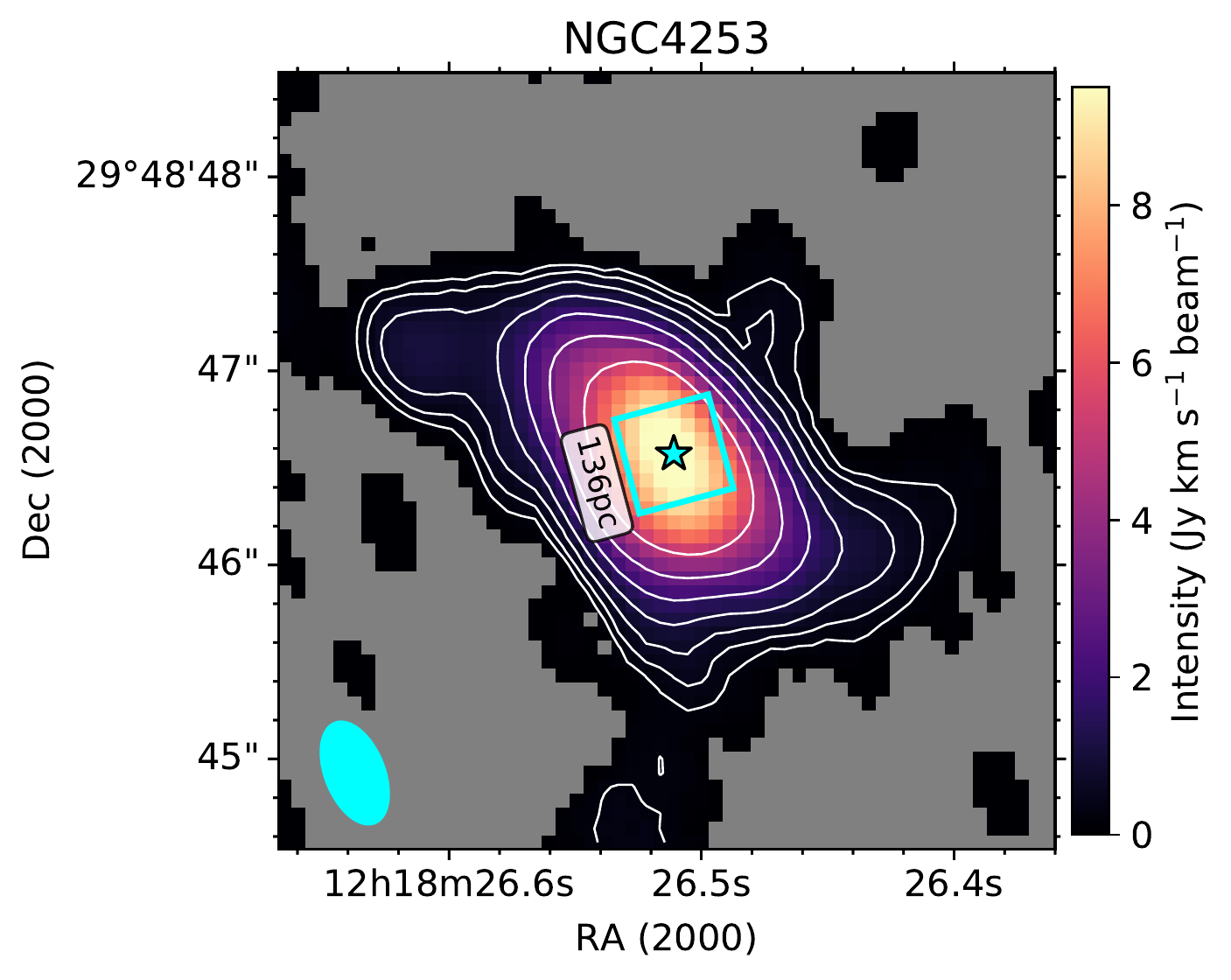}

  \includegraphics[width=6cm]{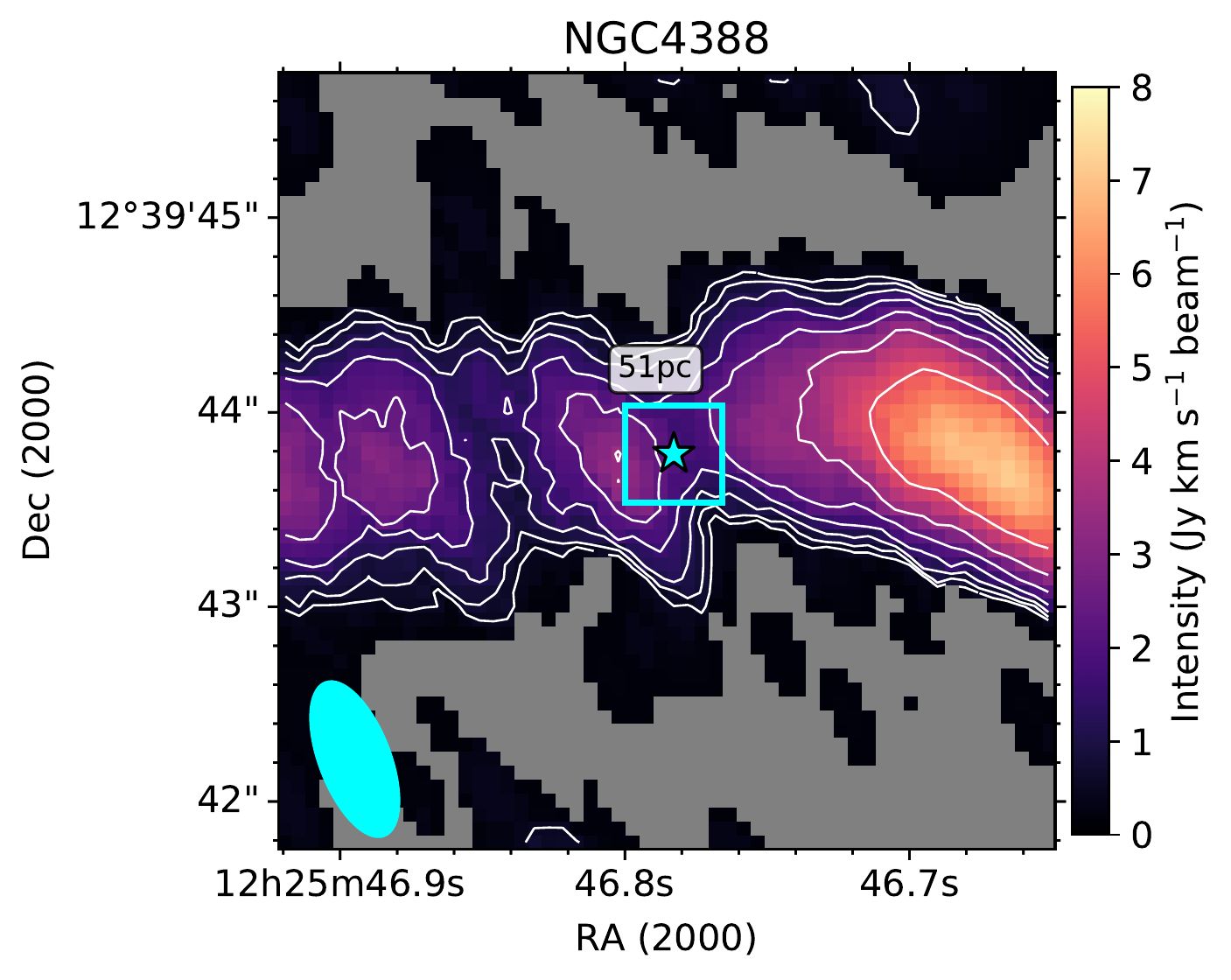}
  \includegraphics[width=6cm]{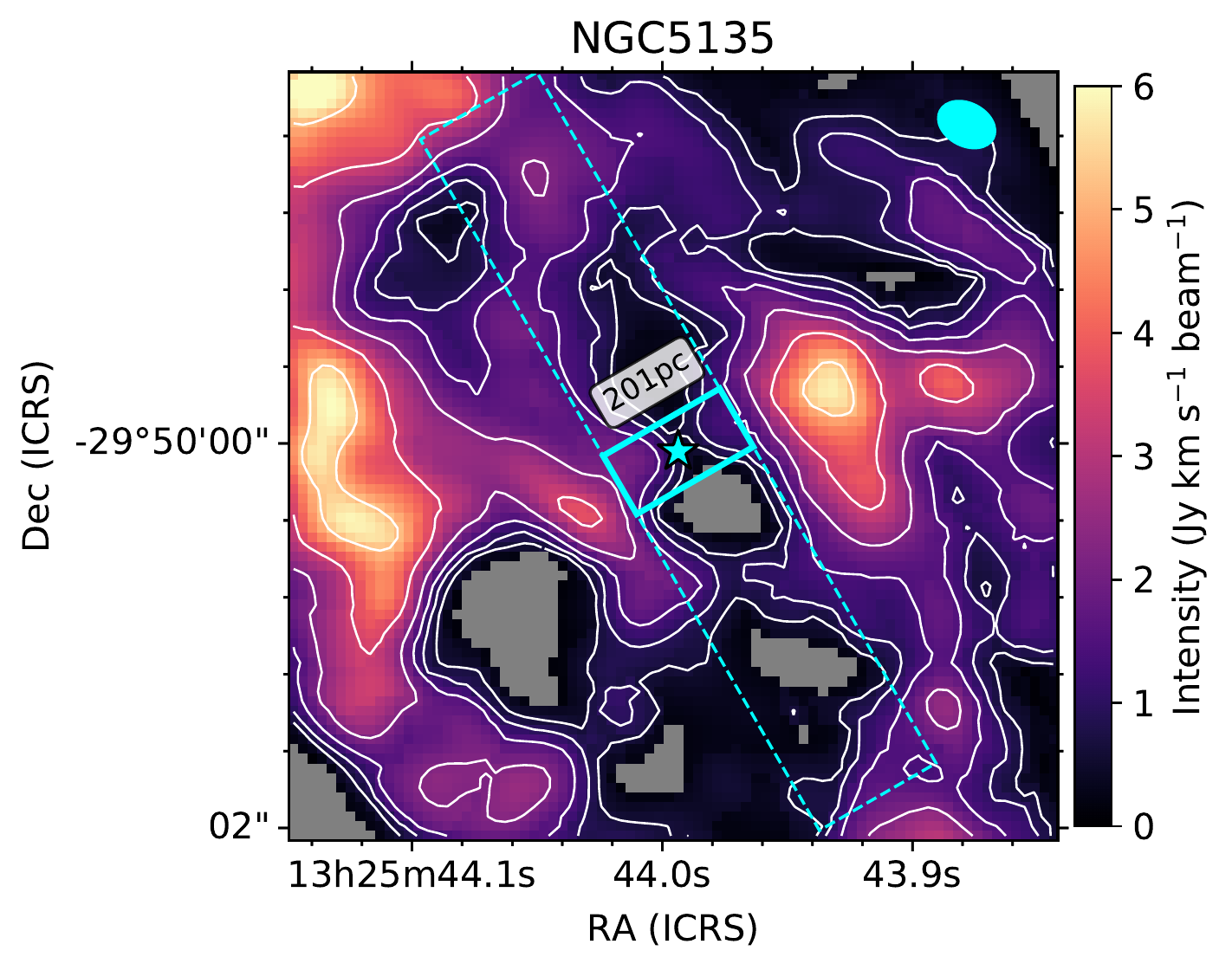}
  \includegraphics[width=6cm]{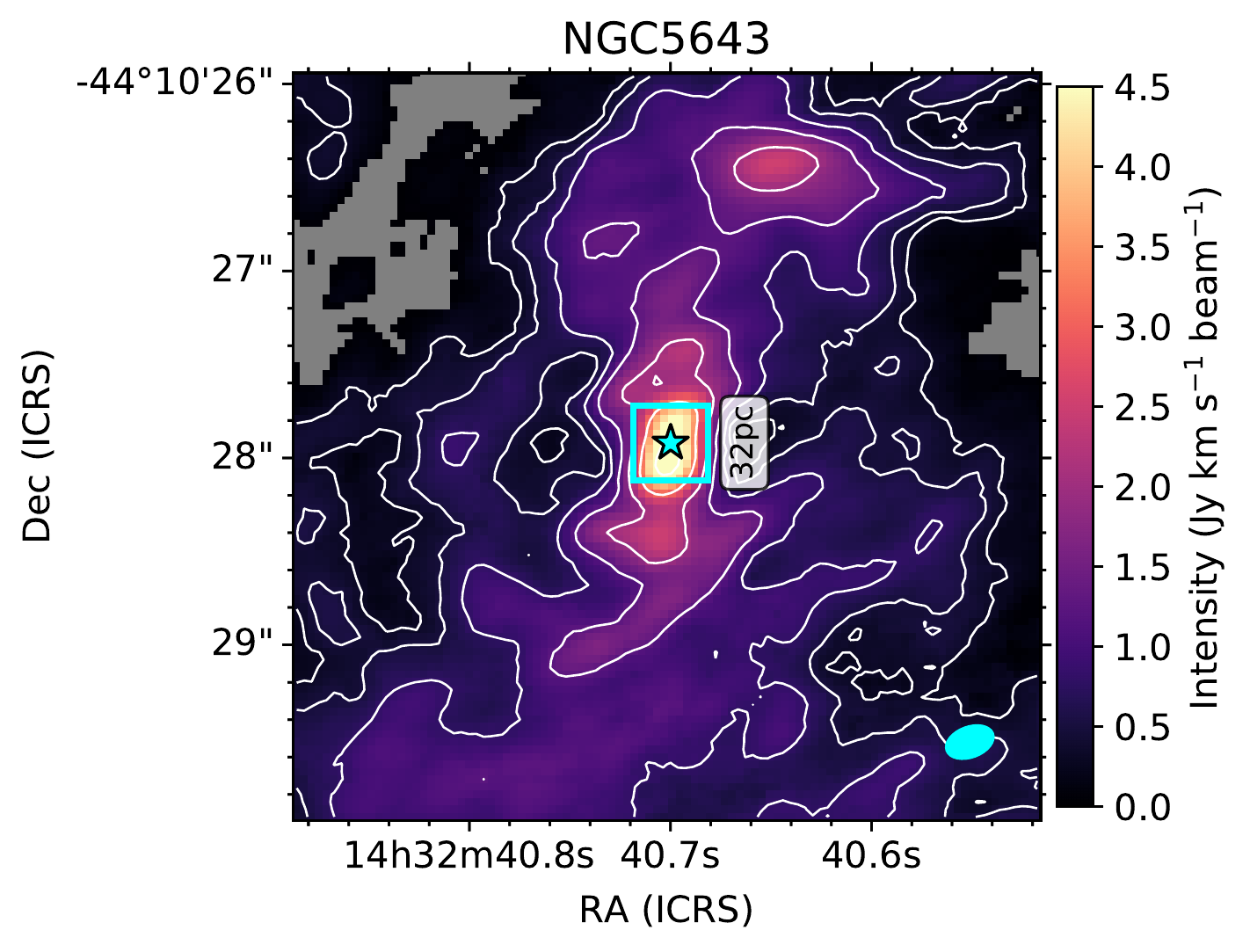}

  \caption{Maps of the integrated CO(2-1) emission (in a linear scale in units of
    Jy km s$^{-1}$ beam$^{-1}$ marked with the color bar on the right-hand side
    of each panel)
    showing  the central $4'' \times 4''$
    regions that are approximately covered by the {\it Spitzer}/IRS SL
    observations. We do not plot the actual orientation of
    the IRS slits. The
    NOEMA CO(2-1) maps (Mrk~1066, NGC~2273, NGC~4253, NGC~4388, and NGC~7465)
    were constructed using a $3\sigma$ detection threshold \citep[see][]{DominguezFernandez2020}, whereas the
    ALMA CO(2-1) maps with 
    $4\sigma$, 
    and $2\sigma$ detection thresholds (see
    text for details).
    The pixels with no detected CO(2-1) emission below the defined
    thresholds are shown in
    gray. The contours are also shown on a linear scale. The filled star
marks the location of the 1.3\,mm continuum peak assumed to be the
    AGN location (see Section~\ref{subsec:AGNpos} and Table~\ref{tab:1p3mm} for
    full details). The filled cyan ellipses
    display the synthesized  beam (size and PA) of the observations for each galaxy
    (see Table~\ref{tab:CO21obs}).
    For each galaxy, the cyan square is the aperture used to extract
    the nuclear CO(2-1) fluxes and simulates the mid-IR nuclear slit. The distance from
    the AGN value $r_{\rm AGN}$ is computed as  half of the
    ground-based slit 
    width. The size of
    the latter is also indicated in parsecs for each galaxy.
    For NGC~5135 and NGC~7582, the dashed lines are the slit orientation to
  indicate the direction of the surface brightness profile extraction.}
              \label{fig:CO21maps}%
    \end{figure*}

  \begin{figure*}
   \setcounter{figure}{0} 
   \centering
  \includegraphics[width=6cm]{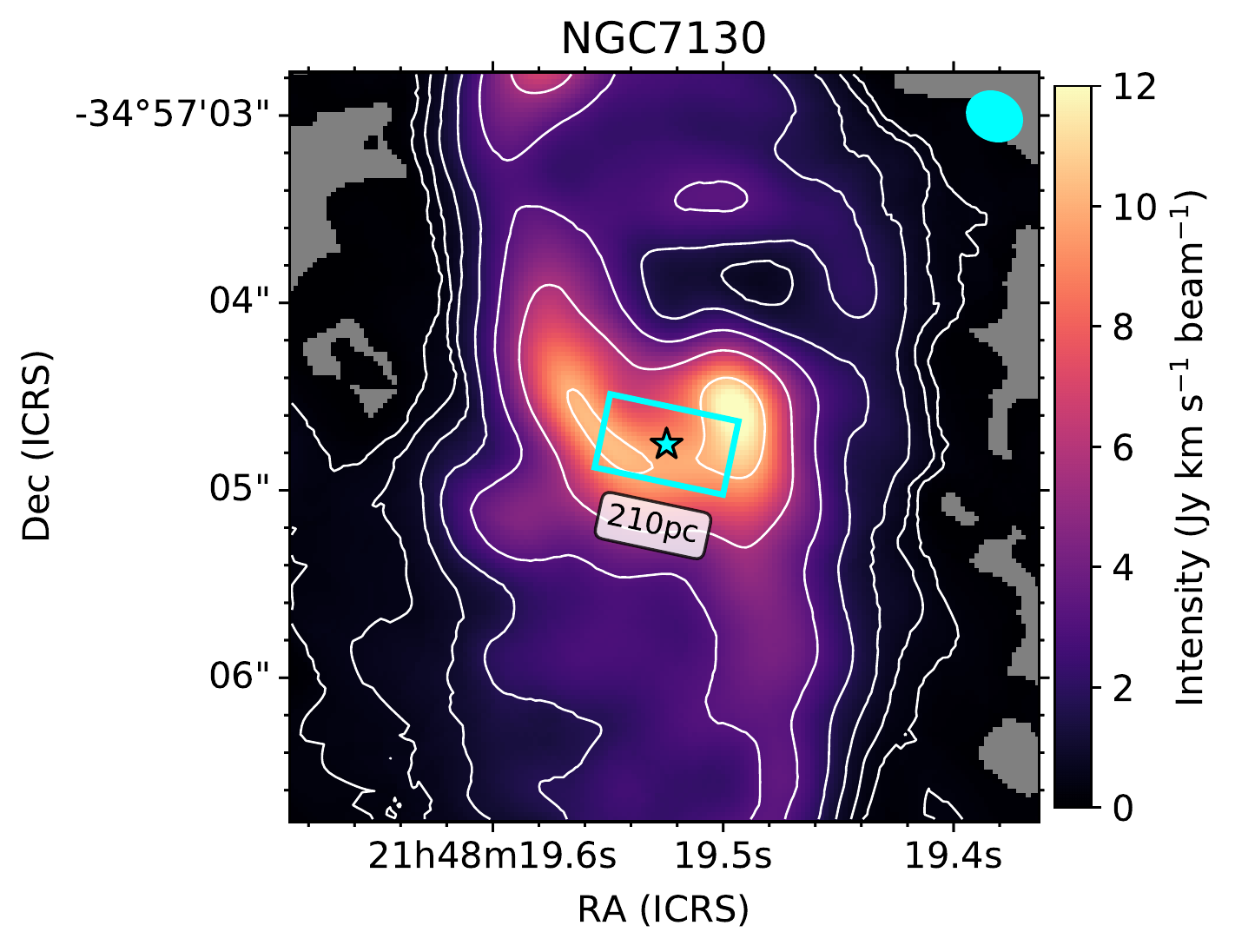}
  \includegraphics[width=6cm]{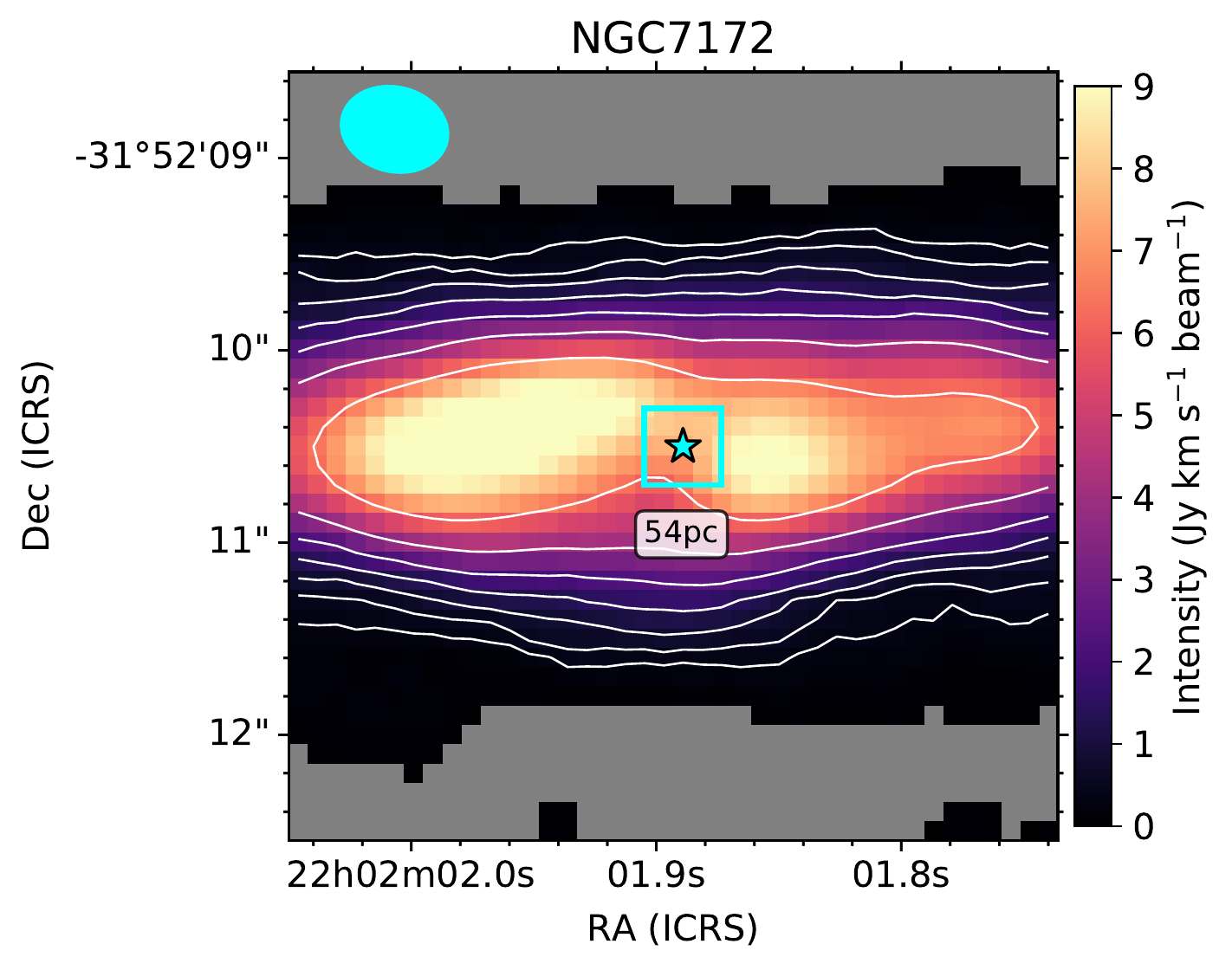}
  \includegraphics[width=6cm]{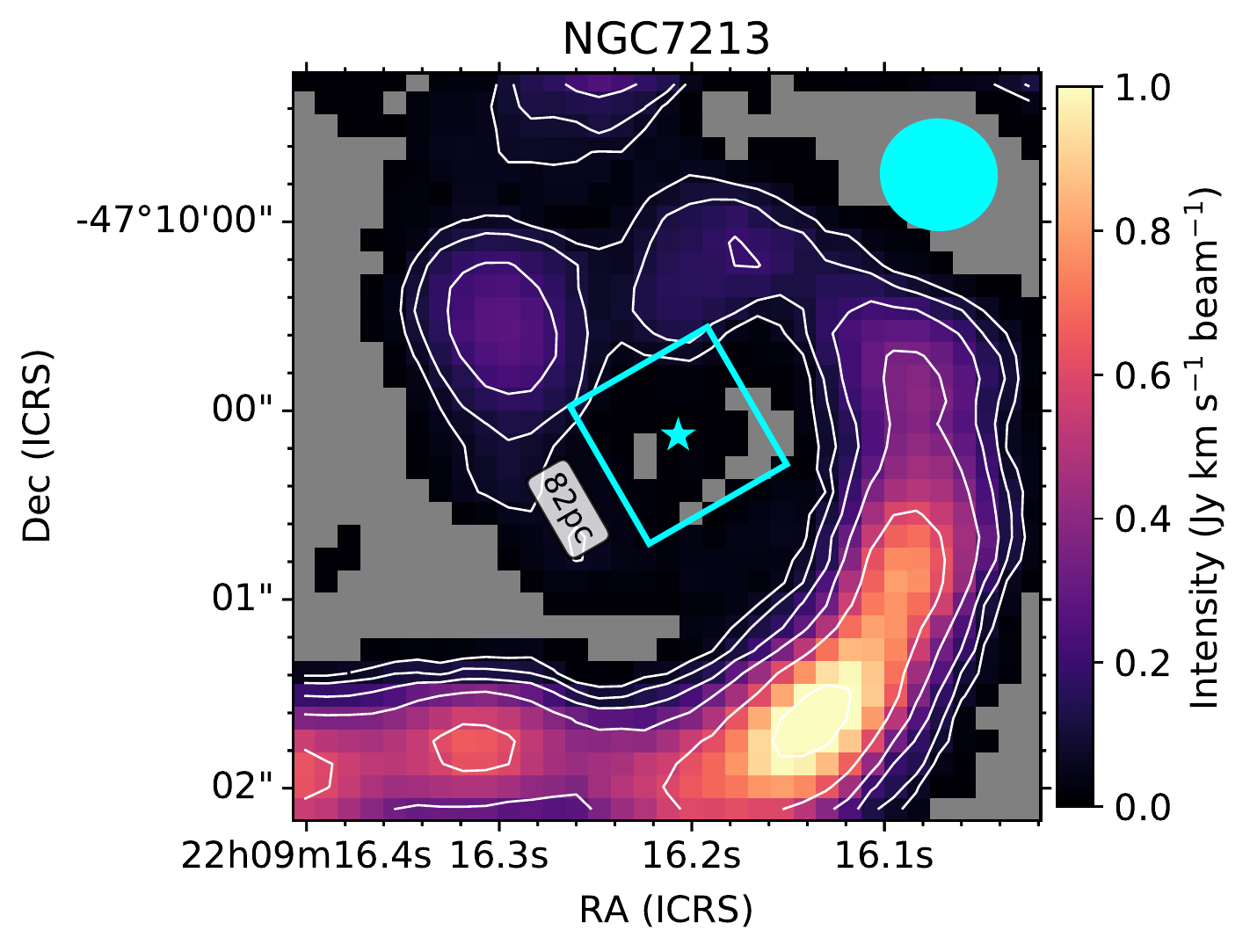}

  \includegraphics[width=6cm]{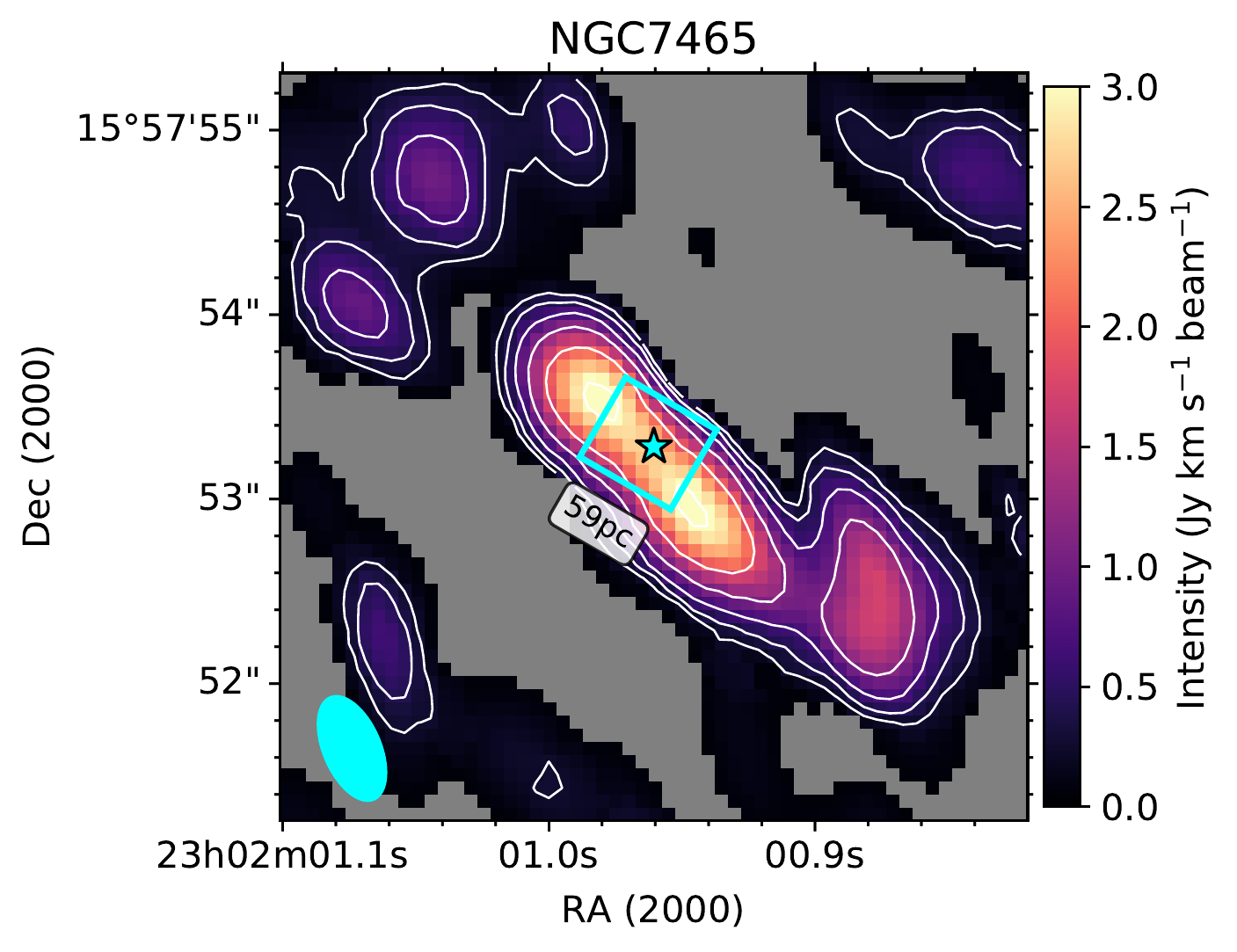}
  \includegraphics[width=6cm]{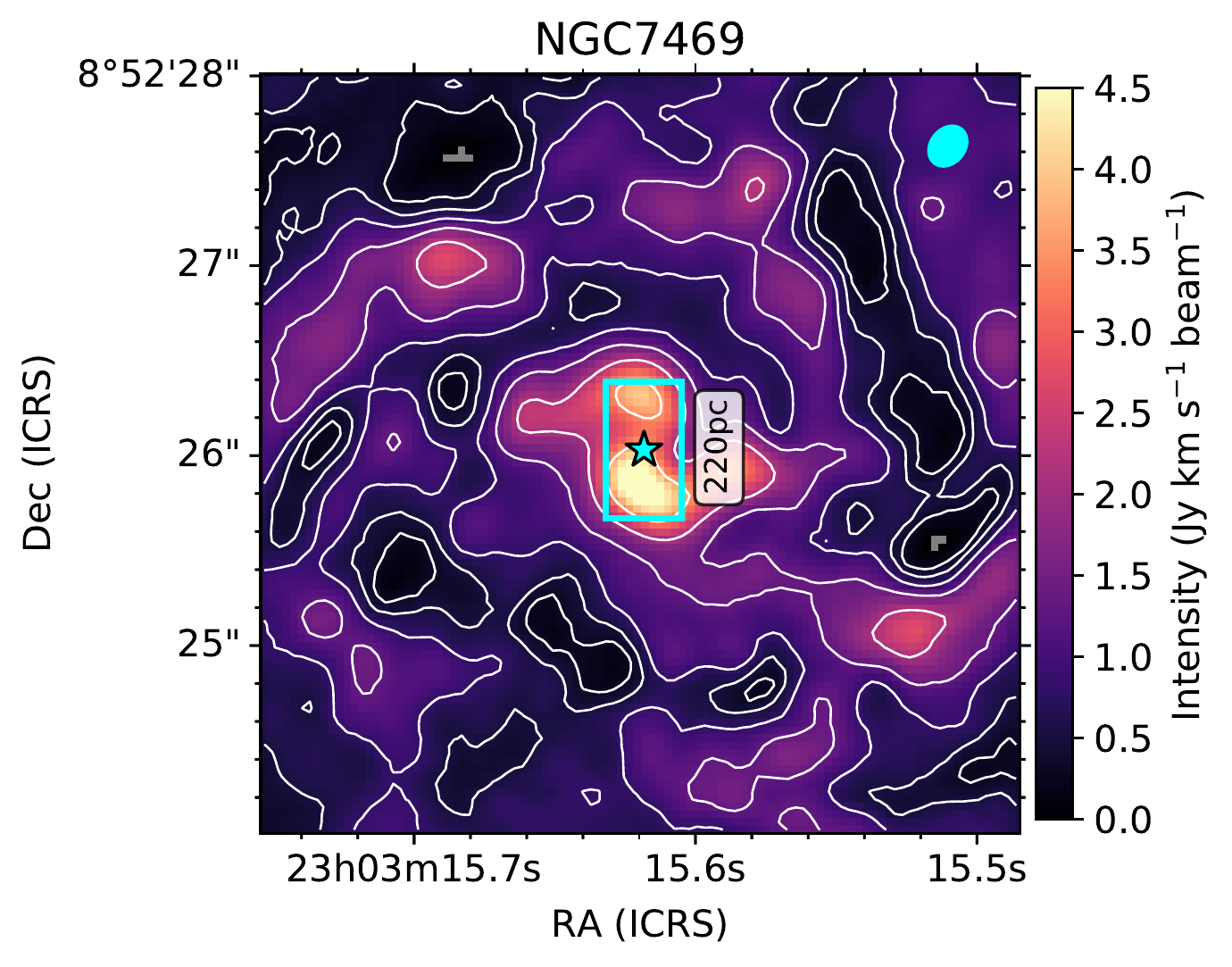}
  \includegraphics[width=6cm]{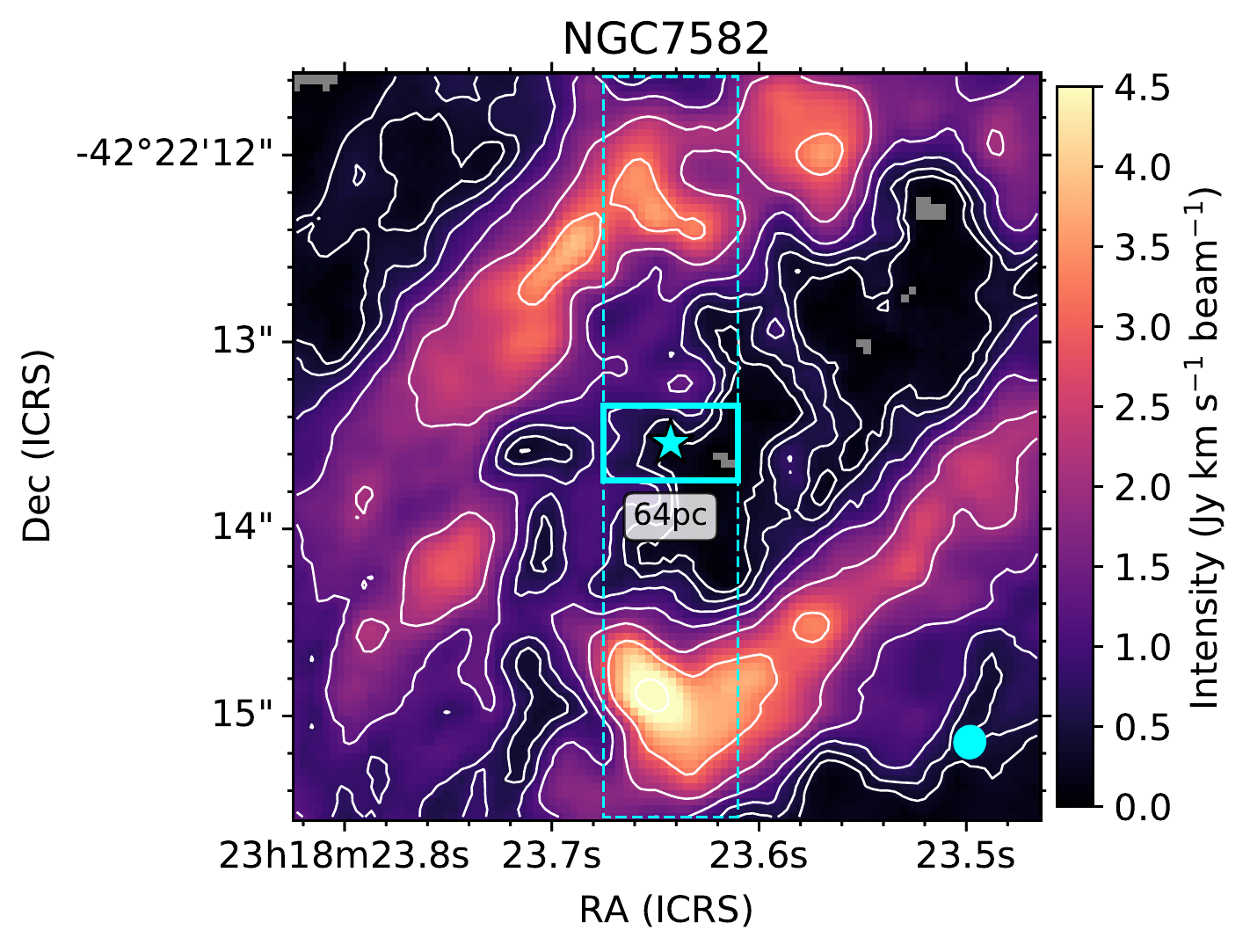}

  \caption{Continued.}
    \end{figure*}

\subsection{Ground-based mid-IR spectroscopy}\label{subsec:mid-IRobs}
All the ground-based mid-IR spectroscopic observations used in this
work have been published previously, and the corresponding references are
given in the last column of Table~\ref{tab:midIR}.
The majority of the mid-IR spectroscopic observations were taken
close to the diffraction limit at $8.7\,\mu$m with angular resolutions in the range
$0.3-0.6\arcsec$ \citep[full width at half-maximum, FWHM, see, e.g.,][]{Hoenig2010, GonzalezMartin2013, AlonsoHerrero2016}.
We also indicate whether the $11.3\,\mu$m PAH feature was detected on the nuclear
scales defined by the slit width.
Of the 22 Seyfert galaxies in our sample, this feature is detected in 12 nuclei.
The nuclear regions probed by the ground-based mid-IR spectra are determined by the slit width.
On nuclear scales, the maximum distance from the AGN where the $11.3\,\mu$m PAH
feature is
detected (or not detected) is therefore given by half of the mid-IR slit width. We refer it to as  
the distance from the AGN, $r_{\rm AGN}$ (see Table~\ref{tab:midIR} for the value
for each galaxy and Section~\ref{subsec:halflive} for a discussion).
The range of $r_{\rm AGN}$ is quite varied, from 12 to 115\,pc. However,
the circumnuclear rings, when present in our sample galaxies (see
Figure~\ref{fig:CO21maps} and Table~\ref{tab:CO21obs}) and in other
galaxies \citep{Comeron2010}, are well beyond the radius defined by the
nuclear mid-IR apertures.

\begin{table*}
\caption{ALMA and NOEMA CO(2-1) observations.}             
\label{tab:CO21obs}      
\centering                          
\begin{tabular}{c  c c c c c c c}        
\hline\hline                 

Galaxy   & Beam             &  PA$_{\rm beam}$  & Pixel & Instrument $\&$ Prog ID &PI &Ref & CO(2-1) \\
          & ($\arcsec \times \arcsec$) & (deg) & ($\arcsec$) & & & & Ring/Mini-spiral?\\
\hline
IC~4518W  &$0.23 \times 0.20$ & -86 & 0.025 & ALMA$\#$2017.1.00255.S &
                                                              Pereira-Santaella & this work & Perturbed\\   
Mrk~1066  &$0.44 \times 0.40$ & 40 & 0.075 & NOEMA W14CB & Alonso-Herrero & 1 & Yes\\
NGC~1365  &$0.25 \times 0.22$ & 71 & 0.05 & ALMA$\#$2013.1.01161.S& Sakamoto
                                                                              &
                                                                                this
                                                                                work
                                                                                   &
                                                                                     Yes\\
NGC~1386  &$0.62 \times 0.45$ & 85 & 0.1& ALMA$\#$2012.1.00474.S& Nagar & 2 & Yes?\\
NGC~1808  &$1.14 \times 0.84$ & 87 & 0.15 & ALMA$\#$2017.1.00984.S & Salak &
                                                                      this work & Yes\\
NGC~2273  &$0.72 \times 0.58$ & 32 & 0.13 & NOEMA W14CB & Alonso-Herrero & 1 & Yes\\
NGC~3081  &$0.57 \times 0.49$ & 83 & 0.1 & ALMA$\#$2015.1.00086.S& Nagar & 2 & Yes \\
NGC~3227  &$0.21 \times 0.16$ & 42 & 0.025 & ALMA$\#$2016.1.00254.S & Alonso-Herrero & 3 & Yes\\
NGC~4253  &$0.55 \times 0.30$ & 22 & 0.071 & NOEMA WP16BP & Alonso-Herrero & 1 & Yes\\
NGC~4388  &$0.84 \times 0.36$ & 21 & 0.071 & NOEMA W14CB& Alonso-Herrero& 1 & ?\\
NGC~5135  &$0.31 \times 0.22$ & 63 & 0.05 & ALMA$\#$2013.1.00243.S & Colina & 4 & Yes\\
NGC~5643  &$0.26 \times 0.16$ & -70& 0.04 & ALMA$\#$2016.1.00254.S & Alonso-Herrero & 5 & Yes\\
NGC~7130  &$0.30 \times 0.25$ & 62 & 0.025 & ALMA$\#$2017.1.00255.S &
                                                              Pereira-Santaella & this work & Yes\\
NGC~7172  &$0.56 \times 0.44$ & 76 & 0.1 & ALMA$\#$2015.1.00116.S& Kohno &
                                                                     this work & ?\\
NGC~7213  &$0.61 \times 0.58$ & 79 & 0.12 & ALMA$\#$2012.1.00474.S & Nagar & 2 & Yes\\
NGC~7465  &$0.59 \times 0.30$ & 23 & 0.071 & NOEMA WP16BP & Alonso-Herrero & 1 & No\\
NGC~7469  &$0.23 \times 0.18$ & -39& 0.04 & ALMA$\#$2017.1.00395.S
                                                                          &D\'{\i}az-Santos
                                                                              &
                                                                                this work & Yes\\
NGC~7582  &$0.17 \times 0.16$ & -16 & 0.04 & ALMA$\#$2016.1.00254.S &
                                                               Alonso-Herrero
                                                                              & this work & Yes\\

\hline
\end{tabular}
\tablefoot{Beam and PA$_{\rm beam}$ are the synthesized beam sizes and
  position angles of the CO(2-1) observations, respectively. ``Pixel'' refers to the pixel
  size of the CO(2-1) flux images. The
  ``Ref''  column lists references of previously published works using the
same dataset. The last column indicates whether a ring or a
mini-spiral is detected in CO(2-1) in the central $\sim 4\arcsec
\times 4\arcsec$.}

\tablebib{1. \cite{DominguezFernandez2020},  2. \cite{Ramakrishnan2019},
3. \cite{AlonsoHerrero2019}, 4. \cite{Sabatini2018}, 5. \cite{AlonsoHerrero2018}. }
\end{table*}

\begin{table*}
\caption{Measurements with {\sc pahfit} using the {\it Spitzer}/IRS spectra.}             
\label{tab:SpitzermidIR}      
\centering                          
\begin{tabular}{c c c c c c c c c  c c c c c c c c c c c c c c c c}        

\hline\hline                 
Galaxy   & \multicolumn{1}{c}{6.2$\,\mu$m PAH} & \multicolumn{3}{c}{$7.7\,\mu$m PAH complex}
& \multicolumn{2}{c}{$11.3\,\mu$m PAH complex}\\  
  &  &  $7.42\,\mu$m & $7.6\,\mu$m &
$7.85\,\mu$m & $11.23\,\mu$m &
$11.33\,\mu$m \\
\hline
IC4518W & $ 124 ^{+ 1 }_{- 1 }$ & $ 664 ^{+ 6 }_{- 6 }$ & $ 50 ^{+ 6 }_{- 6 }$ & $ 243 ^{+ 3 }_{- 3 }$ & $ 7 ^{+ 1 }_{- 1 }$ & $ 131 ^{+ 1 }_{- 1 }$\\
Mrk1066 & $ 367 ^{+ 6 }_{- 6 }$ & $ 853 ^{+ 41 }_{- 38 }$ & $ 359 ^{+ 24 }_{- 26 }$ & $ 557 ^{+ 18 }_{- 18 }$ & $ 129 ^{+ 9 }_{- 9 }$ & $ 326 ^{+ 11 }_{- 11 }$\\
NGC1320 & $ 63 ^{+ 2 }_{- 2 }$ & $ 332 ^{+ 10 }_{- 9 }$ & $<9$ & $ 66 ^{+ 4 }_{- 4 }$ & $ 13 ^{+ 1 }_{- 1 }$ & $ 62 ^{+ 2 }_{- 2 }$\\
NGC1365 & $ 556 ^{+ 4 }_{- 4 }$ & $ 639 ^{+ 21 }_{- 20 }$ & $ 640 ^{+ 11 }_{- 12 }$ & $ 776 ^{+ 8 }_{- 8 }$ & $ 167 ^{+ 2 }_{- 2 }$ & $ 473 ^{+ 3 }_{- 3 }$\\
NGC1386 & $ 50 ^{+ 6 }_{- 6 }$ & $ 654 ^{+ 29 }_{- 30 }$ & $<90$ & $ 263 ^{+ 12 }_{- 12 }$ & $ 17 ^{+ 4 }_{- 4 }$ & $ 127 ^{+ 9 }_{- 9 }$\\
NGC1808 & $ 1175 ^{+ 56 }_{- 16 }$ & $ 1080 ^{+ 99 }_{- 160 }$ & $ 1853 ^{+ 408 }_{- 151 }$ & $ 1958 ^{+ 180 }_{- 512 }$ & $ 518 ^{+ 21 }_{- 13 }$ & $ 1044 ^{+ 27 }_{- 59 }$\\
NGC2110 & $ 30 ^{+ 5 }_{- 5 }$ & $ 98 ^{+ 18 }_{- 22 }$ & $<54$ & $<52$ & $ 13 ^{+ 2 }_{- 3 }$ & $ 48 ^{+ 5 }_{- 6 }$\\
NGC2273 & $ 220 ^{+ 5 }_{- 5 }$ & $ 474 ^{+ 38 }_{- 35 }$ & $ 152 ^{+ 29 }_{- 29 }$ & $ 235 ^{+ 21 }_{- 18 }$ & $ 53 ^{+ 4 }_{- 4 }$ & $ 198 ^{+ 6 }_{- 7 }$\\
NGC2992 & $ 78 ^{+ 12 }_{- 12 }$ & $<464$ & $<169$ & $ 175 ^{+ 27 }_{- 27 }$ & $ 37 ^{+ 11 }_{- 10 }$ & $ 107 ^{+ 20 }_{- 18 }$\\
NGC3081 & $ 21 ^{+ 4 }_{- 4 }$ & $ 202 ^{+ 19 }_{- 22 }$ & $<46$ & $ 45 ^{+ 10 }_{- 10 }$ & $<16$ & $ 40 ^{+ 5 }_{- 4 }$\\
NGC3227 & $ 231 ^{+ 16 }_{- 16 }$ & $ 677 ^{+ 114 }_{- 94 }$ & $<278$ & $ 345 ^{+ 29 }_{- 31 }$ & $ 67 ^{+ 20 }_{- 21 }$ & $ 277 ^{+ 28 }_{- 24 }$\\
NGC4253 & $ 77 ^{+ 3 }_{- 2 }$ & $ 254 ^{+ 12 }_{- 14 }$ & $ 102 ^{+ 8 }_{- 8 }$ & $ 131 ^{+ 6 }_{- 5 }$ & $ 26 ^{+ 1 }_{- 1 }$ & $ 84 ^{+ 2 }_{- 2 }$\\
NGC4388 & $ 108 ^{+ 5 }_{- 4 }$ & $ 840 ^{+ 26 }_{- 27 }$ & $<35$ & $ 208 ^{+ 23 }_{- 20 }$ & $<23$ & $ 120 ^{+ 6 }_{- 8 }$\\
NGC5135 & $ 256 ^{+ 5 }_{- 5 }$ & $ 366 ^{+ 54 }_{- 52 }$ & $ 281 ^{+ 70 }_{- 68 }$ & $ 549 ^{+ 39 }_{- 40 }$ & $ 102 ^{+ 6 }_{- 7 }$ & $ 216 ^{+ 9 }_{- 8 }$\\
NGC5643 & $ 116 ^{+ 5 }_{- 5 }$ & $ 155 ^{+ 33 }_{- 31 }$ & $ 86 ^{+ 22 }_{- 22 }$ & $ 164 ^{+ 11 }_{- 11 }$ & $ 44 ^{+ 3 }_{- 3 }$ & $ 104 ^{+ 6 }_{- 6 }$\\
NGC7130 & $ 203 ^{+ 4 }_{- 4 }$ & $ 478 ^{+ 40 }_{- 25 }$ & $ 255 ^{+ 33 }_{- 31 }$ & $ 346 ^{+ 15 }_{- 16 }$ & $ 73 ^{+ 2 }_{- 3 }$ & $ 181 ^{+ 4 }_{- 4 }$\\
NGC7172 & $ 111 ^{+ 3 }_{- 4 }$ & $ 515 ^{+ 25 }_{- 24 }$ & $ 136 ^{+ 17 }_{- 29 }$ & $ 262 ^{+ 12 }_{- 12 }$ & $ 12 ^{+ 2 }_{- 3 }$ & $ 121 ^{+ 6 }_{- 6 }$\\
NGC7213 & $ 23 ^{+ 6 }_{- 6 }$ & $ 77 ^{+ 19 }_{- 20 }$ & $<29$ & $<49$ & $<28$ & $ 61 ^{+ 6 }_{- 6 }$\\
NGC7582 & $ 193 ^{+ 6 }_{- 6 }$ & $ 596 ^{+ 37 }_{- 35 }$ & $ 194 ^{+ 20 }_{- 22 }$ & $ 275 ^{+ 17 }_{- 21 }$ & $ 91 ^{+ 7 }_{- 9 }$ & $ 242 ^{+ 11 }_{- 10 }$\\
NGC7469 & $ 586 ^{+ 129 }_{- 51 }$ & $ 1778 ^{+ 239 }_{- 142 }$ & $ 515 ^{+ 211 }_{- 152 }$ & $ 1344 ^{+ 242 }_{- 417 }$ & $ 181 ^{+ 24 }_{- 23 }$ & $ 578 ^{+ 107 }_{- 56 }$\\
\hline
\end{tabular}
\tablefoot{The fluxes and errors are in units of $10^{-14}\,{\rm erg\,cm}^{-2}\,{\rm s}^{-1}$. }
\end{table*}

\subsection{Spitzer/IRS observations}\label{subsec:IRSobs}
To study the PAH emission on circumnuclear
scales we use  archival {\it Spitzer} observations
taken with IRS
covering the SL and long-low (LL) spectral ranges
of $5-15\,\mu$m and $15-40\,\mu$m, respectively, with
spectral resolutions $R\sim 60-120$. All the
galaxies in our sample, except for NGC~1320 and NGC~4253,
were observed in the staring mode.  We
downloaded the fully calibrated spectra from the Cornell
Atlas of {\it Spitzer}/IRS Sources \citep[CASSIS, version LR7,][]{Lebouteiller2011} or the {\it Spitzer} archive.
Only NGC~7465 was not observed with IRS.
For the galaxies in this work,
the optimal CASSIS extraction was equivalent to a point-source
extraction. The slit width of the SL observations is $3.7\arcsec$.
Taking the SL module as the basis, we applied a small offset to stitch together
the SL and LL spectra. 
NGC~1320 and NGC~4253 were observed in spectral mapping mode. We used
{\sc cubism} \citep{Smith2007cubism} to fully reduce the observations and
extract the circumnuclear spectra. We chose apertures of $5.9\arcsec
\times 5.9\arcsec$ and $17.3\arcsec \times 17.3\arcsec$ for the SL and LL
modules, respectively. To recover the nuclear flux at each wavelength, we
applied an aperture correction using a standard star.

The IRS PAH feature fluxes of the majority of
the galaxies in our sample are published in other works \citep[see, e.g.,][]{Wu2009, Gallimore2010, DiamondStanic2010, PereiraSantaella2010} but using different methods
and extraction apertures. For consistency, we obtained our
own measurements. Given the complicated nature of the
mid-IR continuum and the different contributions from the AGN and star
forming regions, instead of using a local continuum we
chose to perform a full fit of the $5-40\,\mu$m spectral range.

We used a modified version of the
{\sc pahfit} code \citep{Smith2007} to obtain the
fluxes of different PAH features. This version
fits the mid-IR spectrum with a continuum, PAH emission, and the $10\,\mu$m silicate 
feature in emission or absorption. For the 
feature in emission, the code uses 
the \cite{Ossenkopf1992} cold dust model where the silicate feature
peaks at $\sim 10.1\,\mu$m \citep[see][for more details]{Gallimore2010}.
There are  seven type 1 Seyfert galaxies
in our sample with evidence of having the silicate feature in emission or nearly flat. 
NGC~2110 and NGC~7213
clearly show the silicate feature  in emission, both in the
{\it Spitzer}/IRS  and in the nuclear ground-based spectra
\citep[see][]{Mason2009, Hoenig2010, Esquej2014}. For these, we allowed
for the silicate feature peak to vary. The best fits were at
$10.6\,\mu$m for NGC~7213 and $10.7\,\mu$m for  NGC~2110. 
For the other
type 1 Seyfert galaxies in our sample (NGC~1365, NGC~3081, NGC~3227, NGC~4253, and
NGC~7469), the $10\,\mu$m
silicate feature is nearly flat or slightly in emission, as can be seen in the
nuclear spectra \citep[see][]{Hoenig2010,
  GonzalezMartin2013, Esquej2014}. 
Because the feature is not as prominent
as in the
other two galaxies, we did not vary the emission peak wavelength when
we fit the spectra with {\sc pahfit}.
We are only interested in the brightest PAH
features, specifically, in the ratios between the $11.3\,\mu$m PAH complex
($11.23\,\mu$m and $11.33\,\mu$m features) and the $7.7\,\mu$m PAH complex ($7.42\,\mu$m,
$7.60\,\mu$m, and $7.85\,\mu$m features)
and
between the $6.2\,\mu$m PAH and the $7.7\,\mu$m PAH complex. We list the fluxes (or upper limits)
and corresponding
errors in
Table~\ref{tab:SpitzermidIR}.

\begin{figure}
      \centering
  \includegraphics[width=8cm]{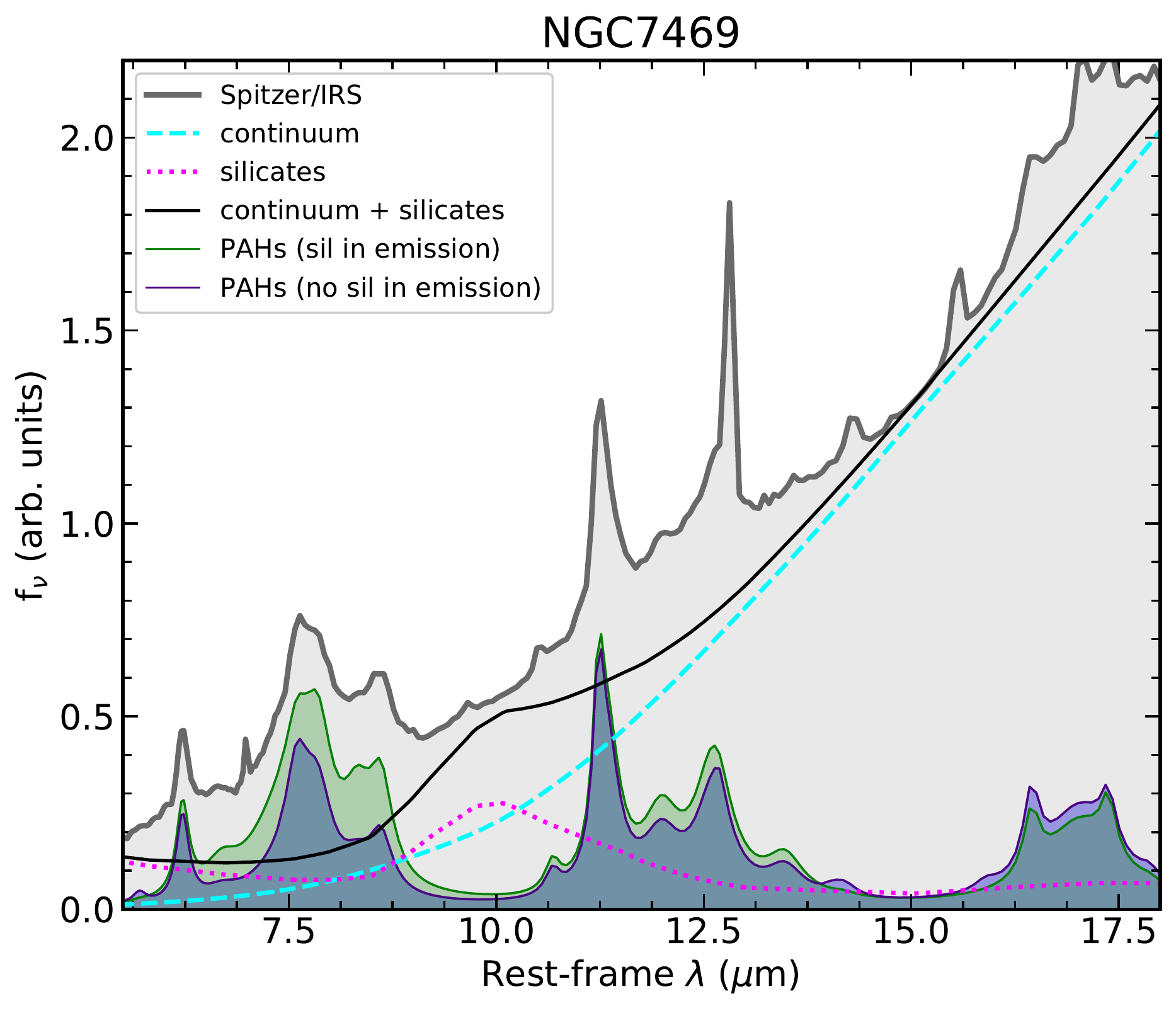}

  \caption{Observed {\it Spitzer}/IRS spectrum (in gray) of NGC~7469. 
   We show the continuum (dashed cyan line) and silicate feature in emission (dotted magenta line) and
    their sum (black line), as well as 
    the PAH feature spectra (green curve) fit
    with {\sc pahfit}. We also show the fit  PAH feature spectrum (blue and purple curve)
    when {\sc pahfit} fit was run without
    the silicate feature in emission. Both
    the $f(11.3\,\mu$m)/$f(7.7\,\mu$m) and $f(6.2\,\mu$m)/$f(7.7\,\mu$m) PAH ratios are
    noticeably lower when the fit includes a component with
    the silicate feature in emission. }\label{fig:NGC7469pahfit}
    \end{figure}

We finally discuss an additional source of uncertainty related to the
continuum fit when the fluxes of the different PAH features are obtained. When
the $10\,\mu$m silicate feature in emission is included, the 
$f(11.3\,\mu$m)/$f(7.7\,\mu$m) and 
$f(6.2\,\mu$m)/$f(7.7\,\mu$m) PAH ratios tend to decrease, as does 
the continuum level in the $6-13\,\mu$m region. We illustrate this in
Figure~\ref{fig:NGC7469pahfit}  for NGC~7469. When we did 
not use this additional component for the silicate feature in emission,
we obtained PAH ratios of $f(11.3\,\mu$m)/$f(7.7\,\mu$m)= 0.39 and 
$f(6.2\,\mu$m)/$f(7.7\,\mu$m)=0.27, whereas we obtained
 $f(11.3\,\mu$m)/$f(7.7\,\mu$m)= 0.21 and 
$f(6.2\,\mu$m)/$f(7.7\,\mu$m)=0.16 when it was included.

\section{Analysis of the ALMA and NOEMA observations}\label{sec:COmeasurements}

Figure~\ref{fig:CO21maps} shows the variety of the nuclear and circumnuclear
CO(2-1) morphologies of the galaxies in our sample over the selected FoV of
$4\arcsec \times 4\arcsec$. A small fraction
of the galaxies show centrally peaked emission with associated  nuclear
mini-spirals (e.g., NGC~1808, NGC~5643, and NGC~7130). Many also show CO(2-1) circumnuclear rings that are probably
associated with the inner Lindblad resonance due to the presence of
a large-scale bar (e.g., NGC~1365, NGC~7469, and NGC~7582, see morphological types
in Table~\ref{tab:Sample}). 
Nevertheless, the vast majority show complicated CO(2-1)
morphologies with several emitting regions where the AGN might 
not be easily identified. In this section  we first determine the AGN position
before we perform the aperture photometry. We  then describe the method
with which we measured the nuclear CO(2-1) fluxes.
We also extract the CO(2-1) surface brightness profiles for
two galaxies with
extended $11.3\,\mu$m PAH emission
\citep[see][]{EsparzaArredondo2018}, circumnuclear rings of SF, and
ionization cones. These galaxies are NGC~5135 and NGC~7582.

\subsection{Identification of the AGN position}\label{subsec:AGNpos}

A number of works have shown that the continuum emission at $1.3\,$mm in Seyfert galaxies is
mostly produced by synchrotron emission with varying contributions from
cold dust emission and free-free emission \citep{Pasetto2019, AlonsoHerrero2019, GarciaBurillo2019}. We therefore
assumed that the peak of the 1.3\,mm can be used to locate the position of the AGN (see also
Section~\ref{sec:appendix}).

Of the galaxies analyzed in this work,  only six show one
1.3\,mm continuum peak in the ALMA maps. 
The remaining seven 
(NGC~1365, NGC~1808,
NGC~3081, NGC~5135, NGC~7130, NGC~7469, and NGC~7582) present 
several bright 1.3\,mm emitting sources over the ALMA FoV.  
In Section~\ref{sec:appendix} we describe in detail for each source
the source that we identified as the AGN. To derive the
coordinates, we obtained the centroid of the source at $1.3\,$mm
in the plane of the sky and converted the pixel position into the International Coordinate
Reference System (ICRS) using the WCS header information and {\sc astropy} routines
\citep{Astropy2018}.
The 1.3\,mm continuum peaks are marked as
a filled star in Figure~\ref{fig:CO21maps} and the ICRS
coordinates are listed in Table~\ref{tab:1p3mm} in the Appendix. 
We refer to \cite{DominguezFernandez2020}
for the analysis of the 1.3\,mm NOEMA maps.

\subsection{CO(2-1) aperture photometry}\label{subsec:phot}

We defined rectangular apertures to
simulate the nuclear extraction aperture of the mid-IR spectra.
One of the aperture sides is well defined
by the mid-IR slit width (see Table~\ref{tab:midIR}). The majority of the ALMA and NOEMA
observations have beam sizes smaller than the slit width, except in the case
of NGC~1808 and NGC~1386 (see Table~\ref{tab:CO21obs}). In these two cases, the
rectangular extraction apertures (see Table~\ref{tab:NucMolGas}) were chosen to be close to the
beam size. 
The other side of the rectangular aperture can be defined in
principle  as the angular resolution FWHM of
the mid-IR observations, assuming that the mid-IR emitting continuum source
was unresolved. 
When possible, we used square apertures, except in the cases where the
slit width was larger than the angular resolution. We also defined the aperture
sizes to contain entire pixels on the CO(2-1) integrated intensity maps and rotated them 
to the orientation of the mid-IR slits, unless they were close to PA$_{\rm slit}=0\deg$ or 
PA$_{\rm slit}=90\deg$. The sizes of the square or rectangular
apertures used for each galaxy are listed in Table~\ref{tab:NucMolGas}
and are overlaid on the CO(2-1)
maps in Figure~\ref{fig:CO21maps}.

We performed the photometry with these square or rectangular
apertures on the CO(2-1) integrated intensity maps. The CO(2-1) fluxes are listed
in Table~\ref{tab:NucMolGas}. The errors in these fluxes are dominated by the photometric
calibration
uncertainty of ALMA and NOEMA, which in band 6 is typically $10-15\%$.

\begin{table*}
\caption{Nuclear molecular gas properties.}             
\label{tab:NucMolGas}      
\centering                          
\begin{tabular}{c c c c c c c}        
\hline\hline                 
Galaxy      & \multicolumn{2}{c}{Nuclear Aperture} & CO(2-1) & $M({\rm H}_2)$  & $\Sigma \,{\rm H}_2$ & $\log N({\rm H}_2)$\\
& ($\arcsec \times \arcsec$) & (${\rm pc} \times {\rm pc}$)& (Jy km s$^{-1}$) 
& (M$_\odot$) & (M$_\odot\,{\rm pc}^{-2}$) & (cm$^{-2}$)\\
\hline

IC4518W & $0.7 \times 0.4$ & $223 \times  128$ & 6.9 & 6.4e+07 & 2.2e+03  &23.1\\
Mrk1066 & $0.5 \times 0.5$ & $112 \times  112$ & 14.0 & 6.2e+07 & 5.0e+03 &23.5\\
NGC1068$^*$ & $0.40 \times 0.16$ & $27 \times  11$ & 1.43 & 5.6e+05 & 2.6e+03 &23.2\\
NGC1320 & $0.5 \times 0.5$ & $82 \times  82$ & 3.0 & 7.1e+06 & 1.1e+03 &22.8\\
NGC1365 & $0.35 \times 0.35$ & $36 \times  36$ & 1.3 & 8.7e+05 & 6.8e+02 &22.6\\
NGC1386 & $0.5 \times 0.5$ & $51 \times  51$ & 6.8 & 4.4e+06 & 1.7e+03 &23.0\\
NGC1808 & $0.9 \times 0.9$ & $60 \times  60$ & 61.7 & 2.4e+07 & 6.7e+03 &23.6\\
NGC2110 & $1.0 \times 1.0$ & $158 \times  158$ & 0.5 & 1.0e+06 & 4.1e+01 &21.4\\
NGC2273 & $0.65 \times 0.65$ & $81 \times  81$ & 6.4 & 8.5e+06 & 1.3e+03 &22.9\\
NGC2992 & $0.5 \times 0.5$ & $87 \times  87$ & 0.7 & 1.9e+06 & 2.5e+02 &22.2\\
NGC3081 & $0.6 \times 0.5$ & $107 \times  90$ & 0.8 & 2.3e+06 & 2.4e+02 &22.2\\
NGC3227 & $0.5 \times 0.5$ & $49 \times  49$ & 8.4  & 7.0e+06 & 2.9e+03 &23.3\\
NGC4253 & $0.5 \times 0.5$ & $136 \times  136$ & 11.6 & 7.7e+07 & 4.2e+03 &23.4\\
NGC4388 & $0.5 \times 0.5$ & $51 \times  51$ & 2.5 & 1.7e+06 & 6.3e+02 &22.6\\
NGC5135 & $0.7 \times 0.35$ & $201 \times  100$ & 3.9 & 2.9e+07 & 1.4e+03 &23.0\\
NGC5643 & $0.35 \times 0.35$ & $32 \times  32$ & 11.0 & 8.1e+06 & 7.8e+03 &23.7\\
NGC7130 & $0.7 \times 0.4$ & $209 \times  120$ & 25.4 & 2.1e+08 & 8.2e+03 &23.7\\
NGC7172 & $0.4 \times 0.4$ & $61 \times  61$ & 4.4 & 9.1e+06 & 2.4e+03 &23.2\\
NGC7213 & $0.84 \times 0.84$ & $86 \times  86$ & 0.03 & 2.8e+04 & 3.8e+00 &20.4\\
NGC7465 & $0.56 \times 0.49$ & $59 \times  51$ & 3.3 & 3.2e+06 & 1.0e+03 &22.8\\
NGC7469 & $0.72 \times 0.4$ & $212 \times  118$ & 21.5 & 1.7e+08 & 6.3e+03 &23.6\\
NGC7582 & $0.72 \times 0.4$ & $63 \times  35$ & 4.6 & 3.1e+06 & 1.4e+03 &22.9\\

\hline
\end{tabular}
\tablefoot{$^*$ The nuclear flux is from \cite{GarciaBurillo2019}
   measured through an elliptical aperture of the full Gaussian sizes indicated.}
\end{table*}

For  NGC~1068 we took the nuclear CO(2-1) flux inside the area
defined by the FWHM size of the Gaussian disk (deconvolved size $0.22\arcsec \times 0.09\arcsec$)
derived by \cite{GarciaBurillo2019}.
Given the limited angular resolution of the ALMA CO(2-1) observations of NGC~2110 
\citep[see][for full details]{Rosario2019}, we extracted the nuclear fluxes
with a $1\arcsec \times 1\arcsec$ aperture.
The nuclear CO(2-1) fluxes for NGC~1320 and NGC~2992  within   
$0.5\arcsec \times 0.5\arcsec$ apertures were provided by J. A. Fern\'andez Ontiveros (private
communication).


For the CO(2-1) surface brightness profiles of NGC~5135 and NGC~7582,
we followed a similar
procedure as for the nuclear apertures. We extracted aperture photometry in
rectangular apertures placed
along the length of the mid-IR slit (plotted in the corresponding
panels in Figure~\ref{fig:CO21maps}), that is, at the PA$_{\rm slit}$ value
listed in Table~\ref{tab:midIR}. The areas used to compute the surface
brightnesses are the projected areas of the simulated slits.
The angular resolution of the ALMA CO(2-1)
observations of NGC~7582 ($0.17\arcsec \times 0.16\arcsec$, see Table~\ref{tab:CO21obs})
is higher than that of the mid-IR spectroscopy and slit width. We thus smoothed
the ALMA CO(2-1) image to an approximately angular resolution of 0.4-0.5$\arcsec$ to match
that of the mid-IR observations. The synthesized beam of the ALMA
observations of NGC~5135 ($0.31\arcsec \times 0.22\arcsec$) is
relatively close to the angular resolution of the mid-IR
spectroscopy \citep{DiazSantos2010}, and thus we did not smooth the
image. In Section~\ref{subsec:surfacebrightnessprofiles}  we discuss the CO(2-1) and
$11.3\,\mu$m PAH surface
brightness profiles of these two galaxies.

 \subsection{Nuclear molecular gas masses}

We used the CO(2-1) transition fluxes to estimate the bulk of the molecular
gas mass in the nuclear regions of our galaxies. In using the CO(2-1)
transition, which is sensitive to gas temperatures of a few tens of
Kelvin and densities $\sim 10^3\,{\rm cm}^{-3}$, and based on our choice of the
2--1/1--0 temperature ratio (see below), we implicitly assumed that the contribution from
highly excited molecular gas is small. To derive the molecular gas masses, we
used the following expression derived by
\cite{Sakamoto1999}:

\begin{equation}
  M_{H_{2}}(M_{\odot}) = 1.18 \times 10^{4} \times D^{2}_{\rm L}
    \times S_{\rm CO(1-0)} \times X_{\rm CO} \label{eq:mass}
,\end{equation}

\noindent where  $D_{\rm L}$ is the luminosity distance in units of Mpc, $S_{\rm CO(1-0)}$ is the CO(1-0) transition flux in units of
Jy\,km\,s$^{-1}$, and $X_{\rm CO}$   is the CO-to-H$_2$ conversion factor in units of
$3\times10^{20}\,{\rm cm}^{-2} \,({\rm K\,km\,s}^{-1})^{-1}$.
We first estimated the CO(1-0) flux from our observations. We took an
$r_{21}=$CO(2–1)/CO(1–0) brightness temperature ratio of one,  assuming a thermally
excited and optically
thick gas \citep{Braine1992}. However, we note that this ratio
is rather uncertain. For instance, \cite{Papadopoulos1998}
found for a sample of Seyfert 1 and Seyfert 2 galaxies (typical beam sizes $32-55\arcsec$)
an average value of $r_{21}=0.7$, which is lower than the typical values measured for
starburst and spiral galaxies. However, these authors also pointed out that
$r_{21}$ is likely to become smaller for larger beam sizes because warmer
gas may be more confined in the nuclear regions, including those of Seyfert galaxies
\citep[see also][]{Rigopoulou2002, Lambrides2019}. 
\cite{Israel2020} compiled $r_{21}$ values in the inner
$22\arcsec$ for a large sample of galaxies that includes seven of our
targets. The $r_{12}$ values range from 0.6 (NGC~5135) to 1.7 (NGC~3227), with an average value of 1.1.
Therefore, the
assumed value of $r_{21}=1$ for the nuclear or circumnuclear
regions of Seyfert galaxies might be a lower limit
\citep[see also][]{Viti2014} and the estimated masses accordingly upper limits.

For the CO-to-H$_2$ conversion factor we used the
Galactic value of
$X_{\rm CO} = 2\times10^{20}\,{\rm cm}^{-2} ({\rm K\,km\,s}^{-1})^{-1}$ \citep{Bolatto2013}.
However, we note that hydrodynamical  simulations coupled with radiation
transfer for clumpy torus models predict a large scatter of this
factor on the nuclear scales of AGN \citep{Wada2018}, adding another
source of uncertainty to the molecular mass calculation.

We list in Table~\ref{tab:NucMolGas} the
molecular gas masses within the regions covered by the mid-IR slits.
  The  molecular gas masses measured in the nuclear regions probed by the mid-IR
  slits in our sample
  span a broad  range  from $3\times 10^4\,M_\odot$ for NGC~7213 to
  $2\times 10^{8}\,M_\odot$ for NGC~7130. However,  most
have nuclear molecular
  gas masses in the range $10^6-10^8\,M_\odot$. These values are higher than the
  typical values measured for tori and nuclear disks
  in Seyfert galaxies and low-luminosity
  AGN \citep{GarciaBurillo2016, GarciaBurillo2019, AlonsoHerrero2018, AlonsoHerrero2019, Combes2019}. This is because the mid-IR slits cover regions 
  a factor of a few larger than the torus diameters ($10-50\,$pc).

\section{Detection of nuclear $11.3\,\mu$m PAH emission}\label{sec:PAHdet}

\subsection{Dependence on  molecular gas}\label{subsec:molgas}

  \begin{figure}
   \centering
  \includegraphics[width=8.5cm]{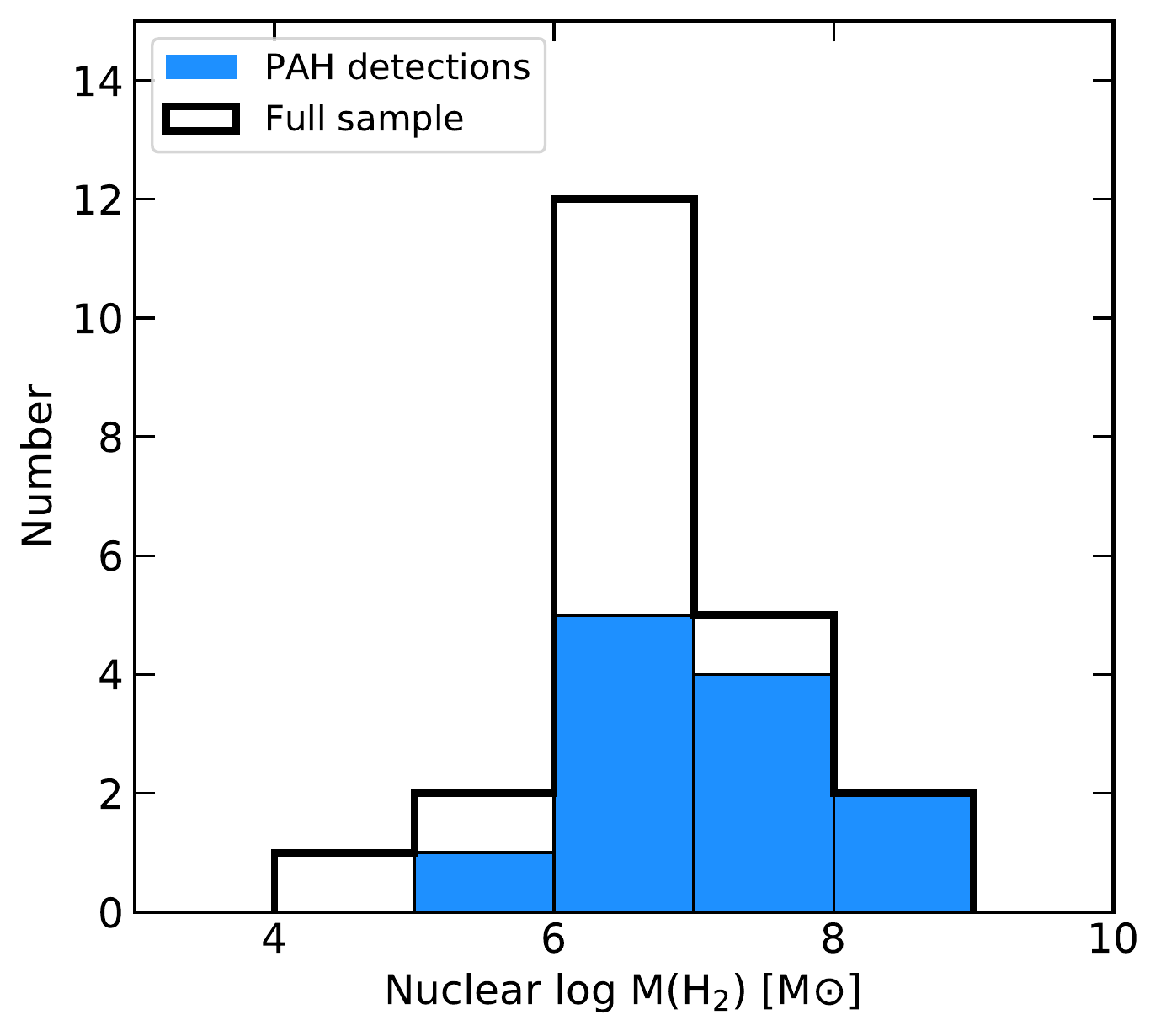}
  \includegraphics[width=8.5cm]{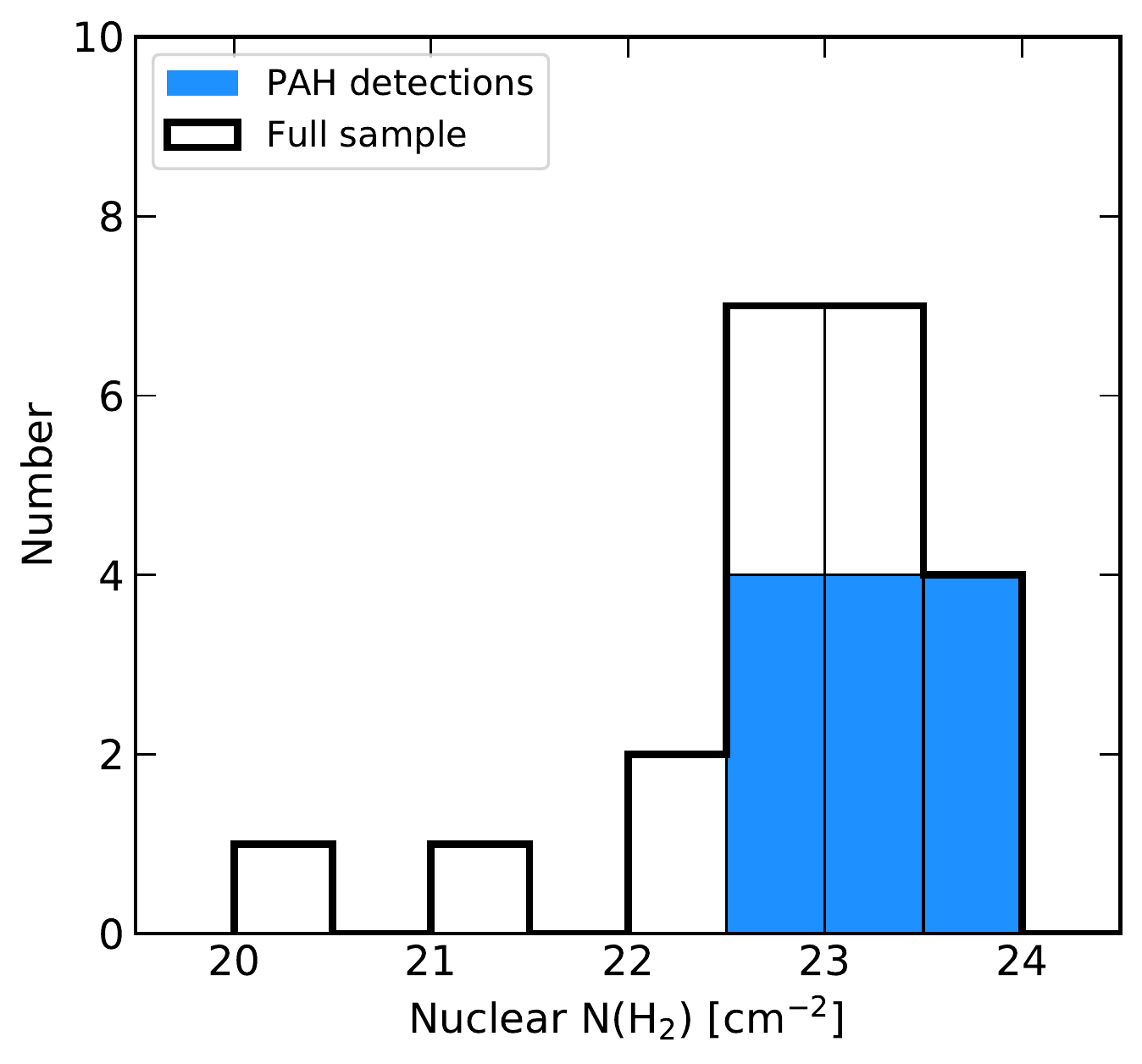}

  \caption{Distributions of the nuclear  molecular gas mass (top
    panel) and the column density averaged over the nuclear
    region covered by the CO(2-1) aperture
    photometry (bottom panel). The thick line histograms
    are the full sample, and the filled blue histograms show
  nuclei with a detection of the $11.3\,\mu$m PAH feature.}\label{fig:histograms}
    \end{figure}

  Figure~\ref{fig:histograms} (top panel) shows the distribution of molecular gas masses
  within the nuclear regions covered by the mid-IR slits for the full sample
    and for those with a nuclear detection of the $11.3\,\mu$m PAH
    feature. 
    While the ranges of molecular gas masses are similar for the two samples,  the median values
    for galaxies with nuclear detections of the  $11.3\,\mu$m PAH feature are higher 
     ($1.6\times 10^7\,M_\odot$) than for those without detections ($2\times 10^6\,M_\odot$).
  Thus, the molecular
  gas mass in the nuclear regions of our sample of Seyfert galaxies
  does  appear to have an effect on the detection of the nuclear
  PAH emission. A Kolmogorov-Smirnov (K-S) test indicates that
there is a probability of  2\% that the two 
distributions of cold molecular gas
are drawn from the same parent population.

  The nuclear mid-IR slits cover a
  range of physical regions from tens to hundreds of parsecs. A more appropriate quantity therefore
  is the 
  molecular gas surface density, $\Sigma \,{\rm H}_2$ averaged over the
  nuclear region. We calculated it
  by dividing the molecular gas mass by the (projected) slit area for
  each galaxy.
  In none of the Seyfert galaxies in our sample with low surface densities
  $\Sigma \,{\rm H}_2 < 10^3\,M_{\odot}\,{\rm pc}^{-2}$ 
  is the $11.3\,\mu$m PAH detected in the nuclear region. On the other hand, Seyfert nuclei with a nuclear detection of the $11.3\,\mu$m PAH feature have
  $\Sigma \,{\rm H}_2 > 10^3\,M_{\odot}\,{\rm pc}^{-2}$, although the reverse is not
  always true.
The galaxies in our sample with the highest nuclear values of the
 H$_2$ surface  density $\Sigma \,{\rm H}_2 \ge 5 \times 10^3\,M_{\odot}\,{\rm pc}^{-2}$, which are
 Mrk~1066, NGC~1808, NGC~5643, NGC~7130,
 and NGC~7469, are also known to show recent (more recent than a few
 hundred million years) SF activity in their nuclear regions
\citep{StorchiBergmann2000, GonzalezDelgado2001,
  Davies2007}. This means that the detection of the nuclear $11.3\,\mu$m
PAH emission is also likely associated with this recent SF activity.

The molecular hydrogen column densities
(see Figure~\ref{fig:histograms} and Table~\ref{tab:NucMolGas}) range from
$N({\rm H}_2) = 3\times 10^{20}\,{\rm cm}^{-2}$ for NGC~7213 to
$N({\rm H}_2) = 5\times 10^{23}\,{\rm cm}^{-2}$ for NGC~5643,
NGC~7130, and NGC~7469.
We emphasize that the $N({\rm H}_2)$ values are averaged over the
nuclear region. For instance, in
NGC~1808  \cite{Combes2019} measured an $N({\rm H}_2)$ ten
times higher  from 
high angular resolution CO(3-2) observations.
The average value  for our sample is
$N({\rm H}_2) = 2\times10^{23}\,{\rm cm}^{-2}$. This is  compatible with
the average values of $N({\rm H})$
estimated from the  H$_2$ $2.12\,\mu$m line by \cite{Hicks2009}.
The galaxies with a nuclear
 $11.3\,\mu$m PAH feature have a median of  $N({\rm H}_2) = 2 \times 10^{23}\,{\rm
   cm}^{-2}$,  compared to a median of $N({\rm H}_2) = 4 \times
 10^{22}\,{\rm cm}^{-2}$ for those without a detection. A K-S test indicates that
there is a probability of 0.8\% that the two $N({\rm H}_2)$
distributions are drawn from the same parent population. We therefore
confirm previous inferences \citep{Esquej2014, AlonsoHerrero2014}
that high molecular gas column densities might
play a role in protecting the PAH molecules in the nuclear regions of Seyfert galaxies.
 

\subsection{Dependence on AGN luminosity and distance from the AGN}\label{subsec:dependencies}

\begin{figure}
   \hspace{-0.3cm}
  \includegraphics[width=9cm]{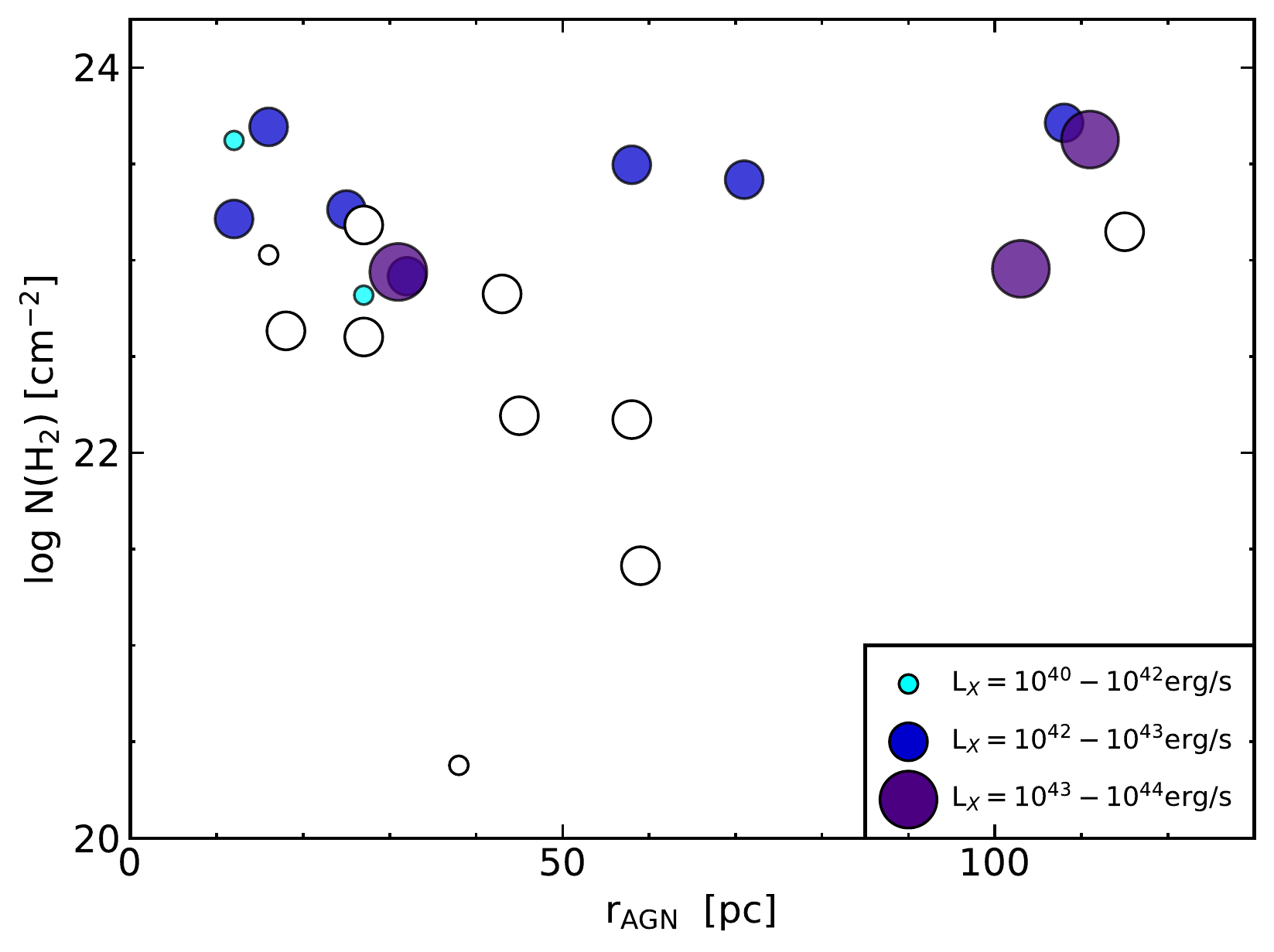}

  \caption{Distance from the AGN  vs. the molecular hydrogen
    column density for nuclear detections (filled circles) and nondetections
    (open circles) of the $11.3\,\mu$m PAH feature. The
  symbol sizes increase for increasing intrinsic AGN luminosities.}\label{fig:rAGNvsNH}
    \end{figure}

We investigated whether an additional dependency of the detection of
  nuclear PAH emission on the AGN luminosity and the
  distance from the AGN might be established \citep[see also][]{Esquej2014,
    AlonsoHerrero2014}. Following \cite{Monfredini2019}, at a given distance from the
  AGN ($r_{\rm AGN}$), the X-ray photon flux on a PAH
  molecule for a given photon energy is

  \begin{equation}
    F_{\rm X} = \frac{L_{\rm X}}{4\,\pi \,r_{\rm AGN}^2 \,h\nu}\,{\rm e}^{-\tau_{\rm X}}\label{eq:flux}
,\end{equation}

  \noindent where $L_{\rm X}$ is the AGN luminosity and $\tau_{\rm X}$ is
  the X-ray optical depth. As discussed in \cite{Monfredini2019}, we only considered the
  molecular gas contribution to the X-ray absorption\footnote{Then it can be expressed in terms of the
  X-ray photoabsorption cross-section $\sigma_{\rm H}(E)$
  and the H$_2$ column density as
  $\tau_{\rm X} = 2\,\sigma_{\rm H}(E) \, N({\rm H}_2)$.}. This is because in the nuclear regions of AGN, 
  the molecular gas is expected to be mostly in the plane of the nuclear disk or torus,
  where we assume the PAH molecules to be located as well.   We show in Figure~\ref{fig:rAGNvsNH} 
  the values of the distance from the AGN as given by the ground-based
  mid-IR observations against the nuclear column density averaged on this
  region. The symbols are color-coded according to the intrinsic
  $2-10\,$ keV AGN luminosity.
  There is little dependence of the detection of nuclear $11.3\,\mu$m PAH emission
  on either AGN luminosity or distance from the AGN. The median values are similar, 
    $r_{\rm AGN}=32\,$pc
    and $\log L(2-10\,{\rm keV})=42.7\,{\rm erg\,s}^{-1}$  for the nuclei with $11.3\,\mu$m PAH detections, and
    $r_{\rm AGN}=41\,$pc
    and $\log L(2-10\,{\rm keV})=42.5\,{\rm erg\,s}^{-1}$ for those without detections.  
  It thus  appears that the
  nuclear column density plays a dominant role,  although having
  a high column density does not imply that PAH emission is
  detected in the nuclear region.

\subsection{Half-life of PAH molecules in the nuclear regions of Seyfert galaxies}\label{subsec:halflive}
In the ISM of galaxies, there is evidence that
the PAH spectra are produced mostly by large molecules
\citep[see, e.g., ][and also Section~\ref{subsec:pahratios}]{Draine2001}.
However, photodissociation cross-sections at hard X-ray energies have
been measured only for molecules with a number of carbons 10 and 16
\citep{Monfredini2019}. Monfredini and collaborators proposed 
to use half-life of a particular PAH molecule as an estimate of the survival time
against dissociation due to the incidence of X-ray photons,

\begin{equation}\label{eq:halflife}
  t_{\rm 1/2} =
  \frac{\ln 2}{F_{\rm X}(E).\sigma_{\rm ph-d}(E)}
,\end{equation}

\noindent where $\sigma_{\rm ph-d}$ is the photodissociation
cross-section for a given energy $E$. The half-life includes the 
dependencies we investigated
in Section~\ref{subsec:dependencies}. In other words, 
the half-life of a particular PAH molecule is longer
at larger distances from the AGN and at higher protecting column densities, and
shorter for more luminous AGN and PAHs with 
larger photodissociation cross-sections
(Equations~\ref{eq:flux} and \ref{eq:halflife}). Therefore, we can test whether
$t_{\rm 1/2}$ is related to the nuclear detection of the
$11.3\,\mu$m PAH feature. However, we note that from this
experimental work, the prediction would be that in regions with 
little protection from molecular material most of the emission 
should arise from multiply ionized PAHs. Models of emission from
ionized PAH molecules predict that the $11.3\,\mu$m feature is
decreased relative to the emission of the $7.7\,\mu$m
PAH feature, but it is not absent \citep{Draine2001}. 

  \begin{figure}
   \hspace{-0.3cm}
  \includegraphics[width=8.5cm]{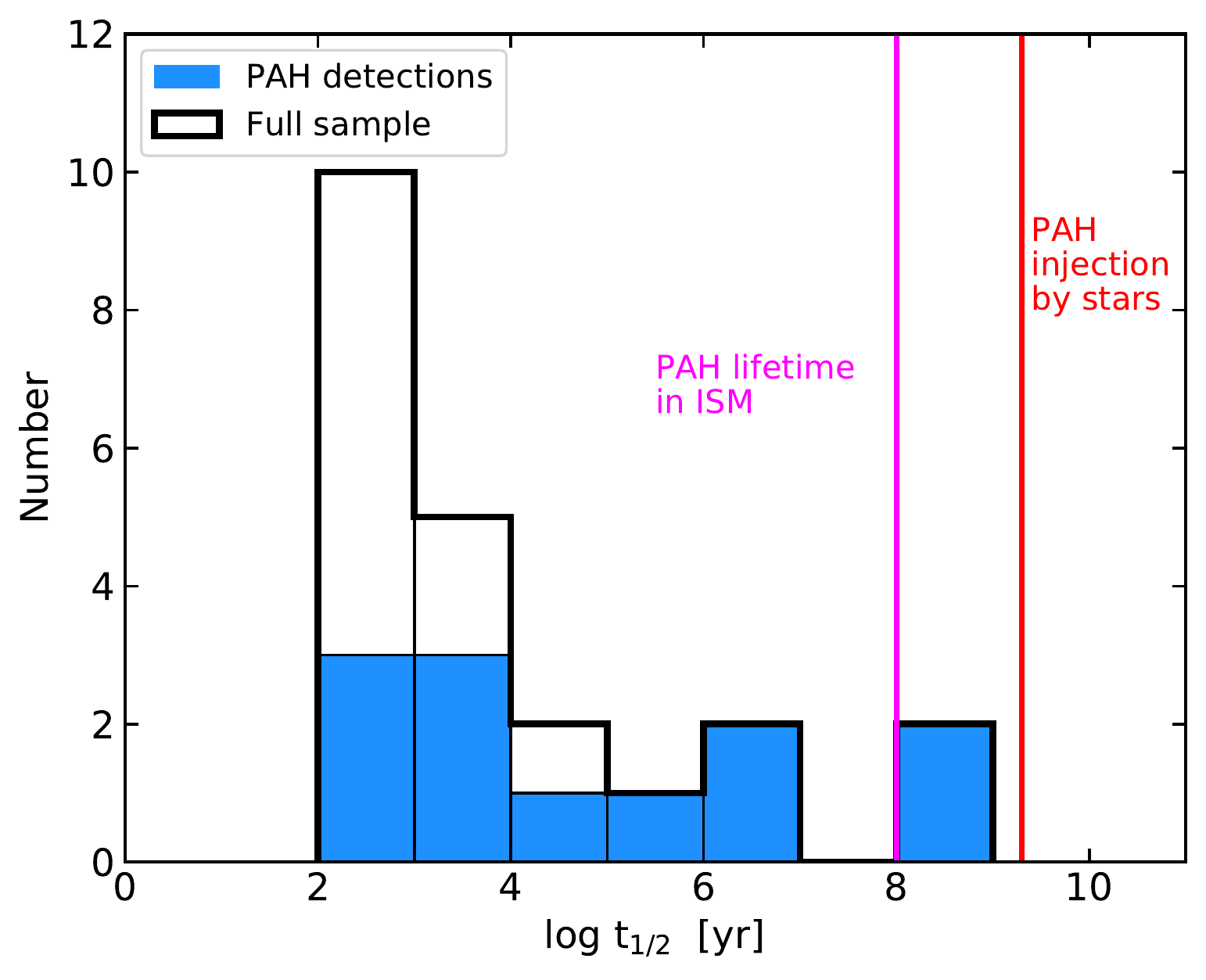}

  \caption{Computed half-lives for the illustrative
    PAH molecule naphthalene in 
    the nuclear regions of our Seyfert galaxies. The thick line
    histogram is the full sample, and the filled
histogram are the galaxies with a nuclear detection of the 
$11.3\,\mu$m PAH feature. We indicate the typical lifetimes of PAHs
in the diffuse ISM \citep[see][]{Tielens2013} and
the timescale for PAH injection into the ISM by evolved
  stars \citep{Jones1994}.}\label{fig:halflives}
    \end{figure}

We chose the 
naphthalene molecule (C$_{10}$H$_8$) as an illustrative example to
compute the half-lives of PAHs
in the nuclear regions of our Seyfert galaxies.
This PAH molecule among those studied by \cite{Monfredini2019} has 
the smallest photodissociation
cross-section when exposed to hard X-ray photons at $E=2.5\,$keV. 
Thus, naphthalene provides upper limits to the 
half-lives (see Equation~\ref{eq:halflife}) expected in the environments of active nuclei,
where the larger molecules might be common.
In addition to the photodissociation cross-section of naphthalene, 
we used the nuclear H$_2$ column densities
(Table~\ref{tab:NucMolGas}), $2-10\,$keV intrinsic luminosities, and
distances from the AGN. The main difference with similar calculations 
reported by \cite{Monfredini2019} is that in our case, the column densities
are measured over the same region as is probed by the ground-based mid-IR slit.

The PAH
half-lives for the nuclear regions in our sample of Seyfert galaxies span  a
broad range (Figure~\ref{fig:halflives}).
Galaxies with short nuclear PAH half-lives of less than approximately
1000\,yr include low-luminosity AGN with low column densities (e.g.,
NGC~7213) as well as luminous Seyfert galaxies without enough protecting 
material in their nuclear regions (NGC~7582). The 
 longest estimated half-lives ($>10^6\,$yr) are for the low-luminosity AGN NGC~1808 \citep[see
  also][]{Monfredini2019}, as well as NGC~5643, NGC~7130, and NGC~7469. These four
nuclei also show the 
highest nuclear $N$(H$_2$) values.   For comparison we show
the values of the PAH injection time by AGB stars into the ISM from
\cite{Jones1994}
and the lifetime of $\sim 10^{8}\,$yr of PAHs in the diffuse ISM \citep[see][and
references therein]{Tielens2013}. We note, however, that the estimated PAH
lifetimes in different environments with hot gas that are exposed to X-rays
can be as short as those calculated for our sample of
Seyfert nuclei \citep{Micelotta2010}. However, PAH emission is detected in them, indicating
that the PAH molecules are not completely destroyed.

In Seyfert nuclei without a nuclear detection of the
$11.3\,\mu$m PAH feature the half-lives for the
naphthalene molecule tend to be shorter than in 
those with nuclear detections (Figure~\ref{fig:halflives}). We remark that these half-lives are only indicative of
the general behavior. Observationally, 
they are affected by large uncertainties, especially
those coming from the determination of  mass of molecular hydrogen and thus
the nuclear H$_2$ column densities
(Section~\ref{subsec:molgas}). Moreover, we also used 
the values of $N({\rm H}_2)$ averaged over the nuclear regions. However,
  the nuclear regions of some galaxies contain clumps with higher
  column densities. 
We also ignored the effects of dehydrogenation of PAH molecules when
exposed to X-rays \footnote{Although these effects have not been studied observationally
in AGN, predictions from density functional theory calculations
indicate that the most noticeable changes would be around the $3.3\,\mu$m PAH region
and  the $20-30\,\mu$m
region \citep{Buragohain2018}.}.
Finally, we emphasize that we only computed the half-lives for the
small PAH molecule naphthalene. We cannot rule out, however, that larger PAH molecules may  behave differently upon absorption of a hard
X-ray photon (see Rigopoulou et al. 2020, in prep.).


To summarize, in the majority of the nuclear regions we studied,
the computed half-lives of the PAH molecule naphthalene are short compared to their
formation or re-formation times, even in cases of nuclear detections of the
$11.3\,\mu$m PAH feature. On the other hand, the experimental work of
\cite{Monfredini2019} also showed that
ionized and multiply ionized PAHs are formed when PAH molecules are exposed to hard X-ray photons. Ionized PAHs tend to show reduced but not completely absent,$11.3\,\mu$m PAH emission
when compared to the $7.7\,\mu$m PAH complex \citep[see, e.g.,][]{Allamandola1999, Draine2001}.
Thus, even in the harsh environments of these active nuclei, the PAH molecules
are not completely destroyed, and it is likely that unless the column densities
are extremely high, most of the emission might be dominated by ionized PAHs.

    \begin{figure*}[ht!]
      \centering
      \vspace{-0.25cm}
  \includegraphics[width=14.5cm]{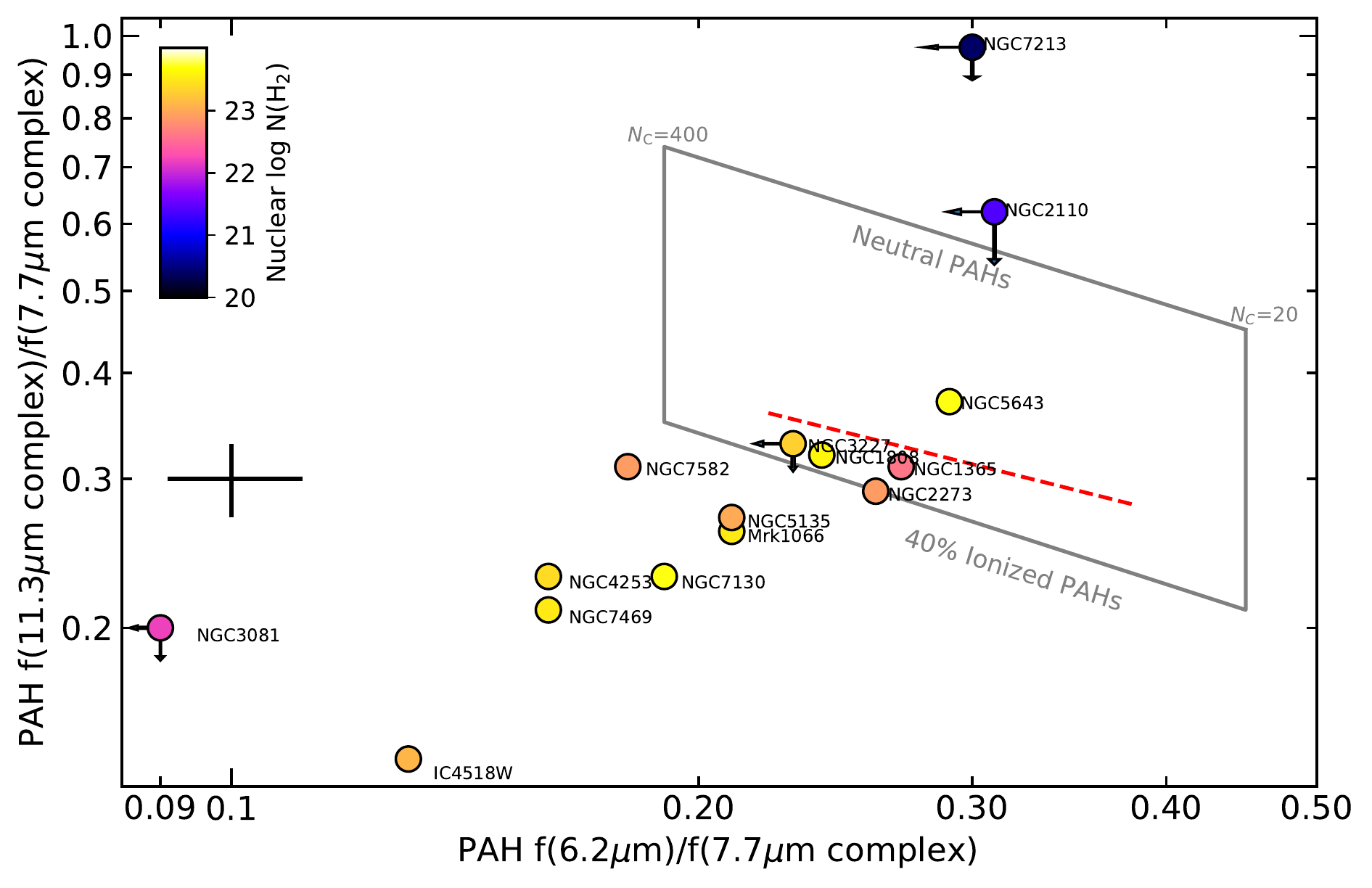}

  \caption{{\it Spitzer}/IRS PAH ratios  for the circumnuclear regions of the
    Seyfert galaxies in our sample (circles) except for those in highly inclined hosts
(see text). For NGC~3081 and NGC~7213, we 
only obtained an upper limit for the
$11.23\,\mu$m PAH. We assumed that its flux is approximately $1/4\text{}$ of that of the  $11.33\,\mu$m PAH,
as in other type 1 Seyfert galaxies. The
    symbols are color-coded with the nuclear H$_2$ column
    density. The plotted error bar in each direction represents
    the effect of a 10\% uncertainty in
    the $7.7\,\mu$m PAH complex flux on the measured PAH ratios.
    The lines are PAH model tracks (Rigopoulou et al. 2020) 
    for molecules in a Galaxy-like ISRF (solid gray lines), 
    an intense ISRF equivalent to $\times 10000$ the Galaxy (dashed red line),
    and numbers of carbon atoms from $N_{\rm C}=20$ to $N_{\rm C}=400$.
For the Galaxy-like
    ISRF, the top line
is for neutral PAHs and the bottom line for a 40\% fraction of
ionized PAHs. The number of carbon
  atoms for the PAH molecules increases from right to left.}\label{fig:PAHratios}
    \end{figure*}

    \begin{figure}[ht!]
      \centering
  \includegraphics[width=8cm]{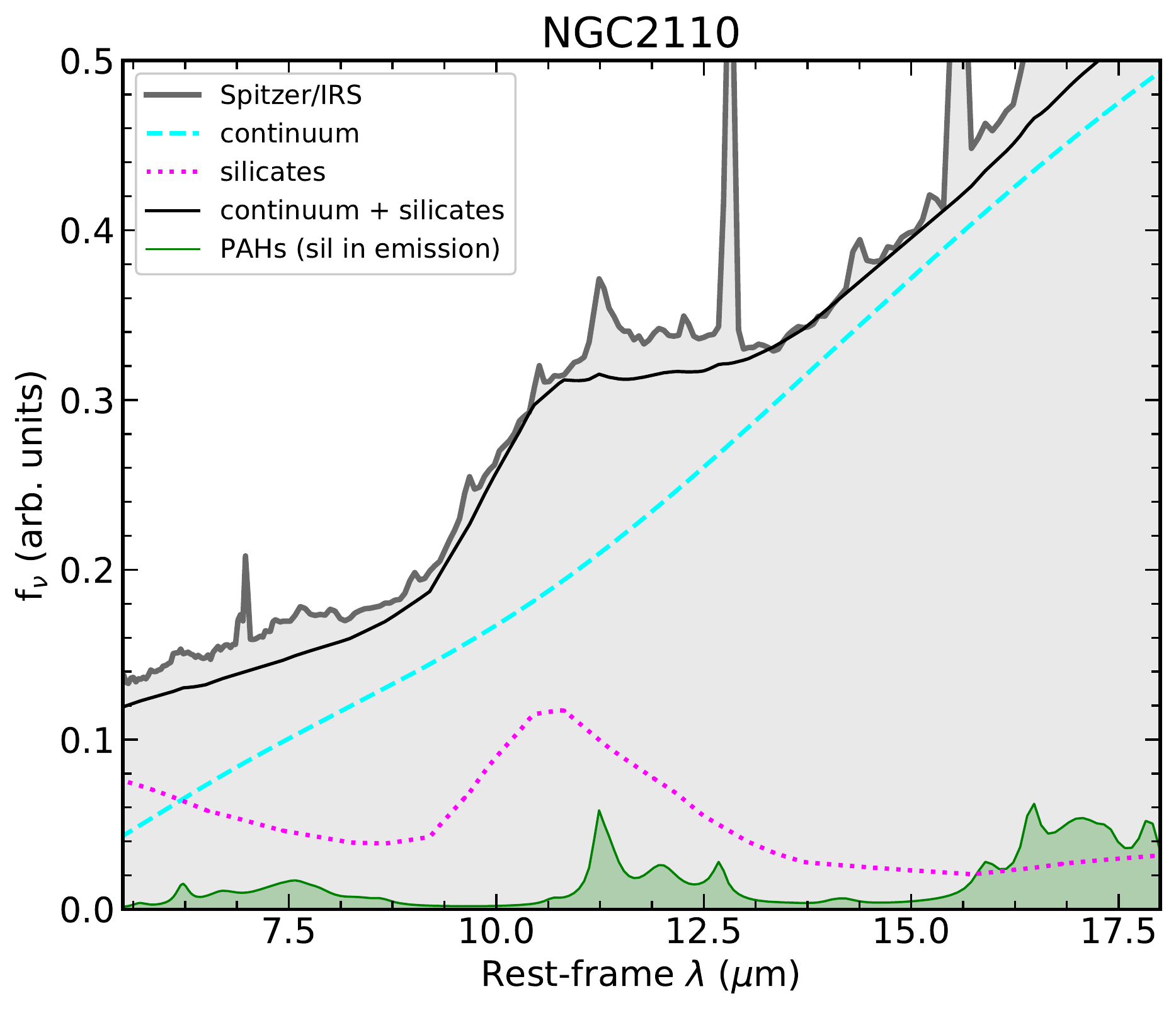}
  \includegraphics[width=8cm]{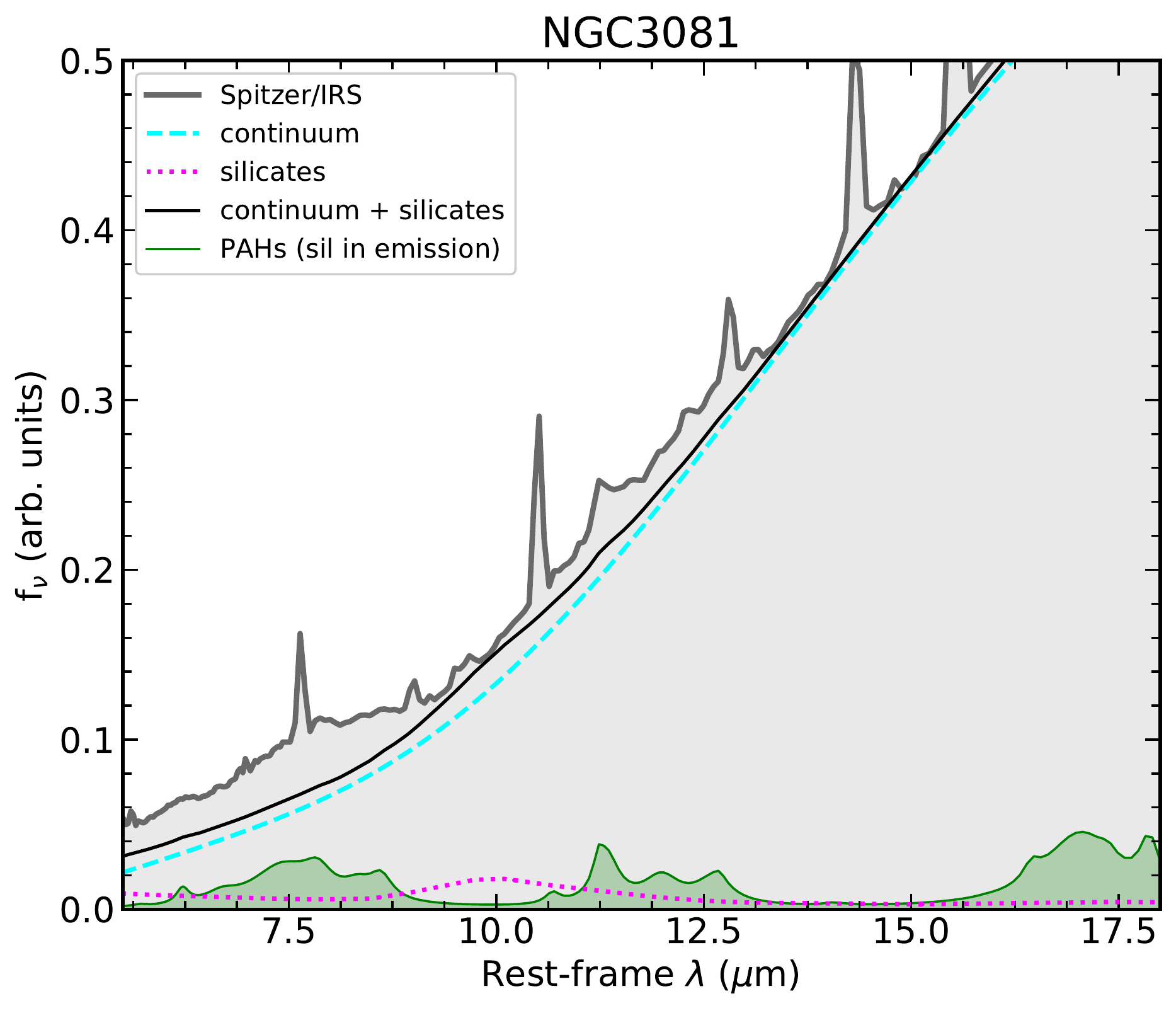}
  \includegraphics[width=8cm]{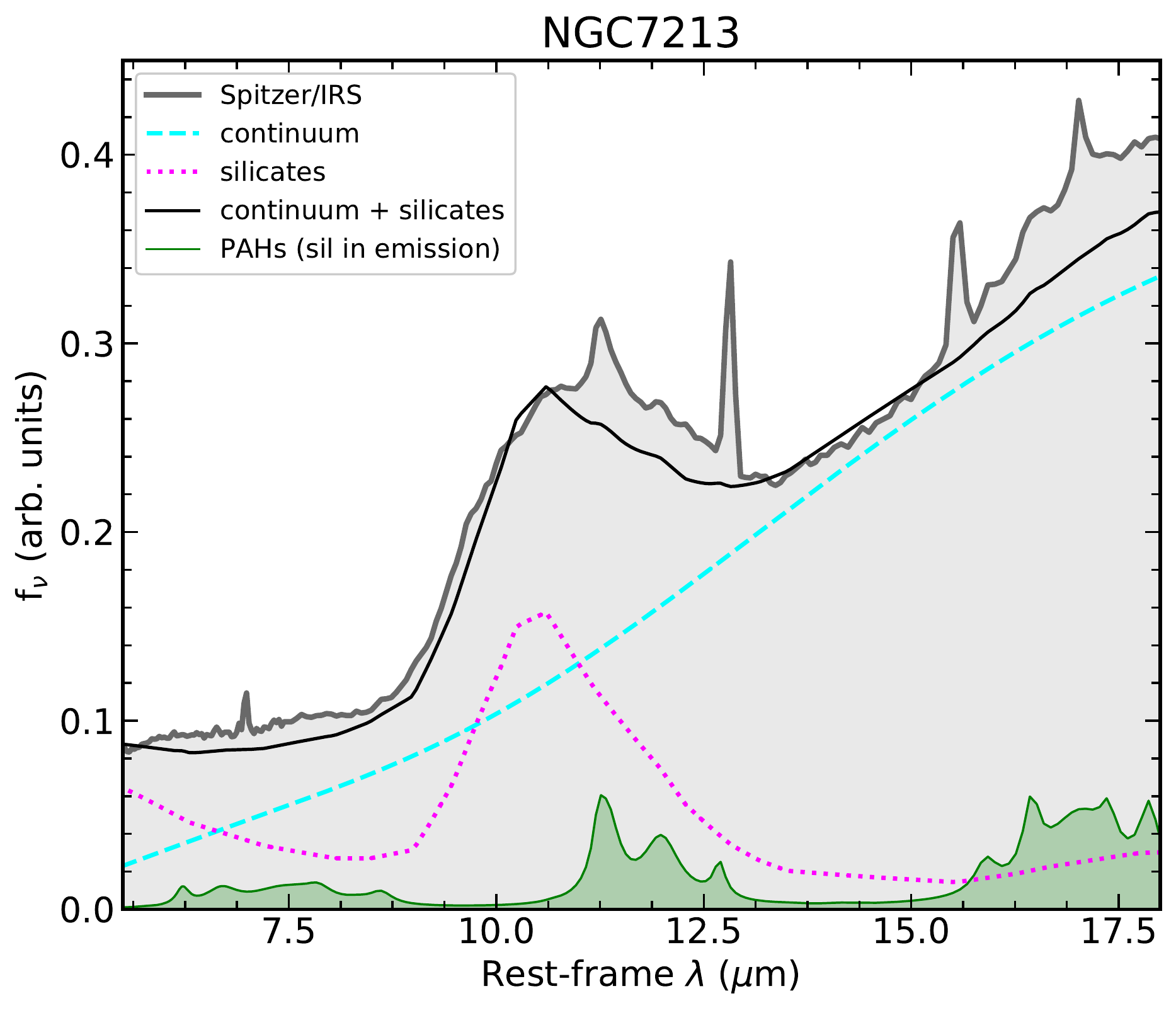}

  \caption{{\it Spitzer}/IRS spectra (gray)
    of the three galaxies with the lowest nuclear $N$(H$_2$). We show
    the continuum (dashed cyan lines) and silicate feature in emission (dotted magenta lines), and
    their sum (black line) as well as 
    the PAH feature spectra (green curves) fitted
    with {\sc pahfit}.}\label{fig:lowNHpahfit}
    \end{figure}

 \section{Circumnuclear PAH  and CO(2-1) emission}\label{sec:circumnuclear}
  
\subsection{Spitzer/IRS PAH emission}\label{subsec:pahratios}

In the previous section we argued that in the nuclear regions of
our sample of Seyfert galaxies the PAH emission may
originate preferentially from small and/or ionized PAH molecules. The exception
would be nuclei with
high H$_2$ column densities, where a higher fraction of the emission would come
from neutral PAHs. 
The $f(11.3\,\mu$m)/$f(7.7\,\mu$m)  versus $f(6.2\,\mu$m)/$f(7.7\,\mu$m) PAH ratio
diagram
has been used to investigate  the fraction of
neutral and ionized PAH
emission in galaxies \citep[see, e.g.,][]{Galliano2008, Lambrides2019},
including nearby Seyfert galaxies \citep{DiamondStanic2010}
and Galactic sources, by comparing with
theoretical predictions for PAH molecules \citep[see, e.g.,][]{Draine2001}.

The  {\it Spitzer}/IRS PAH ratios (Figure~\ref{fig:PAHratios}) are for the central $4\arcsec \times
4\arcsec$ regions (except for NGC~4253, see
Section~\ref{subsec:IRSobs}), which  cover  physical
regions between $\sim 270\,{\rm pc} \times 270\,{\rm pc}$ for NGC~1808 and
$1.3\,{\rm kpc} \times 1.3\,{\rm kpc}$ for IC~4518W.
We excluded from this figure highly inclined galaxies (NGC~1320, NGC~1386,
NGC~2992, NGC~4388, and NGC~7172) because their PAH ratios
may be {\it \textup{contaminated}} by emission from line-of-sight regions in their disks.
For reference, galaxies not dominated by AGN emission generally exhibit
PAH ratios  $f(6.2\,\mu$m)/$f(7.7\,\mu$m) $\gtrapprox
0.4$ to $\sim 0.7$ and
$f(11.3\,\mu$m)/$f(7.7\,\mu$m)$\gtrapprox 0.25$ to $\sim 0.4$ \citep{Lambrides2019}.
We also show PAH
model tracks Rigopoulou et al. 2020, in prep.) for an
interstellar radiation field (ISRF) similar to that of our Galaxy and an intense ISRF.
For the Galaxy-like ISRF, we show one track for
neutral PAHs and another track with a 40\% fraction of ionized PAHs.
The observed {\it Spitzer}/IRS PAH ratios for the majority of our
sample are close to this last track and would
indicate a larger fraction of the emission coming from ionized PAHs
and/or a different ISRF 
than in purely star-forming galaxies.


%
There is no clear dependence of the {\it Spitzer}/IRS PAH
ratios on the nuclear column density (see
Figure~\ref{fig:PAHratios}). In most galaxies in our sample, the CO(2-1) and presumably the PAH emissions 
are not centrally peaked. Thus, the nuclear $N({\rm H}_2$) is not
representative of the column densities in the circumnuclear regions.
NGC~5643 is the galaxy with the highest {\it Spitzer}/IRS
PAH ratios in our sample, which would indicate a higher contribution from neutral PAHs. 
For this galaxy in the central $4\arcsec \times 4\arcsec$,  most of  the CO(2-1) emission
comes from the nuclear region (see Figure~\ref{fig:CO21maps}). At the peak of the CO(2-1)
emission, which is $0.2\arcsec$ from the AGN position, the H$_2$ column density
is even higher \citep[$N({\rm H}_2) \sim 7 \times 10^{23}\,{\rm
    cm}^{-2}$, see][]{AlonsoHerrero2018} than the average value over the
nuclear region. 
IC~4518W shows the lowest {\it Spitzer}/IRS PAH ratios. The CO(2-1)
(see Figure~\ref{fig:CO21maps}), mid-IR, and Pa$\alpha$ morphologies 
over the central $4\arcsec \times 4\arcsec$ of IC~4518W are
complicated, showing in addition to the Sy 2 nucleus, a number of circumnuclear
regions with ongoing or recent SF activity \citep{AlonsoHerrero2006,
  DiazSantos2008, EsparzaArredondo2018}. 
Because the $11.3\,\mu$m PAH is not detected in
the nuclear spectrum,  most of the PAH emission 
in the {\it Spitzer}/IRS spectrum originates in the circumnuclear regions where
the radiation field is thought to be intense, so that most of the emission
appears to be dominated by ionized PAHs.


NGC~2110, NGC~3081, and
NGC~7213 present the lowest values of the nuclear $N({\rm H}_2)$ as well as 
the faintest PAH emission
in the {\it Spitzer}/IRS spectra (see Table~\ref{tab:SpitzermidIR}). Figure~\ref{fig:lowNHpahfit} shows that
NGC~2110 and NGC~7213 have similar fitted PAH spectra with 
faint emission from the $7.7\,\mu$m PAH complex with respect to that of the
$11.3\,\mu$m PAH complex, thus
suggesting some large contribution from neutral PAH.
Neither galaxy appears to
have strong (circum-) nuclear SF activity \citep{Schnorr-Mueller2014NGC2110, Schnorr-Mueller2014NGC7213}. The CO(2-1) emitting regions are
$1-2\arcsec$ away from the AGN position \citep[see Figure~\ref{fig:CO21maps} and ][]{Rosario2019}, where the PAH molecules would be more
protected from the AGN radiation. NGC~3081, on the other hand, shows a fitted PAH
spectrum  with a stronger $7.7\,\mu$m PAH complex compared to the
6.2 and $11.3\,\mu$m PAH, suggesting a larger
contribution from ionized molecules.
The CO(2-1) emitting region at $\sim 2\arcsec$
southeast of the AGN has optical line emission excited by SF \citep{Schnorr-Mueller2016},
and thus it is possible that the PAH emission is mostly originating from this region. In
Figure~\ref{fig:PAHratios}, NGC~3081 appears to be far from the PAH model tracks. However, \cite{Zhen2017, Zhen2018}
showed experimentally that singly and doubly ionized large PAH molecules
show a mid-IR spectra with
faint or even absent 6.2 and $11.3\,\mu$m PAH emission with respect to the $7.7\,\mu$m PAH
complex.

\subsection{Surface brightness profiles}\label{subsec:surfacebrightnessprofiles}


Observationally, there appears to be a spatial correspondence
between cold molecular gas as traced by  CO transitions 
and the PAH emission based on the observed radial 
distributions on large scales (hundreds of parsec)
in  nearby disk galaxies \citep{Regan2006, Bendo2010} and  the
galaxy-integrated emission of infrared luminous
galaxies \citep{Cortzen2019}. Even in the
powerful SF-driven outflow of M82, the PAH emission is observed to
be spatially correlated with that of the CO(2-1) emission
\citep{Leroy2015}.

\begin{figure}

\centering
  \includegraphics[width=9cm]{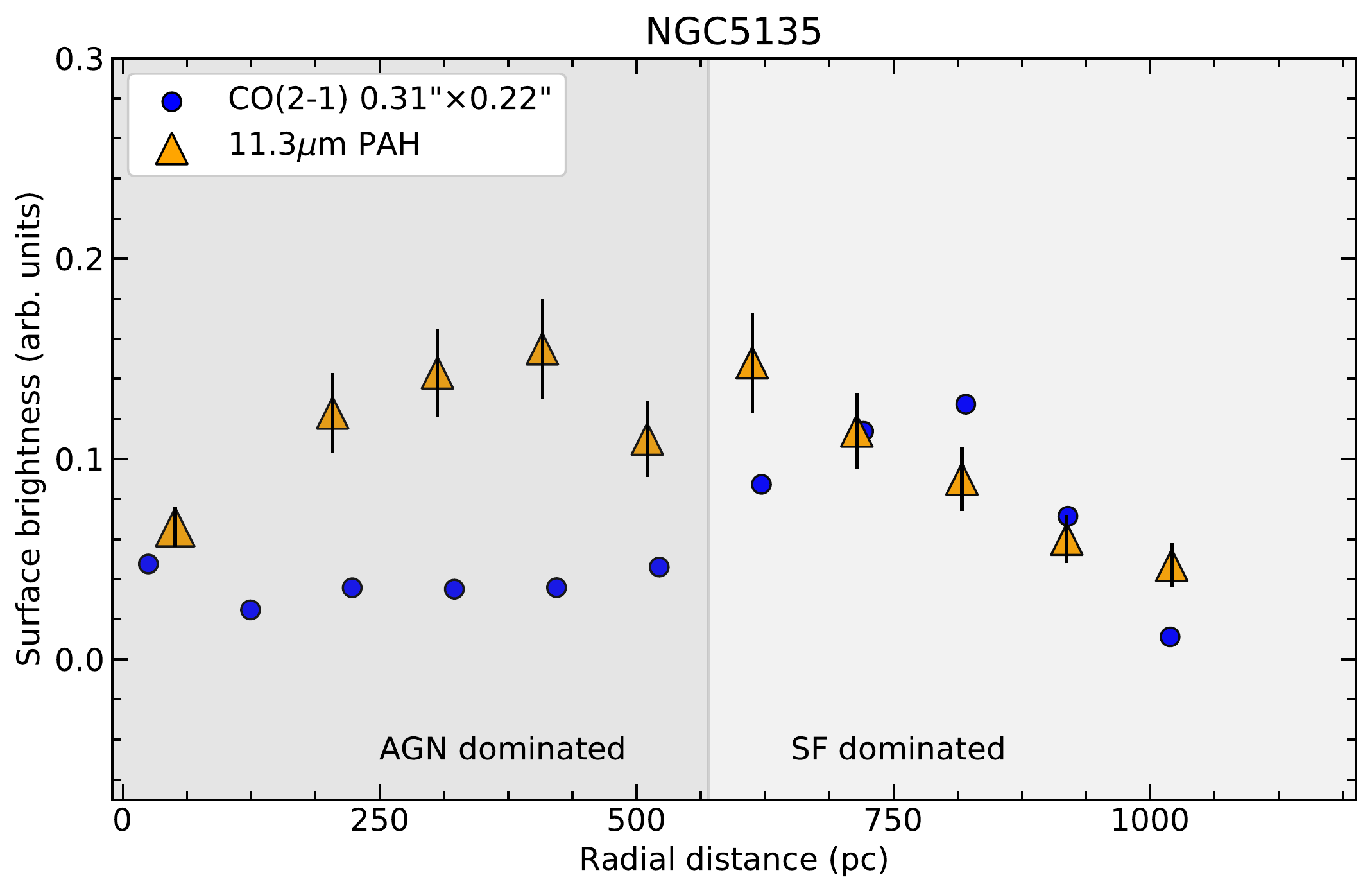}

  \includegraphics[width=9cm]{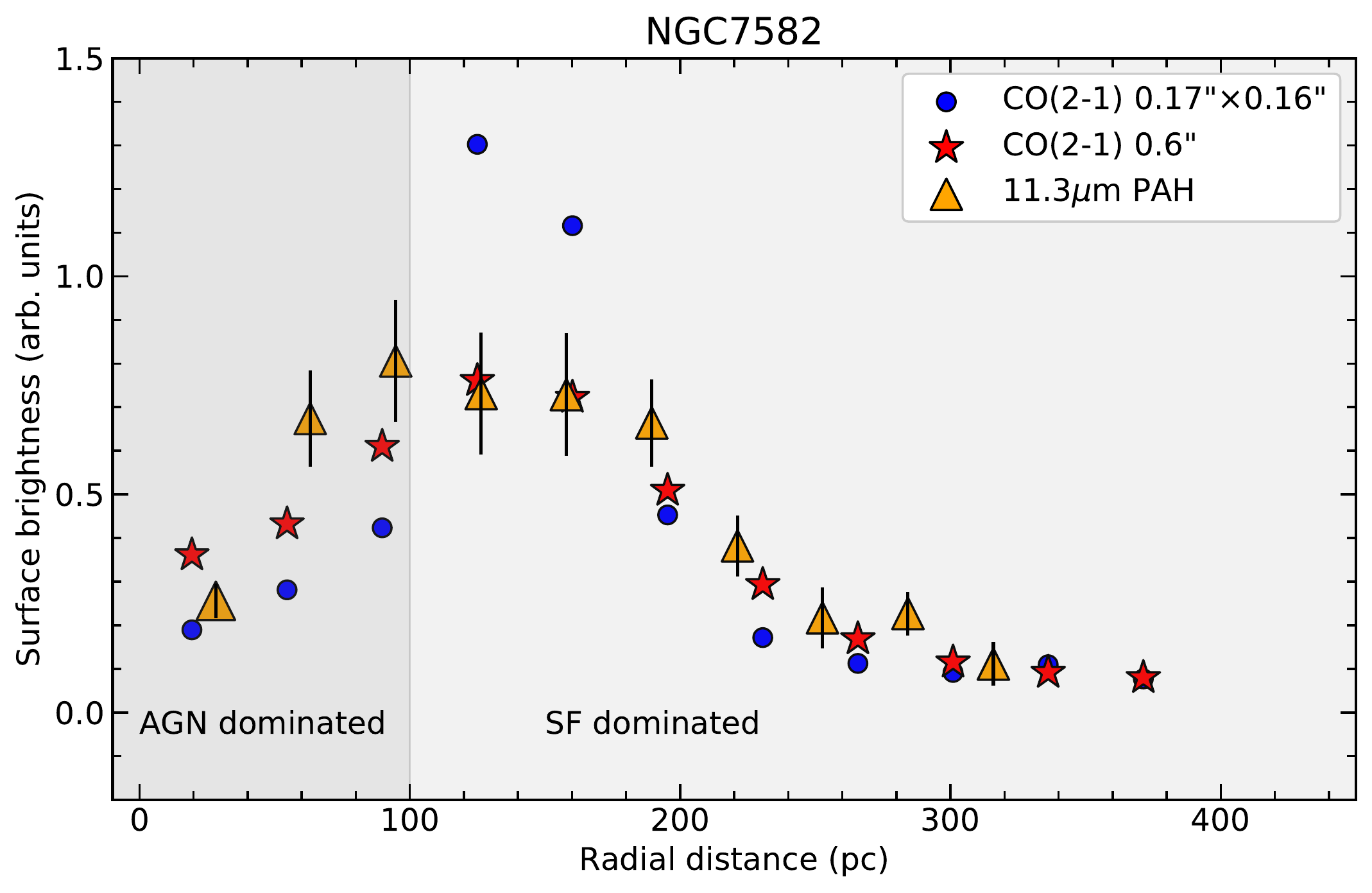}

  \caption{Radial surface brightness profiles (arbitrary units) of the
    CO(2-1) emission (circles for the original ALMA resolution and
    stars for the Gaussian-smoothed image of NGC~7582) and $11.3\,\mu$m
    PAH emission \citep[orange triangles,
    from][]{EsparzaArredondo2018}
    for NGC~5135 (top
    panel) and NGC~7582 (bottom panel, converted into the assumed
    distance in this work). The CO(2-1) radial profiles were
    extracted at the orientation of the
    mid-IR slits (see Figure~\ref{fig:CO21maps},
    Table~\ref{tab:midIR}, and
    Section~\ref{subsec:phot}). The large triangles indicate the nuclear
    $11.3\,\mu$m PAH emission measurement
    \citep[see][for further details]{EsparzaArredondo2018}. We also
    show the approximate separation between AGN- and SF-dominated
    regions (see text for more details).}\label{fig:surfacebrightnessprofiles}
    \end{figure}

For the two selected galaxies with extended
  PAH emission from ground-based observations (Section~\ref{subsec:phot}), we show in
Figure~\ref{fig:surfacebrightnessprofiles} the surface brightness profiles of the CO(2-1) emission
extracted along the orientation of the
mid-IR slit, that is, PA$_{\rm slit}= 30\deg$  for NGC~5135 and
PA$_{\rm slit}= 0\deg$ for NGC~7582. The $11.3\,\mu$m PAH surface brightness
profiles \citep{EsparzaArredondo2018} are also shown in
arbitrary units and  scaled approximately
at the location of the SF rings. In these regions we expect the
CO(2-1) and PAH emission to trace each other through the SF laws.
For both galaxies, the CO(2-1) and $11.3\,\mu$m PAH surface brightnesses
appear to be spatially correlated at projected radial distances $r \gtrapprox
100\,$pc out to $r\sim 500\,$pc for NGC~7582 and $r \gtrapprox
600\,$pc out to $r \simeq 1\,$kpc for NGC~5135. These scales are
approximately coincident with
the location of the  circumnuclear rings of SF,  as traced by hydrogen recombination
lines \cite[see][and also Section~\ref{subsec:AGNpos}]{Riffel2009, Ricci2018,
  AlonsoHerrero2006, Bedregal2009}.

At smaller radial distances
closer to the AGN, some significant differences start to
appear. The mid-IR slit of NGC~5135 was placed exactly at the orientation of the
ionization cone, which extends for approximately 2$\arcsec \sim 570\,$pc
from the AGN mostly to the northeast  \citep{Bedregal2009}.
Along this direction, 
both the observed morphology (see Figure~\ref{fig:CO21maps}) and 
the surface brightness profile compared to that of the PAH feature are suggestive of a
CO(2-1) emission deficit, especially in the central 3$\arcsec \simeq
900\,$pc.  However,
the emission from  the H$_2$ $2.12\,\mu$m line,  
 high-excitation lines, 
 and the hard X-rays  peaks
 at the AGN position and is extended over similar physical scales in this
 direction
 \citep{Bedregal2009, DiazSantos2010, Colina2012, EsparzaArredondo2018}.
At projected radial distances 
between $r\simeq 1\arcsec \sim 290\,$pc and $r\simeq 2\arcsec \simeq
580\,$pc, 
the $11.3\,\mu$m PAH
emission is bright.  If this region is an X-ray dominated region (XDR), then we would expect
 that most of
 the CO emission would originate from higher-J transitions
 \citep[see, e.g.,][]{vanderWerf2010} rather than
 from the CO(2-1) transition. Another possibility is
 that the PAH emission might additionally be excited
 by AGN-produced UV photons, although these regions
 are far from the area of influence of the AGN \citep[][]{Jensen2017}. 
 UV photons from nonionizing stars
 detected in the central $\sim 2\arcsec$ of this galaxy
 \citep{StorchiBergmann2000,
   GonzalezDelgado2001} could
 excite the PAH molecules but would not produce
  Pa$\alpha$ emission.

The orientation of the mid-IR slit of NGC~7582 was north-south. This means that
the slit was not placed along the axis of the ionization cone of
this galaxy ($\sim 225\deg$), although the cone is wide angled. 
Out to projected radial distances of $r\simeq 100\,$pc there is an excess of 
$11.3\,\mu$m PAH emission when compared to that of CO(2-1), although
not as pronounced as in NGC~5135. In this
region of NGC~7582, the optical line ratios
\citep{Ricci2018} indicate a significant contribution from the AGN radiation
field. This could explain the decreased CO(2-1) emission
in this region. In the circumnuclear
ring of SF, the individual regions have higher CO(2-1) fluxes and thus higher $N({\rm H}_2)$.
However, the PAH molecules there are likely exposed to SF-produced UV photons, and therefore the PAH
emission would have a strong component from ionized molecules, as  is apparent from
the {\it Spitzer}/IRS PAH ratios (see Figure~\ref{fig:PAHratios}). 

Finally, in NGC~1808, the $8.6\,\mu$m PAH emission relative to the $11.3\,\mu$m PAH emission increases at
larger radial distances ($r\sim 75\,$pc)
where there is also X-ray emission \citep{Sales2013}.
While our CO(2-1) data have only moderate angular
resolution, the  ALMA CO(3-2) observations 
\citep{Combes2019} resolved the central region in a nuclear spiral 
with high H$_2$ column density at the AGN position. This would
provide the molecules with sufficient shielding, resulting
in photodissociation half-lives for the PAH comparable to the timescales
for PAH injection into the ISM. At larger radial distances, the decreasing
H$_2$ column density would explain the observed 
increased fraction of emission produced by ionized PAHs.

\section{Summary and prospects for JWST}\label{sec:discussion}

We analyzed the properties of the cold molecular gas and PAH emission in the nuclear
($\sim 24-230\,$pc) and circumnuclear ($\sim 250\,$pc to  $\sim 1.3\,$kpc scales)
regions in a sample of 22 nearby Seyfert galaxies. We used ALMA and NOEMA
high angular resolution observations of the CO(2-1) transition
to trace the cold molecular gas, and 
ground-based and {\it Spitzer}/IRS mid-IR spectroscopy for the PAH emission.
Out of the 22 galaxies in our sample, the
$11.3\,\mu$m PAH feature was detected in the nuclear regions of 12, while PAH emission
is always detected within the $\sim 4\arcsec$ IRS apertures.

The majority of galaxies in our sample have 
molecular gas masses in the range $10^6-10^8\,M_\odot$ over the
nuclear region sampled by the ground-based mid-IR slits. These masses are
higher than the typical values for nuclear disks and tori of Seyfert galaxies \citep{AlonsoHerrero2018,
  AlonsoHerrero2019, Combes2019, GarciaBurillo2019}. However, this is explained by the fact that our nuclear apertures,
which were dictated by the ground-based mid-IR slits, are a factor of a few larger than the Seyfert tori.
In Seyfert galaxies with a nuclear detection of the $11.3\,\mu$m PAH feature, the 
median values of the molecular gas masses ($1.6 \times 10^7\,M_\odot$) and
H$_2$ column densities ($N({\rm H}_2) = 2 \times 10^{23}\,{\rm cm}^{-2}$)
are higher than in those without
a detection. This might indicate that the molecular gas plays a role in
protecting PAH molecules in these regions. Additionally, nuclei
with the largest H$_2$ surface brightnesses also host recent SF
activity, which in turn might be responsible for
the detection of the $11.3\,\mu$m PAH emission.
Nevertheless,  the nuclear 
column density  $N({\rm H}_2)$ appears to be the dominant factor for this PAH detection
when compared to
 the molecular gas mass, the distance from the AGN $r_{\rm AGN}$ (given
 by half the mid-IR slit width), and the AGN luminosity.

 Using the small PAH molecule naphthalene as an illustrative case, we computed its half-live when
 exposed to 2.5\,keV hard X-ray photons in the nuclear regions of our sample. The flux of X-ray photons in turn
 is a function of the protecting column density  $N({\rm H}_2)$, as well as $r_{\rm AGN}$ and $L_{\rm X}({\rm AGN})$.
 Except in the cases of high column densities and/or low-luminosity AGN, the computed half-lives for
 naphthalene are considerably shorter than the PAH injection
 timescales in the ISM ($\sim 10^9\,$yr), even for nuclei with detections
 of the $11.3\,\mu$m PAH feature. However, in nuclei without a $11.3\,\mu$m
 PAH detection, the computed half-lives are shorter than in those with a detection.

 On circumnuclear scales, the observed {\it Spitzer}/IRS  PAH ratios of most of the galaxies in our sample
 (excluding those in highly inclined galaxies) indicate a significant contribution from ionized 
 PAHs. In general, we do not find that nuclei with the highest H$_2$ column densities are closer
 to  neutral PAH model tracks. The reason for this is that the majority of galaxies we studied does not
 show centrally peaked CO(2-1) emission, and in some of them, the brighest CO(2-1) emission traces
 circumnuclear sites of strong SF activity. Thus, with the {\it Spitzer}/IRS apertures
 we cannot easily distinguish between AGN and SF-formation dominated regions for our sample of
 Seyfert galaxies. 

 We also analyzed in detail the surface brightness profiles of the ground-based $11.3\,\mu$m
 PAH emission and CO(2-1) for two galaxies (NGC~5135 and NGC~7582)
 in our sample with circumnuclear rings of SF and
 ionization cones. In the rings of SF, the CO(2-1) and  $11.3\,\mu$m
 PAH emissions trace each other
 spatially, as expected from SF laws. In the AGN-dominated regions of these galaxies, we detected 
 not only emission from the  $11.3\,\mu$m PAH, but also
 an excess with respect to that of the CO(2-1) emission. This could be due
   to additional excitation of the PAH molecules by AGN-produced UV photons or,  more likely,
to  a deficit of CO(2-1) emission in these XDR. 

The medium-resolution spectrograph (MRS) of the {\it JWST} Mid-Infrared Instrument 
 \citep[MIRI, ][]{Rieke2015, Wright2015}
 will provide high-sensitivity observations at angular resolutions
(FWHM$\sim 0.3$$\arcsec$ at $8\,\mu$m) similar to
those of the ground-based mid-IR spectroscopic observations used in
this work. Moreover,   the MIRI-MRS $\sim 5-28\,\mu$m
spectral range covers  a large number of PAH features,
fine-structure lines, and molecular hydrogen lines. The improved  spatial resolution with respect
to {\it Spitzer}/IRS (a factor of ten) will
allow a better estimate of the AGN continuum, for instance, by
comparing the nuclear emission
with predictions from clumpy torus models. Moreover, using different
combinations of fine-structure lines observed with MIRI-MRS, it will be
possible to separate
AGN-dominated and SF-dominated regions in nearby Seyfert galaxies. In AGN-dominated regions with
low H$_2$ column densities it should be possible to detect a decreasing ionized PAH fraction
at increasing distances from the AGN. Spatially resolved MIRI-MRS
observations can be used to  determine differences in the ionized PAH spectra in
AGN and star-forming regions. In nuclear regions with high column densities and/or low-luminosity
AGN, we predict
an increased fraction of emission from neutral PAHs (e.g., the nuclei
of NGC~1808 and NGC~5643). The mid-IR rotational H$_2$
lines will provide  independent estimates of the masses and
column densities of the warm molecular gas and avoid the
uncertainties of the  CO-to-H$_2$ conversion factors near AGN. 

\begin{acknowledgements}

  We thank an anonymous referee for providing comments that helped improve
    the paper. 
  We are grateful to J. A. Fern\'andez Ontiveros for providing us with 
  ALMA measurements for two galaxies in our sample prior to publication.
  AA-H and SG-B acknowledge support
through grant PGC2018-094671-B-I00 (MCIU/AEI/FEDER,UE). AA-H and MP-S work
was done under project No. MDM-2017-0737 Unidad de Excelencia "Mar\'{\i}a
de Maeztu"- Centro de Astrobiolog\'{\i}a (INTA-CSIC).
MPS acknowledges support from the Comunidad de Madrid, Spain, through Atracci\'on
de Talento Investigador Grant 2018-T1/TIC-11035.
DRi acknowledges support from the University of Oxford John Fell Fund.
DRi and IGB acknowledge support from STFC through grant ST/S000488/1.
CRA and SGB acknowledge
support from grant AYA2016-76682-C3-2-P (MCIU/AEI/FEDER,UE).
TDS acknowledges support from the CASSACA and CONICYT fund CAS-CONICYT Call 2018. 
DE-A acknowledges support from a CONACYT scholarship.
DE-A and OG-M
acknowledge support from UNAM PAPIIT project IA103118.
OG-M acknowledge financial support of the UNAM PAPIIT project IN105720. 
SFH was supported by the EU Horizon 2020 framework program via the ERC Starting Grant
DUST-IN-THE-WIND (ERC-2015-StG-677117). CRA also
acknowledges support from the Spanish Ministry of Science, Innovation and
Universities  under grant with reference RYC-2014-15779. DRo acknowledges the support of
the UK Science and Technology Facilities Council (STFC) through grant ST/L00075X/1.

  This paper makes use of the following ALMA data: listed in Table~\ref{tab:CO21obs}. ALMA is a partnership of ESO (representing its member states), NSF (USA) and NINS (Japan), together with NRC (Canada), MOST and ASIAA (Taiwan), and KASI (Republic of Korea), in cooperation with the Republic of Chile. The Joint ALMA Observatory is operated by ESO, AUI/NRAO and NAOJ.

  This work is based on observations carried out under project numbers listed in Table~\ref{tab:CO21obs}
  with the IRAM NOEMA Interferometer. IRAM is supported by INSU/CNRS (France), MPG (Germany) and IGN (Spain).
  
  This work is based [in part] on archival data obtained with the Spitzer Space Telescope, which
  is operated by the Jet Propulsion Laboratory, California Institute of Technology under a contract
  with NASA. Support for this work was provided by NASA.
  This research has made use of the NASA/IPAC Extragalactic Database (NED),
which is operated by the Jet Propulsion Laboratory, California Institute of Technology,
under contract with the National Aeronautics and Space Administration.
\end{acknowledgements}

%
%

   \bibliographystyle{aa} 
   \bibliography{aalonsoherrero} 

\appendix

\section{ALMA 1.3\,mm continuum of individual sources}\label{sec:appendix}

As explained in Section~\ref{subsec:AGNpos}, we used the  ALMA 1.3\,mm  continuum  peaks
(see Table~\ref{tab:1p3mm}) 
to identify the AGN positions. Six galaxies have only one peak in the field of view, and the peak
coordinates agree well with the radio peaks listed in the literature as follows:
IC~4518W in \cite{Condon1996}, NGC~1386 and NGC~7172 in \cite{Thean2000}, and
NGC~7213 in \cite{Bransford1998}. NGC~3227 and NGC~5643 are discussed in detail in
\cite{AlonsoHerrero2019} and \cite{AlonsoHerrero2018}, respectively. The remaining
seven show several peaks in the ALMA 1.3\,mm continuum maps and are discussed below.

\begin{figure}
   \centering
  \includegraphics[width=8cm]{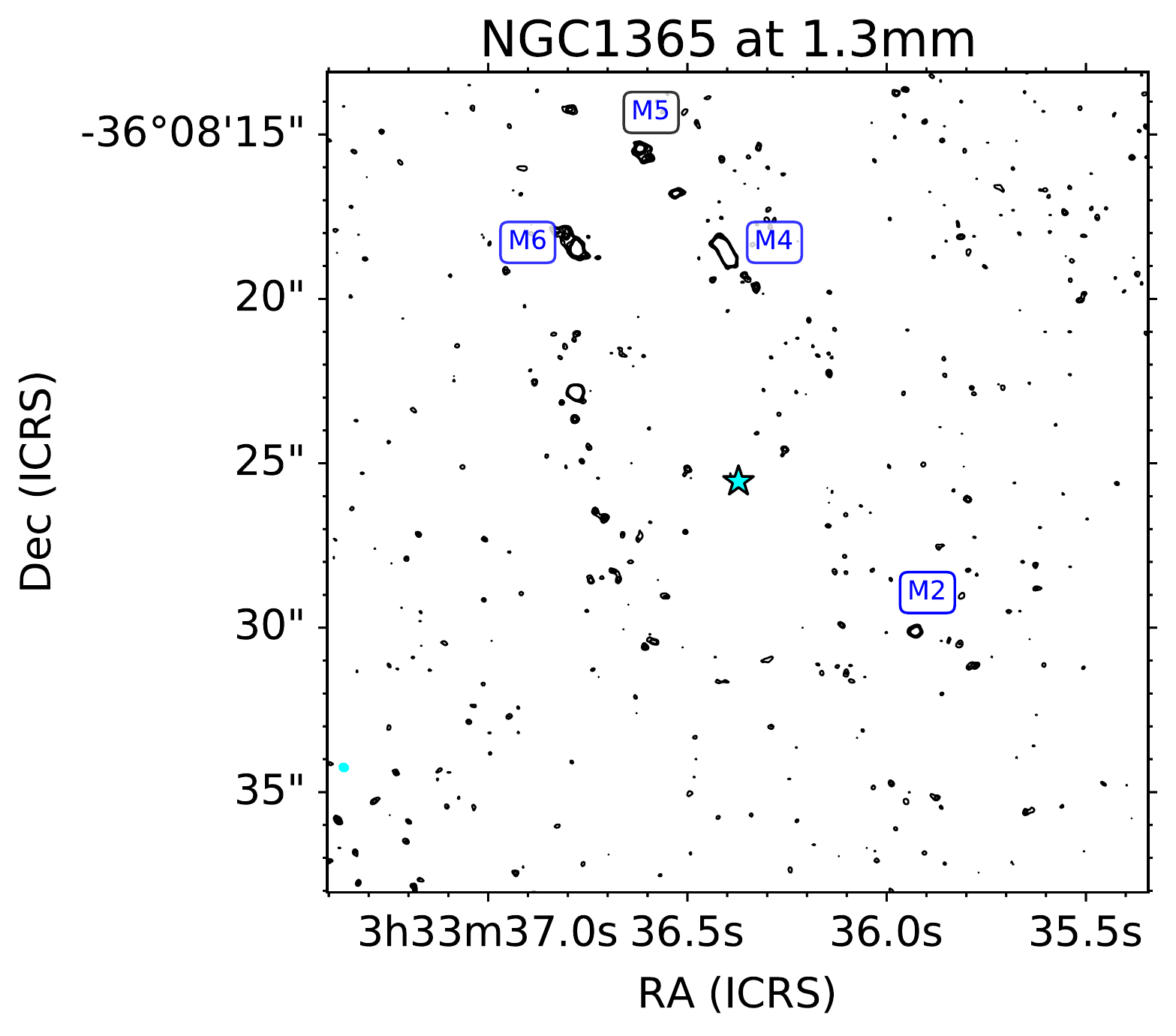}
  \includegraphics[width=8cm]{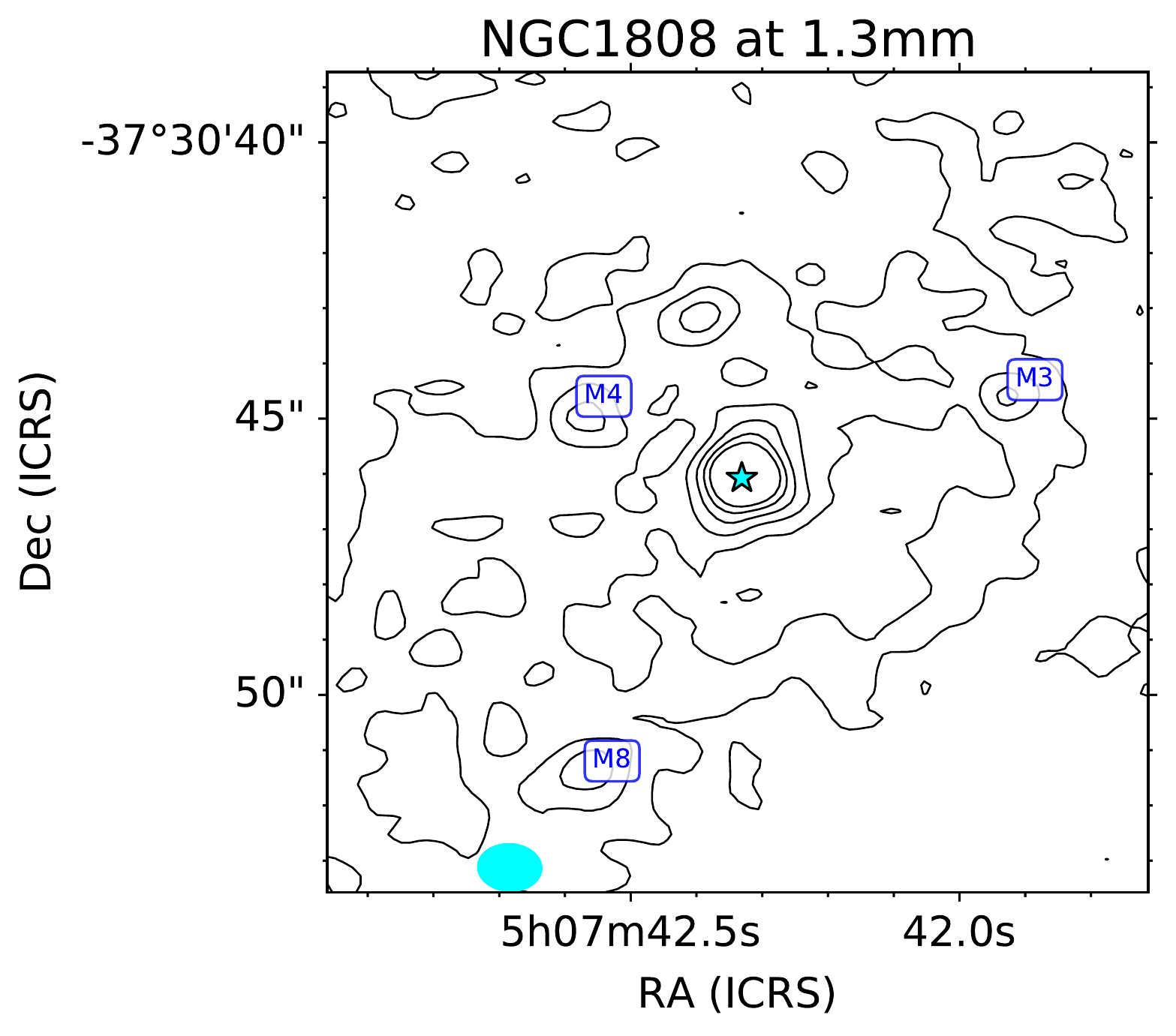}

  \caption{Large-scale ALMA 1.3\,mm continuum emission maps of NGC~1365
    with a field of view of $25'' \times 25''$ (upper panel) and NGC~1808 with a field of view of $15''\times
    15''$ (lower panel). The contours are shown in a linear scale. The filled cyan ellipses
    represent the synthesized beam sizes and orientations. 
    We show the tentative identification of
    some mid-IR emitting sources detected by
  \cite{Galliano2005}.}\label{fig:contNGC1365NGC1808}
    \end{figure}

NGC~1365 --- This Seyfert 1.8 galaxy has a bright circumnuclear
ring of SF in the central $20\arcsec$ ($\simeq 2\,$kpc).
Figure~\ref{fig:contNGC1365NGC1808} shows the 1.3\,mm continuum map with a field of view of $25\arcsec \times
25\arcsec$ where several sources are identified
with mid-IR  sources 
\citep{Galliano2005, AlonsoHerrero2012}. However, at the AGN location, which is located approximately
$0.4\arcsec$ west and $7.1\arcsec$ south of the M4 source \citep{AlonsoHerrero2012},
the 1.3\,mm emission is faint. The coordinates of the 1.3\,mm continuum peak (see
Table~\ref{tab:1p3mm}) agree  with those  measured from the
ALMA $850\,\mu$m continuum peak \citep{Combes2019}. In the
central $4\arcsec \times 4\arcsec \simeq 400\,{\rm pc}
\times 400\,{\rm pc}$ there is faint 1.3\,mm emission at the AGN
position, although it does not coincide with any of the central peaks
detected in the CO(2-1) integrated emission map.

  \begin{table*}
\caption{ALMA 1.3\,mm continuum observations.}             
\label{tab:1p3mm}      
\centering                          
\begin{tabular}{c c r r r}        
\hline\hline                 

Galaxy    & Beam             &  PA$_{\rm beam}$  & RA (ICRS) & Dec (ICRS)\\
          & ($\arcsec \times \arcsec$) & (deg) \\
\hline
IC~4518W  & $0.22\times 0.19$ & $-$88 & 14:57:41.18 & $-$43:07:55.49\\   
NGC~1365  & $0.24\times 0.22$ & 68    & 03:33:36.38 & $-$36:08:25.52\\
NGC~1386  & $0.61\times 0.43$ & $-$85 & 03:36:46.19 & $-$35:59:57.14\\
NGC~1808  & $1.08 \times 0.81$ & 89   & 05:07:42.33 & $-$37:30:46.05\\
NGC~3081  & $0.55\times 0.47$ & $-$84 & 09:59:29.54 & $-$22:49:34.80\\
NGC~3227  & $0.20\times 0.15$ & 48    & 10:23:30.57 & 19:51:54.27 \\
NGC~5135  & $0.27\times0.21$  & 64    & 13:25:43.99 & $-$29:50:00.03\\
NGC~5643  & $0.16\times 0.10$ & $-$67 & 14:32:40.70 & $-$44:10:27.90 \\
NGC~7130  & $0.30\times 0.25$ & 62    & 21:48:19.52 & $-$34:57:04.76\\
NGC~7172  & $0.51\times 0.41$ & 78   & 22:02:01.89 & $-$31:52:10.49\\
NGC~7213  & $0.62\times0.58$ & 84     & 22:09:16.21 & $-$47:10:00.14\\
NGC~7469  & $0.23\times0.17$ & $-$41  & 23:03:15.62 &  08:52:26.05 \\
NGC~7582  & $0.17\times0.16$ & $-$1   & 23:18:23.64 & $-$42:22:13.55\\

\hline
\end{tabular}
\tablefoot{Beam and PA$_{\rm beam}$ are the synthesized beam sizes and
  position angles of the ALMA 1.3\,mm continuum observations,
  respectively. The listed coordinates are assumed to correspond to the
  AGN position.}

\end{table*}

\begin{figure*}
   \centering
  \includegraphics[width=6cm]{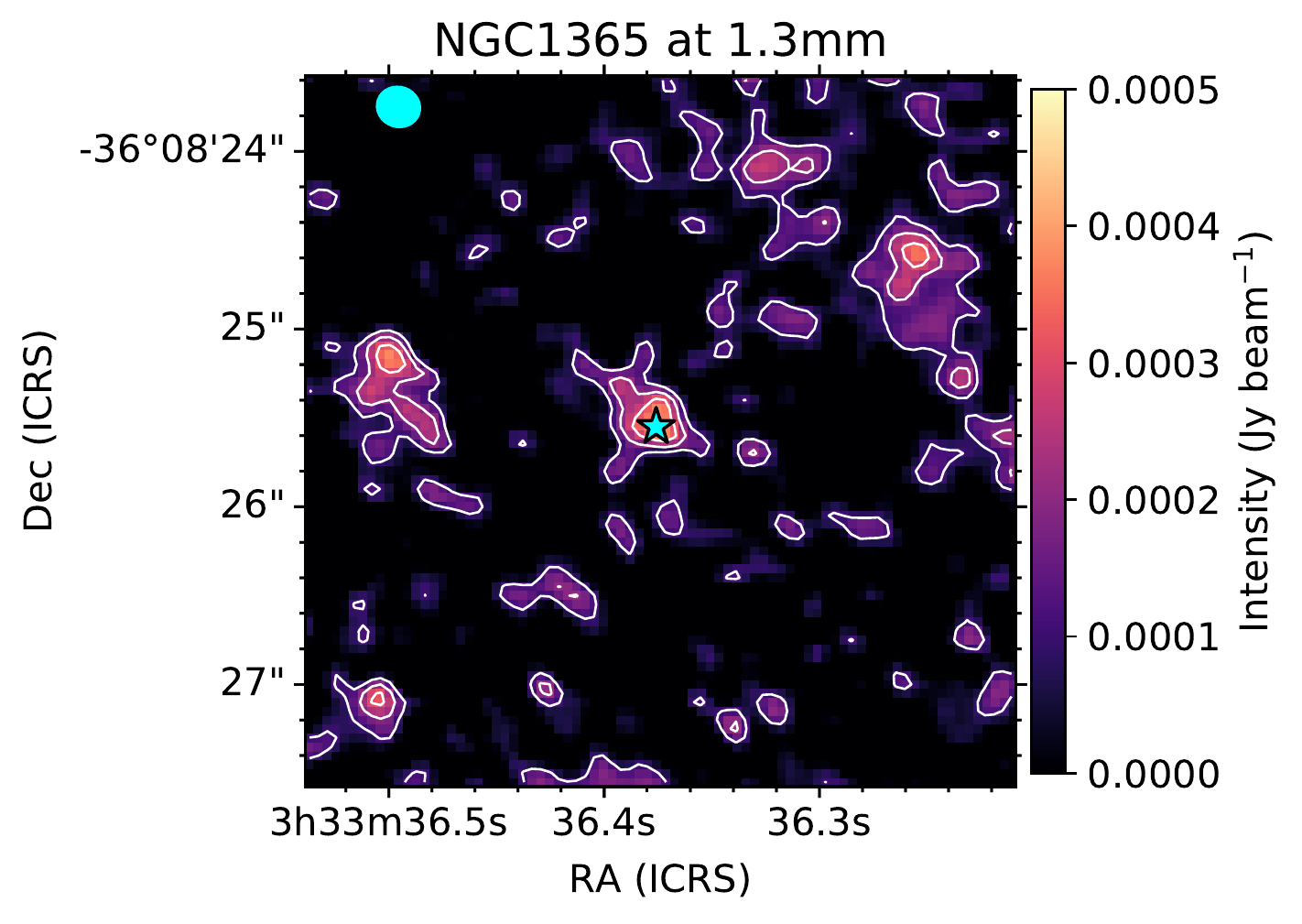}
  \includegraphics[width=6cm]{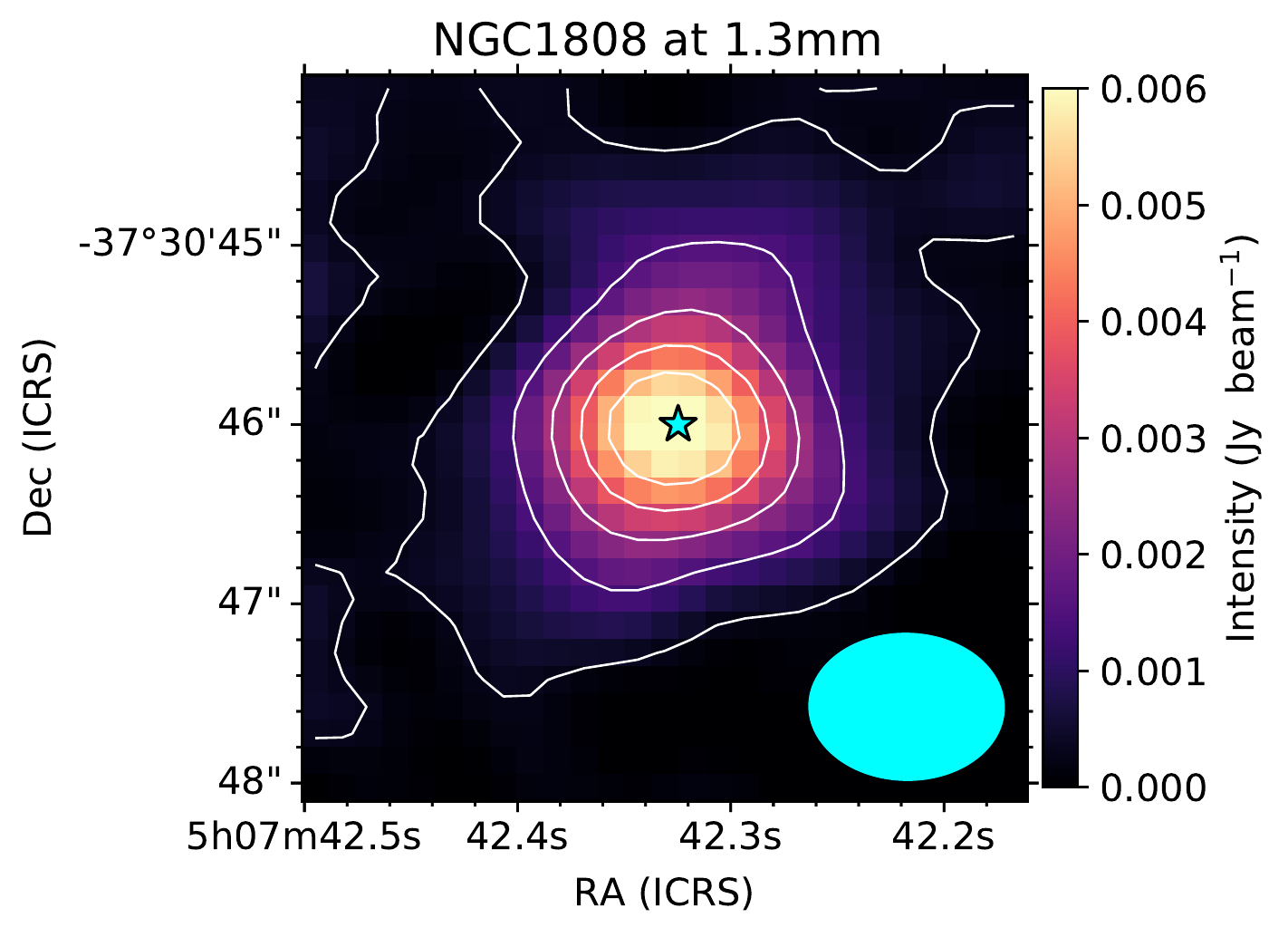}
  \includegraphics[width=6cm]{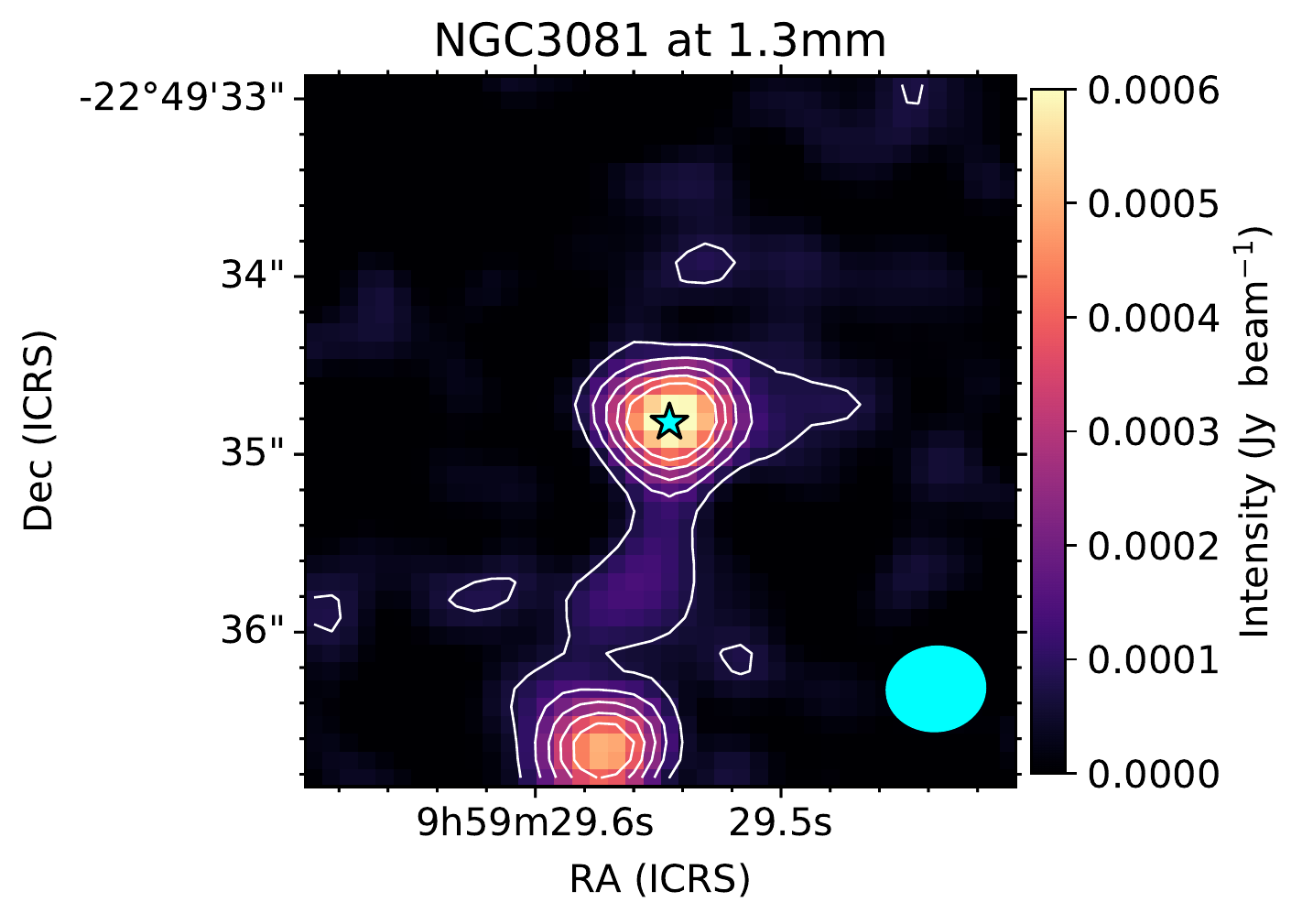}

  \includegraphics[width=6cm]{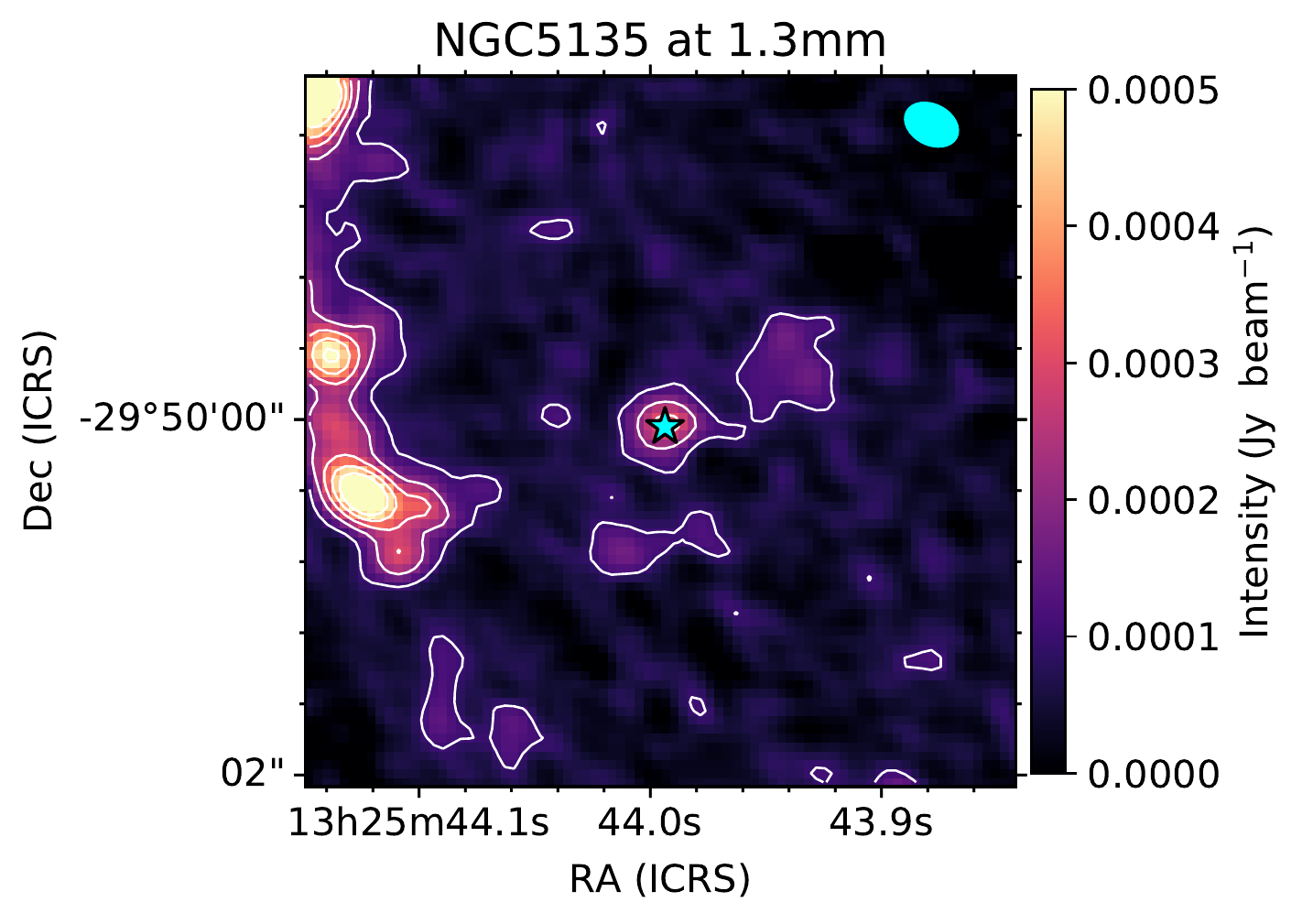}
  \includegraphics[width=6cm]{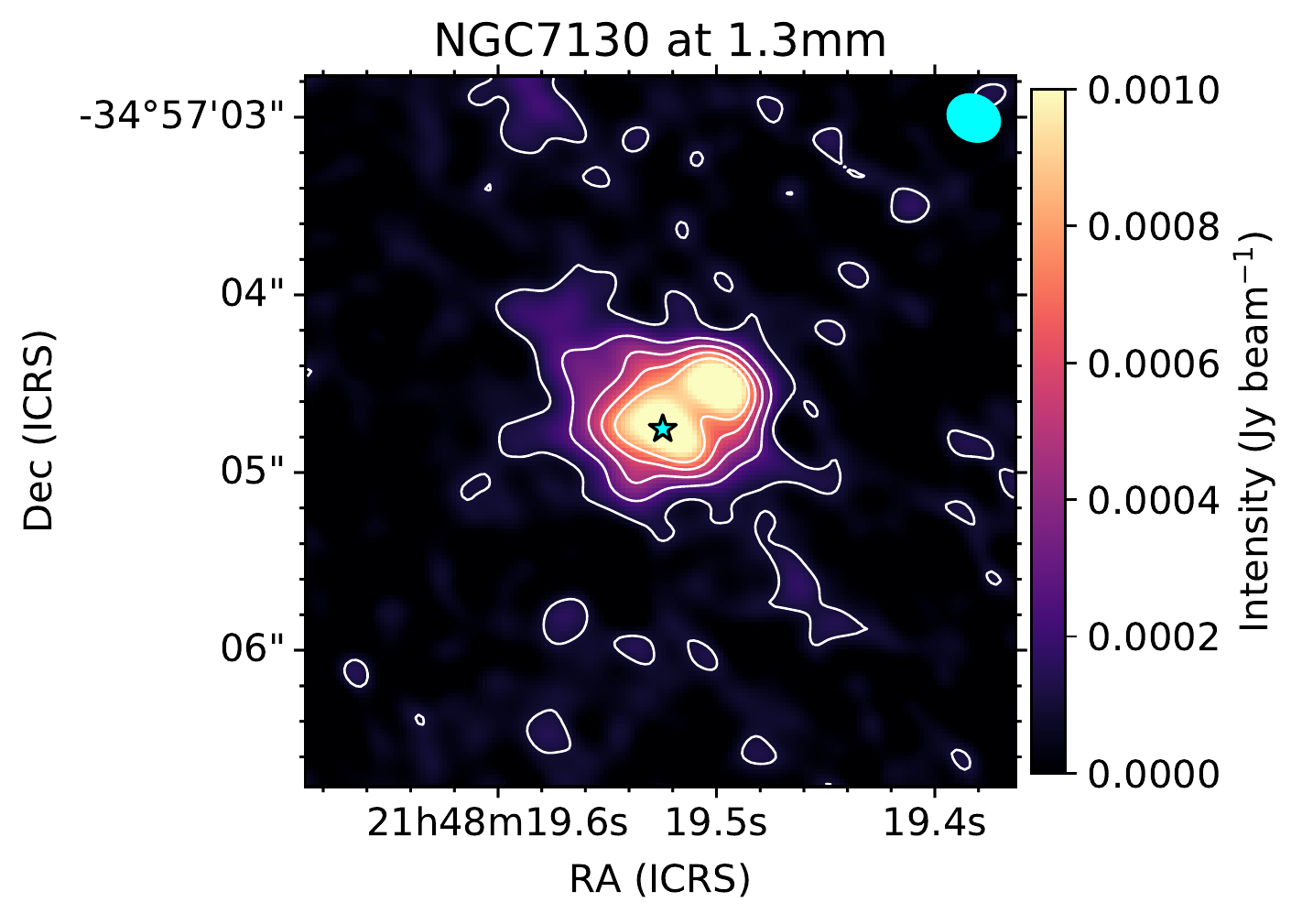}
  \includegraphics[width=6cm]{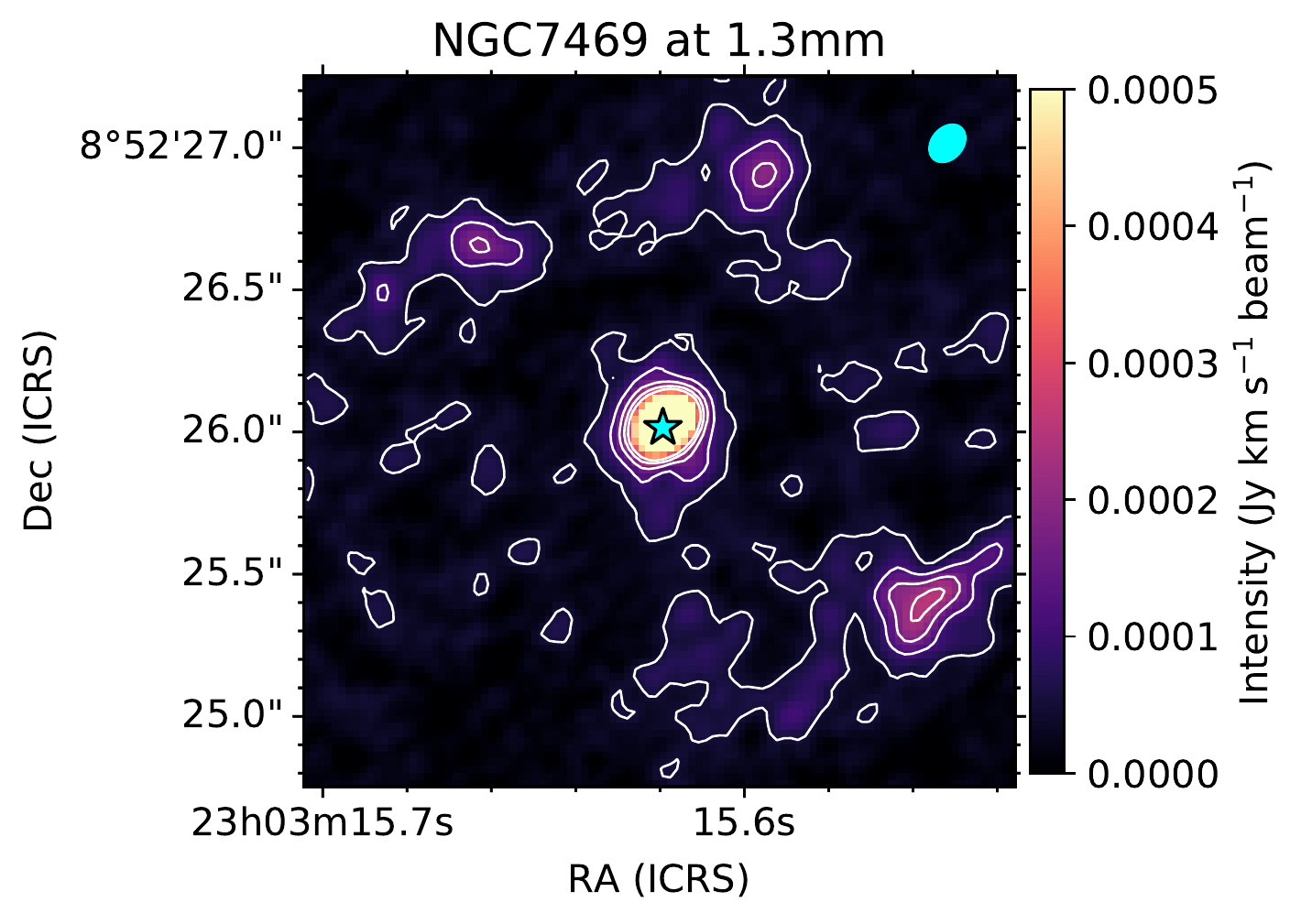}
  
  \includegraphics[width=6cm]{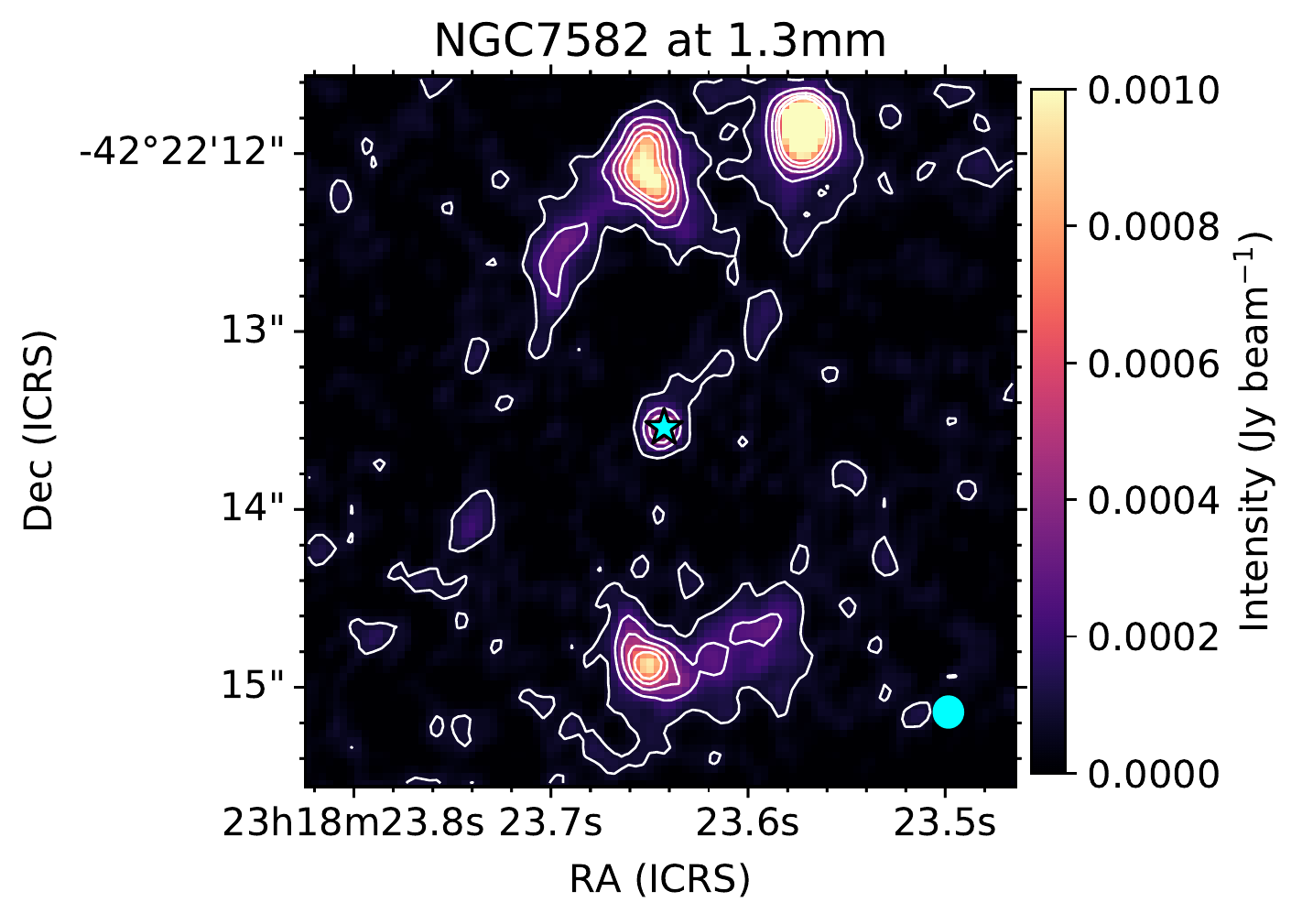}
  \caption{ALMA maps of the 1.3\,mm continuum emission with
    a field of view of $4''\times4''$ as in Figure~\ref{fig:CO21maps} 
    for the targets
    with several continuum sources that are discussed in this appendix. The contours are shown
    in a linear scale. The filled star marks the AGN
  position, and the cyan filled ellipses are the synthesized beams.}\label{fig:cont4x4}
    \end{figure*}

  NGC~1808 --- This galaxy is a
  hot-spot galaxy \citep[see, e.g.,][]{Galliano2005}
  classified as a low-luminosity Seyfert galaxy.  
The 1.3\,mm continuum continuum sources in the central
$15\arcsec\simeq 1\,$kpc (see Figure~\ref{fig:contNGC1365NGC1808})
are associated with the AGN and other hot-spot sources 
detected in the ALMA 100-115\,GHz continuum map \citep{Salak2016}
and in the mid-IR \citep{Galliano2005}. The brightest 1.3\,mm source corresponds to the
AGN position. In the central
$4\arcsec \times 4\arcsec \simeq 270\,{\rm pc} \times 270\,{\rm pc}$
region  (see Figure~\ref{fig:cont4x4}), the only  1.3\,mm source is the AGN, and
its coordinates (Table~\ref{tab:1p3mm}) agree
well with those derived by \cite{Combes2019}.
The peak of the 1.3\,mm continuum  does not
coincide with the peak of the CO(2-1) emission at the resolution we worked with.

NGC~3081 --- The 1.3\,mm continuum map shows two bright peaks of similar flux
(see Figure~\ref{fig:cont4x4}) in the central
$4\arcsec \times 4\arcsec \simeq 715\,{\rm pc} \times 715\,{\rm pc}$.
The coordinates of the northern source (Table~\ref{tab:1p3mm})
are approximately coincident with the 8.4\,GHz radio continuum source
\citep{Mundell2009} and the kinematical center of
the CO(2-1) velocity field \citep[see][]{Ramakrishnan2019}. Thus, we  assume
that it corresponds to the AGN. The 1.3\,mm southern source is
located approximately $2\arcsec$ 
from the AGN. None of the  two 1.3\,mm peaks coincides with the CO(2-1) peaks
in the central $4\arcsec \times 4\arcsec$
(Figures~\ref{fig:CO21maps} and \ref{fig:cont4x4}).

NGC~5135 --- This Seyfert 2 galaxy is also classified as a luminous
infrared galaxy (LIRG) and shows strong SF
activity in the central $10\arcsec \simeq 2.9\,$kpc. There is
a bright unresolved source, which is identified as the
AGN position, surrounded by several star clusters
\citep{AlonsoHerrero2006}. The ALMA
band 6 observations used in this work were discussed in detail by
\cite{Sabatini2018}. The authors identified source A  (see coordinates in
Table~\ref{tab:1p3mm}) as the  AGN
position. In the
CO(2-1) map the AGN location does not coincide with any of the bright
peaks.

NGC~7130 --- This is a  Seyfert 2 galaxy classified as a LIRG with
a nuclear (inner $1-2\arcsec \simeq 300-600\,$pc) starburst  \citep{GonzalezDelgado1998,  
AlonsoHerrero2006, DiazSantos2010}. 
At 1.3\,mm and 434\,$\mu$m \citep[see][]{Zhao2016}
there are two sources separated  by approximately $0.4\arcsec$ (120\,pc).
The northwestern source is the brightest. However, neither coincides with the radio emission peak
at 2.3\,GHz. The radio peak is closer to the southeastern source (see
Table~\ref{tab:1p3mm}). We therefore
assume that it coincides approximately with the AGN position.

NGC~7469 --- This galaxy is classified as a Seyfert 1.5 and also shows a
bright circumnuclear SF ring  in the central
$4\arcsec \times 4\arcsec \simeq 1180\,{\rm pc} \times 1180\,{\rm pc}$.
The 1.3\,mm continuum map shows a bright nuclear source associated with
the AGN as well as several  sources  (see Figure~\ref{fig:cont4x4}) associated with the
SF ring \citep{Colina2001, DiazSantos2007}. The continuum peak does not coincide with the
nuclear CO(2-1) regions that appear to have the morphology of a mini-spiral
(see Figure~\ref{fig:CO21maps}).

NGC~7582 --- This Seyfert  galaxy shows broad lines in  near-infrared
hydrogen recombination lines. It presents a  circumnuclear ring of
SF with an approximate radius of $2\arcsec$ with bright emitting regions
\citep{Wold2006, Riffel2009}.
The AGN is detected at 1.3\,mm. Over the 
central $4\arcsec \times 4\arcsec \sim 350\,{\rm pc} \times 350\,{\rm pc}$,
the continuum and CO(2-1) regions in the  ring are bright
(see Figure~\ref{fig:cont4x4}). At the AGN location, we do not detect
a compact CO(2-1) emitting region, although there is significant emission there
 (see Section~\ref{subsec:phot}).

\end{document}